\pdfoutput=1

\documentclass[aps,reprint,amsmath,amssymb,graphicx,letterpaper,nofootinbib,numbers,longbibliography,floatfix]{revtex4-1}

\usepackage{amsmath}
\usepackage{amsfonts}
\usepackage{amssymb}
\usepackage{mathrsfs}
\usepackage{graphicx}
\usepackage{color}
\usepackage{longtable}
\usepackage{multirow}
\usepackage{eurosym}
\usepackage[dvipsnames]{xcolor}
\usepackage{pstricks}
\usepackage{fancyhdr}
\usepackage{titlesec}
\usepackage{enumitem}
\usepackage{tikz}
\usepackage{float}
\usepackage{booktabs}
\usepackage{natbib}
\usepackage[breaklinks]{hyperref}
\usepackage[
  top=2cm,
  bottom=3cm,
  left=1.5cm,
  right=1.5cm,
  headsep=10pt,
  footskip=40pt,
  asymmetric,
]{geometry}

\newcommand{\mnras}{MNRAS}
\newcommand{\apjl}{ApJ}

\newcommand{\aap}{SAP}
\newcommand{\jcap}{JCAP}

\def\pasa{pA\&A}

\newcommand{\ie}{{\it i.e.}}

\newcommand{\eg}{{\it e.g.}}

\newcommand{\etc}{{\it etc.}}

\newcommand{\fig}{Fig.}
\newcommand{\Fig}{Fig.}

\newcommand{\Ref}{Ref.}
\newcommand{\Refs}{Refs.}

\newcommand{\figu}[1]{\fig~\ref{fig:#1}}

\newcommand{\bi}{\begin{itemize}}
\newcommand{\ei}{\end{itemize}}

\newcommand{\HA}{{\sc HorizonAntenna}}

\newcommand{\mybox}[4]{
    \vspace*{-0.4cm}
    \begin{figure}[H]
        \centering
    \begin{tikzpicture}
        \node[anchor=text,text width=\columnwidth-0.7cm, draw, rounded corners, line width=1pt, fill=#3, inner sep=3mm] (big) {\\#4};
        \node[draw, rounded corners, line width=.5pt, fill=#2, anchor=west, xshift=3mm] (small) at (big.north west) {#1};
    \end{tikzpicture}
    \end{figure}
}

\definecolor{grand_red}{RGB}{154,61,38}
\definecolor{grand_orange}{RGB}{252,145,39}
\definecolor{grand_brown}{RGB}{168,127,106}

\titleformat{\section}{\normalfont\fontsize{13}{15}\bfseries}{\thesection}{1em}{}
\titleformat{\subsection}{\normalfont\fontsize{12}{15}\bfseries}{\thesubsection}{1em}{}
\titleformat{\subsubsection}{\normalfont\fontsize{10}{15}\bfseries}{\thesubsubsection}{1em}{}
  
\fancypagestyle{bib}{
 \fancyhead[L]{References}
 \fancyhead[R]{}
}

\fancypagestyle{exsummary}{
 \fancyhead[L]{}
 \fancyhead[R]{}
}

\fancypagestyle{cover}{
 \fancyhead[L]{}
 \fancyhead[R]{}
 \fancyfoot[L]{}
 \fancyfoot[R]{}
 \fancyfoot[C]{}
}

\begin{document}


\pagestyle{plain}
{
\begin{figure}[p]
 \vspace*{-2.01cm}
 \hspace*{4.72cm}  
 \makebox[\linewidth]{
  \includegraphics[width=1.165\textwidth]{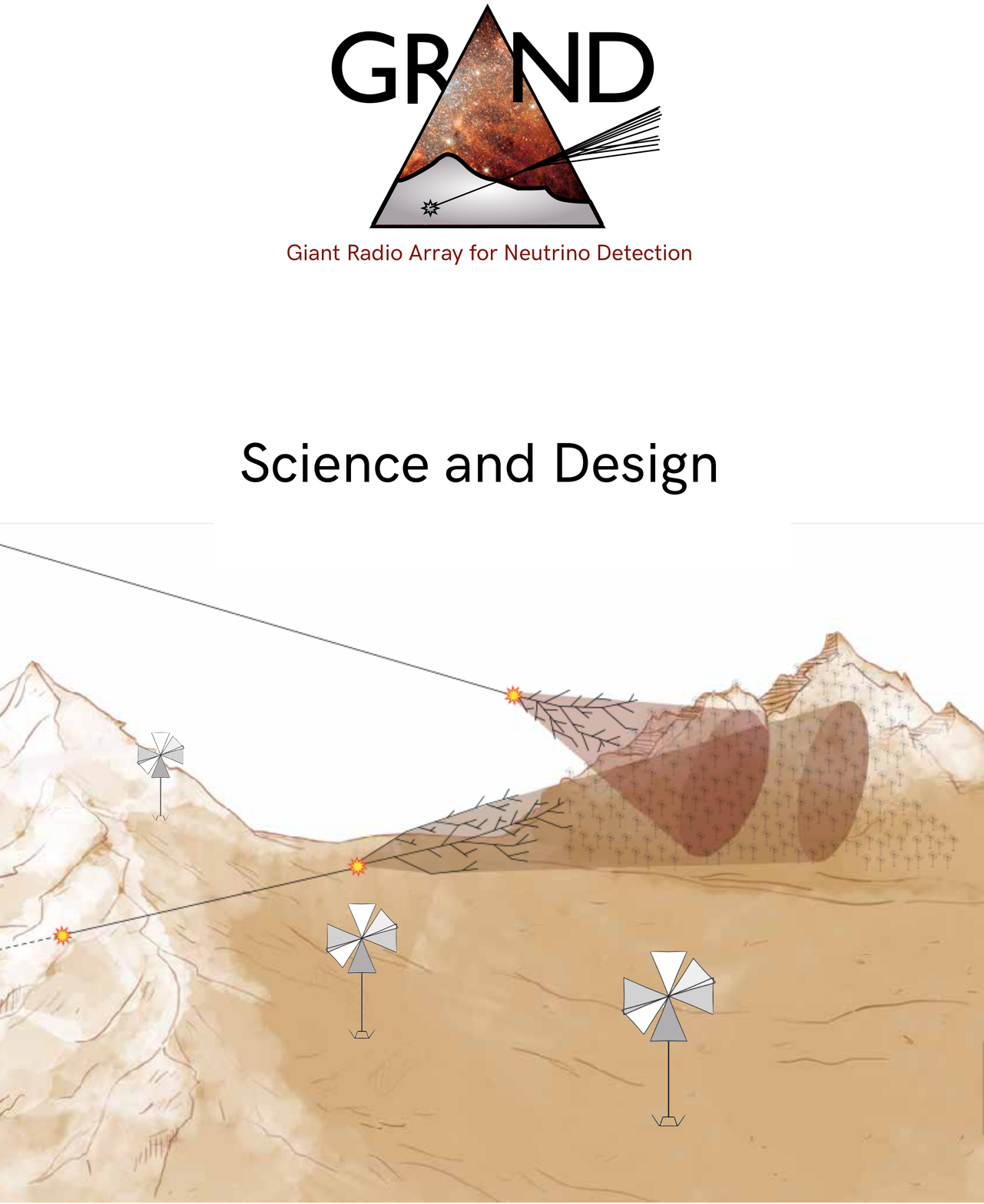}
 }
\end{figure}
}

\clearpage
\newpage

\onecolumngrid


\pagestyle{fancy}
\thispagestyle{cover}

\vspace*{8cm}

\begin{flushleft}
 \fontsize{120}{30}\selectfont
 Giant Radio Array for Neutrino Detection:\\
 \bigskip
 Science and Design
\end{flushleft}

\vspace*{11.3cm}

\begin{flushright}
 \large
 July 2019 (v2.3)
\end{flushright}

\clearpage
\newpage


\pagestyle{fancy}

\section*{Executive summary}
\label{section:ExecutiveSummary}
\addcontentsline{toc}{section}{Executive summary}

The {\bf Giant Radio Array for Neutrino Detection (GRAND)}\footnote{\href{http://grand.cnrs.fr}{http://grand.cnrs.fr}} is a planned large-scale observatory of ultra-high-energy (UHE) cosmic particles --- cosmic rays, gamma rays, and neutrinos with energies exceeding $10^8$~GeV.  Its ultimate goal is to solve the long-standing mystery of the origin of UHE cosmic rays.  It will do so by detecting an unprecedented number of UHECRs and by looking with unmatched sensitivity for the undiscovered UHE neutrinos and gamma rays associated to them.  Three key features of GRAND will make this possible: its large exposure at ultra-high energies, sub-degree angular resolution, and sensitivity to the unique signals made by UHE neutrinos.

\smallskip

The strategy of GRAND is to detect the radio emission coming from large particle showers that develop in the terrestrial atmosphere --- {\it extensive air showers} --- as a result of the interaction of UHE cosmic rays, gamma, rays, and neutrinos.  To achieve this, GRAND will be the largest array of radio antennas ever built.  The relative affordability of radio antennas makes the scale of construction possible.  GRAND will build on years of progress in the field of radio-detection and apply the large body of technological, theoretical, and numerical advances, for the first time, to the radio-detection of air showers initiated by UHE neutrinos.

\smallskip

The design of GRAND will be modular, consisting of several independent sub-arrays, each of 10\,000 radio antennas deployed over 10\,000 km$^2$ in radio-quiet locations.  A staged construction plan ensures that key techniques are progressively validated, while simultaneously achieving important science goals in UHECR physics, radioastronomy, and cosmology early during construction.  

\smallskip

Already by 2025, using the first sub-array of 10\,000 antennas, GRAND could discover the long-sought cosmogenic neutrinos --- produced by interactions of ultra-high-energy cosmic-rays with cosmic photon fields --- if their flux is as high as presently allowed, by reaching a sensitivity comparable to planned upgraded versions of existing experiments.  By the 2030s, in its final configuration of 20 sub-arrays, GRAND will reach an unparalleled sensitivity to cosmogenic neutrino fluxes of $4 \cdot 10^{-10}$~GeV~cm$^{-2}$~s$^{-1}$~sr$^{-1}$ within 3 years of operation, which will guarantee their detection even if their flux is tiny. Because of its sub-degree angular resolution, GRAND will also search for point sources of UHE neutrinos, steady and transient, potentially starting UHE neutrino astronomy.  Because of its access to ultra-high energies, GRAND will chart fundamental neutrino physics at these energies for the first time.

\smallskip

GRAND will also be the largest detector of UHE cosmic rays and gamma rays.  It will improve UHECR statistics at the highest energies ten-fold within a few years, and either discover UHE gamma rays or improve their limits ten-fold. Further, it will be a valuable tool in radioastronomy and cosmology, allowing for the discovery and follow-up of large numbers of radio transients --- fast radio bursts, giant radio pulses --- and for precise studies of the epoch of reionization. 

\smallskip

Following the discovery of high-energy astrophysical neutrinos, gravitational waves, and the multi-wavelength, multi-messenger detection of neutron-star mergers, we stand today at the threshold of a new era in astroparticle physics.  Several exciting high-energy astroparticle experiments are planned, both extensions of existing cosmic-ray and neutrino experiments --- AugerPrime, TA$\times$4, IceCube-Gen2 --- and new experiments --- LSST, CTA, LISA.  At the ultra-high-energy front, GRAND completes the picture.

\medskip

In this document, we present the science goals, detection strategy, preliminary design, performance goals, and construction plans for GRAND.


\clearpage
\newpage


\section*{Author list}

\vspace{-0.4cm}

\begin{center}
Jaime \'Alvarez-Mu\~niz$^1$,
Rafael Alves Batista$^{2,3}$,
Aswathi Balagopal V.$^4$,
Julien Bolmont$^5$,
Mauricio Bustamante$^{6,7,8,\dagger}$, \\
Washington Carvalho Jr.$^9$,
Didier Charrier$^{10}$,
Isma\"el Cognard$^{11,12}$,
Valentin Decoene$^{13}$,
Peter B.\ Denton$^6$, \\
Sijbrand De Jong$^{14,15}$,
Krijn D.\ De Vries$^{16}$,
Ralph Engel$^{17}$,
Ke Fang$^{18,19,20}$,
Chad Finley$^{21,22}$,
Stefano Gabici$^{23}$, \\
QuanBu Gou$^{24}$,
Junhua Gu$^{25}$,
Claire Gu\'epin$^{13}$,
Hongbo Hu$^{24}$,
Yan Huang$^{25}$,
Kumiko Kotera$^{13,\ast}$,
Sandra Le Coz$^{25}$, \\
Jean-Philippe Lenain$^{5}$,
Guoliang L\"u$^{26}$,
Olivier Martineau-Huynh$^{5,25,\ast}$,
Miguel Mostaf\'a$^{27,28,29}$,
Fabrice Mottez$^{30}$, \\
Kohta Murase$^{27,28,29}$,
Valentin Niess$^{31}$,
Foteini Oikonomou$^{32,27,28,29}$,
Tanguy Pierog$^{17}$,
Xiangli Qian$^{33}$,
Bo Qin$^{25}$, \\
Duan Ran$^{25}$,
Nicolas Renault-Tinacci$^{13}$,
Markus Roth$^{17}$,
Frank G.\ Schr\"oder$^{34,17}$, 
Fabian Sch\"ussler$^{35}$, 
Cyril Tasse$^{36}$, \\
Charles Timmermans$^{14,15}$,
Mat\'ias Tueros$^{37}$,
Xiangping Wu$^{38,25,\ast}$,
Philippe Zarka$^{39}$,
Andreas Zech$^{30}$, \\
B.\ Theodore Zhang$^{40,41}$,
Jianli Zhang$^{25}$,
Yi Zhang$^{24}$,
Qian Zheng$^{42,24}$,
Anne Zilles$^{13}$
\end{center}

\vspace*{-0.3cm}

\begin{center}
\small
$^1$Departamento de F\'isica de Part\'iculas \& Instituto Galego de F\'\i sica de Altas Enerx\'\i as, \\Universidad de Santiago de Compostela, 15782 Santiago de Compostela, Spain \\
$^2$Instituto de Astronomia, Geof\'isica e Ci\^{e}ncias Atmosf\'ericas, Universidade de S\~{a}o Paulo -- Rua do Mat\~{a}o, 1226, 05508-090,\\ S\~{a}o Paulo-SP, Brazil\\
$^3$Department of Physics -- Astrophysics, University of Oxford, DWB, Keble Road, OX1 3RH, Oxford, UK \\
$^4$Institute of Experimental Particle Physics (ETP), Karlsruhe Institute of Technology (KIT), \\
D-76021 Karlsruhe, Germany \\
$^5$Sorbonne Universit\'e, Universit\'e Paris Diderot, Sorbonne Paris Cit\'e, CNRS,\\
Laboratoire de Physique Nucl\'eaire et de Hautes Energies (LPNHE), 4 place Jussieu, F-75252, Paris Cedex 5, France \\
$^6$Niels Bohr International Academy and DARK, Niels Bohr Institute, 2100 Copenhagen, Denmark \\
$^7$Discovery Center, Niels Bohr Institute, 2100 Copenhagen, Denmark \\
$^8$Center for Cosmology and AstroParticle Physics (CCAPP) \& Dep.\ of Physics, Ohio State University, Columbus, OH 43210, USA \\
$^9$Universidade de Santiago de Compostela, 15782 Santiago de Compostela, Spain \\
$^{10}$SUBATECH, Institut Mines-Telecom Atlantique -- CNRS/IN2P3 -- Universit\'e de Nantes, Nantes, France \\
$^{11}$Laboratoire de Physique et Chimie de l'Environnement et de l'Espace LPC2E CNRS-Universit\'e d'Orl\'eans,\\
F-45071 Orl\'eans, France \\
$^{12}$Station de Radioastronomie de Nan\c{c}ay, Observatoire de Paris, CNRS/INSU F-18330 Nan\c{c}ay, France \\
$^{13}$Sorbonne Universit\'e, UMR 7095, Institut d'Astrophysique de Paris, 98 bis bd Arago, 75014 Paris, France \\
$^{14}$Institute for Mathematics, Astrophysics and Particle Physics (IMAPP), Radboud Universiteit, Nijmegen, Netherlands \\
$^{15}$Nationaal Instituut voor Kernfysica en Hoge Energie Fysica (Nikhef), Netherlands \\
$^{16}$IIHE/ELEM, Vrije Universiteit Brussel, Pleinlaan 2, 1050 Brussels, Belgium \\
$^{17}$Institute for Nuclear Physics (IKP), Karlsruhe Institute of Technology (KIT), D-76021 Karlsruhe, Germany \\
$^{18}$Einstein Fellow, Stanford University, Stanford, CA 94305, USA \\
$^{19}$Department of Astronomy, University of Maryland, College Park, MD 20742-2421, USA \\
$^{20}$Joint Space-Science Institute, College Park, MD 20742-2421, USA \\
$^{21}$Oskar Klein Centre, Stockholm University, SE-10691 Stockholm, Sweden \\
$^{22}$Department of Physics, Stockholm University, SE-10691 Stockholm, Sweden \\
$^{23}$AstroParticule et Cosmologie (APC), Univ.\ Paris Diderot, CNRS/IN2P3, CEA/Irfu, Obs.\ de Paris, Sorbonne Paris Cit\'e, France \\
$^{24}$Institute of High Energy Physics, Chinese Academy of Sciences, 19B YuquanLu, Beijing 100049, China \\
$^{25}$National Astronomical Observatories, Chinese Academy of Sciences, Beijing 100012, China \\
$^{26}$School of Physical Science and Technology, Xinjiang University, Urumqi, 830046, China \\
$^{27}$Center for Particle and Gravitational Astrophysics, Pennsylvania State University, University Park, PA 16802, USA \\
$^{28}$Department of Physics, Pennsylvania State University, University Park, PA 16802, USA \\
$^{29}$Department of Astronomy \& Astrophysics, Pennsylvania State University, University Park, PA 16802, USA \\
$^{30}$LUTH, Obs.\ de Paris, CNRS, Universit\'e Paris Diderot, PSL Research University, 5 place Jules Janssen, 92190 Meudon, France \\
$^{31}$Clermont Universit\'e, Universit\'e Blaise Pascal, CNRS/IN2P3, Lab.\ de Physique Corpusculaire, 63000 Clermond-Ferrand, France \\
$^{32}$European Southern Observatory, Karl-Schwarzschild-Str.\ 2, D-85748 Garching bei M\"unchen, Germany \\
$^{33}$Department of Mechanical and Electrical Engineering, Shandong Management University, Jinan 250357, China \\
$^{34}$Bartol Research Institute, Department of Physics and Astronomy, University of Delaware, Newark, DE 19716, USA \\
$^{35}$IRFU, CEA, Universit\'e Paris-Saclay, F-91191 Gif-sur-Yvette, France \\
$^{36}$GEPI, Observatoire de Paris, CNRS, Universit\'e Paris Diderot, 5 place Jules Janssen, F-92190 Meudon, France \\
$^{37}$Instituto de F\'isica La Plata, CONICET, Boulevard 120 y 63 (1900), La Plata, Argentina \\
$^{38}$Shanghai Astronomical Observatory, Chinese Academy of Sciences, 80 Nandan Road, Shanghai 200030, China \\
$^{39}$LESIA, Observatoire de Paris, CNRS, PSL/SU/UPD/SPC, Place J. Janssen, 92195 Meudon, France \\
$^{40}$Department of Astronomy, School of Physics, Peking University, Beijing 100871, China \\
$^{41}$Kavli Institute for Astronomy and Astrophysics, Peking University, Beijing 100871, China \\
$^{42}$School of Engineering and Computer Science, PO Box 600, Victoria University of Wellington, Wellington 6140, New Zealand
\end{center}

\vspace*{0.1cm}

\noindent
$\ast$: Spokesperson / $\dagger$: Corresponding author (mbustamante@nbi.ku.dk)

\clearpage
\newpage


\twocolumngrid

\tableofcontents{}

\clearpage
\newpage



\section{Introduction}
\label{section:PhysicsCase}

Ultra-high-energy cosmic rays (UHECRs) --- extraterrestrial charged particles with energies of EeV $\equiv 10^{18}$~eV and above --- have been observed for more than fifty years, yet their origin is unknown \cite{Kotera2011}.  They are likely extragalactic in origin and purportedly made in powerful cosmic accelerators, though none has been identified.  As long as the sources of UHECRs remain undiscovered, our picture of the high-energy Universe will be incomplete.  

The direct strategy to discover UHECR sources is to look for localized excesses in the distribution of arrival directions of detected UHECRs.
Yet, identifying sources in this way is challenging: our incomplete knowledge of the properties of UHECRs --- notably, their mass composition --- and of the effect of Galactic and intergalactic magnetic fields on their propagation prevents us from precisely retracing their trajectories back to their sources.  The situation is worse at the highest energies because of decreasing statistics.  This is partially due to the opaqueness of the Universe to UHECRs: their interaction on the cosmic microwave background (CMB) dampens their energy.  As a result, few UHECRs above 40~EeV reach the Earth from distances beyond 100~Mpc --- the GZK horizon\ \cite{Greisen:1966jv, Zatsepin:1966jv}.  Thus, not only do individual UHECRs not point back at their sources, but limited statistics at the highest energies hinder studies of their properties and sources.

The indirect strategy is to look for EeV gamma rays and neutrinos made by UHECRs.  
Unaffected by cosmic magnetic fields, they point back at their sources.  Though still undetected, their existence is guaranteed: {\it cosmogenic} EeV gamma rays and neutrinos should be produced in the same interactions on the CMB responsible for the opaqueness to UHECRs, and their fluxes echo the properties of UHECRs and their sources.  UHE gamma rays and neutrinos may also be produced by UHECRs interacting inside their sources: detecting a directional excess in their number would be the smoking gun of a UHECR source. 

Yet, like for UHECRs, interactions with the CMB make the Universe opaque to UHE gamma rays: they do not reach Earth from beyond 10~Mpc.  Instead, they cascade down to GeV--TeV, where they are more difficult to disentangle from gamma rays produced in unrelated phenomena.

Only for UHE neutrinos is the Universe transparent: they travel unimpeded, their energies unaffected by interactions, on trajectories that point back directly at their points of production, even if they lie beyond the GZK horizon.  Thus, they could reveal the most energetic and distant UHECR sources.  However, predictions of the cosmogenic neutrino flux are uncertain and allow for tiny fluxes.  Thus, the assured discovery of cosmogenic neutrinos is contingent on the existence of a detector with exquisite sensitivity.

The Giant Radio Array for Neutrino Detection (GRAND) is a proposed large-scale observatory designed to discover and study the sources of UHECRs.  It will combine the direct and indirect strategies, by collecting unprecedented UHECR statistics and looking for UHE gamma rays and neutrinos, with sensitivity to even pessimistic predictions of their cosmogenic fluxes and angular resolution sufficient to discover point sources.  

Upon arriving at Earth, UHE cosmic rays, gamma rays, and neutrinos initiate large particle showers in the atmosphere --- {\it extensive air showers}.  Their propagation through the geomagnetic field results in radio emission that can be detected far from the shower, since it undergoes little attenuation in the atmosphere.
In GRAND, a large number of antennas will autonomously detect the ground footprint of the radio emission, tens of km in size, in the 50--200~MHz band.  To achieve this, the design of GRAND will be modular, with up to 20 independent and separate arrays, each made up of 10\,000 radio antennas deployed over 10\,000~km$^2$.  The large number of antennas will allow to collect large cosmic-rays statistics, reach sensitivity to low fluxes of neutrinos and gamma rays, and achieve high pointing accuracy.

\begin{figure}[tb]
 \centering
 \includegraphics[width=0.8\columnwidth]{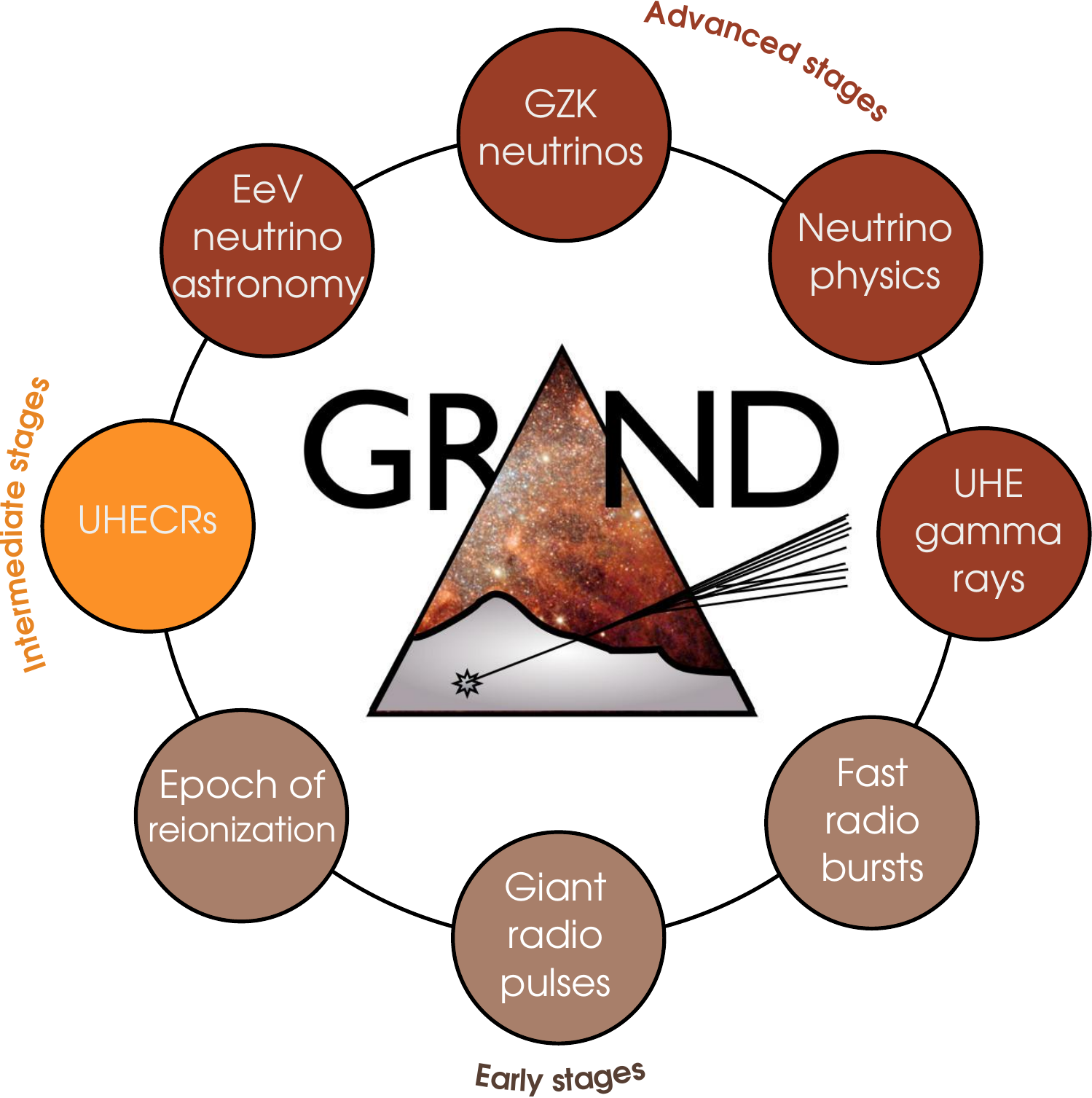}
 \caption{The science goals of GRAND, grouped according to the detector construction stage at which they first become accessible}
 \label{fig:science_diagram}
\end{figure}

Figure \ref{fig:science_diagram} shows the science goals of GRAND, including also studies in cosmology and radioastronomy.  Some of them will be achievable already in early and intermediate construction stages.  Later, we examine each one in detail.  

It is timely to plan and build GRAND now, following a stream of discoveries in neutrino physics at progressively higher energies, culminating in the recent discovery of PeV astrophysical neutrinos\ \cite{Aartsen:2013bka,Aartsen:2013jdh,Aartsen:2014gkd,Aartsen:2015rwa,Aartsen:2016xlq}.  Further, the multi-messenger detection of neutron-star merger GW170814\ \cite{GBM:2017lvd} has shown that addressing the challenges of high-energy astronomy will require combining observations from different experiments.  In the multi-messenger era, UHE neutrinos will be the key to testing the absolute highest energies.

Sections \ref{section:uhe_messengers} and \ref{section:cosmology_radioastronomy} present the GRAND science goals.  Sections \ref{section:detector_design} and \ref{section:construction_stages} detail the detection strategy, design, and construction plans.  Section \ref{section:Summary} summarizes and concludes.


\section{Ultra-high-energy messengers}
\label{section:uhe_messengers}

\mybox{{\bf At a glance}}{Tan!70}{grand_brown!20}
{
 \begin{itemize}[leftmargin=*]
  \item
   While propagating, UHECRs make UHE neutrinos and gamma rays of energies of $10^9$~GeV and higher
  \item
   Detecting UHE neutrinos is the best way to probe the high-energy end of the UHECR spectrum and the most distant UHECR sources
  \item
   GRAND will detect the radio signals made at Earth by UHE cosmic rays, and by UHE gamma rays and neutrinos made by them, even if their flux is tiny
  \item
   With sub-degree angular resolution, GRAND could discover the first sources of UHE neutrinos
  \item
   Detecting UHE neutrinos will open up a new regime for fundamental neutrino physics
  \item
   GRAND will collect large statistics of UHECRs
  \item
   GRAND could discover the first UHE gamma rays even if UHECRs are heavy nuclei
 \end{itemize}
}

\begin{figure}[!t]
 \centering
 \includegraphics[width=\columnwidth]{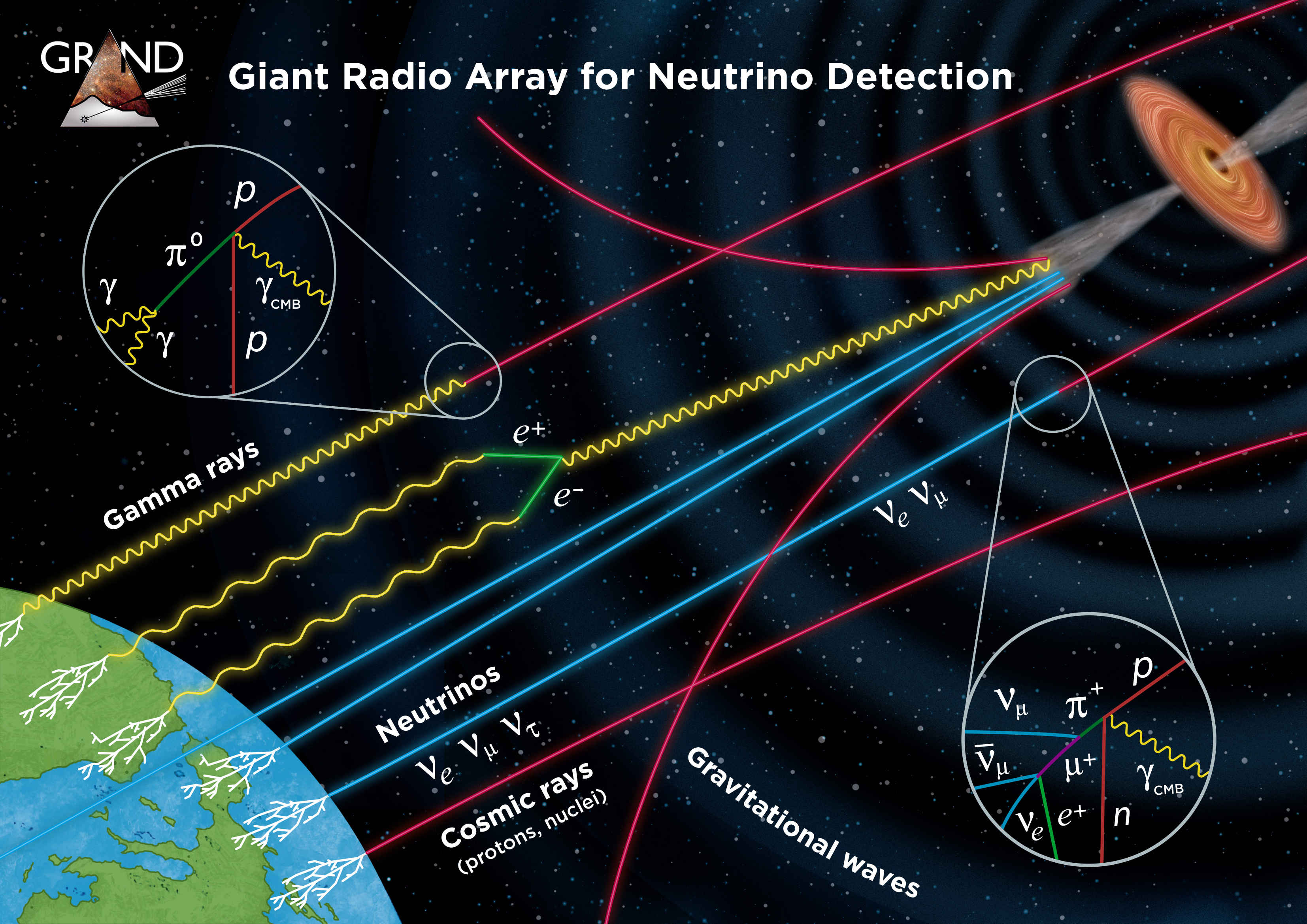}
 \caption{\label{fig:grand_propagation}Propagation of ultra-high-energy (UHE) cosmic rays from the astrophysical sources to the Earth.  Via interactions on cosmic photo backgrounds, cosmic rays create UHE gamma rays --- which cascade down in energy --- and UHE neutrinos --- which oscillate during propagation.  At Earth, all three UHE messengers may induce extensive air showers in the atmosphere.}
\end{figure}

Figure\ \ref{fig:grand_propagation} sketches the propagation of UHECRs --- and associated secondary particles --- from their sources to Earth.  A complete picture of UHECR sources will come from jointly studying cosmic rays, neutrinos, and gamma rays across all available energies. 

\begin{figure}[t!]
 \centering
 \includegraphics[trim = 0.2cm 0.5cm 1.6cm 1.0cm, clip=true, width=\columnwidth]{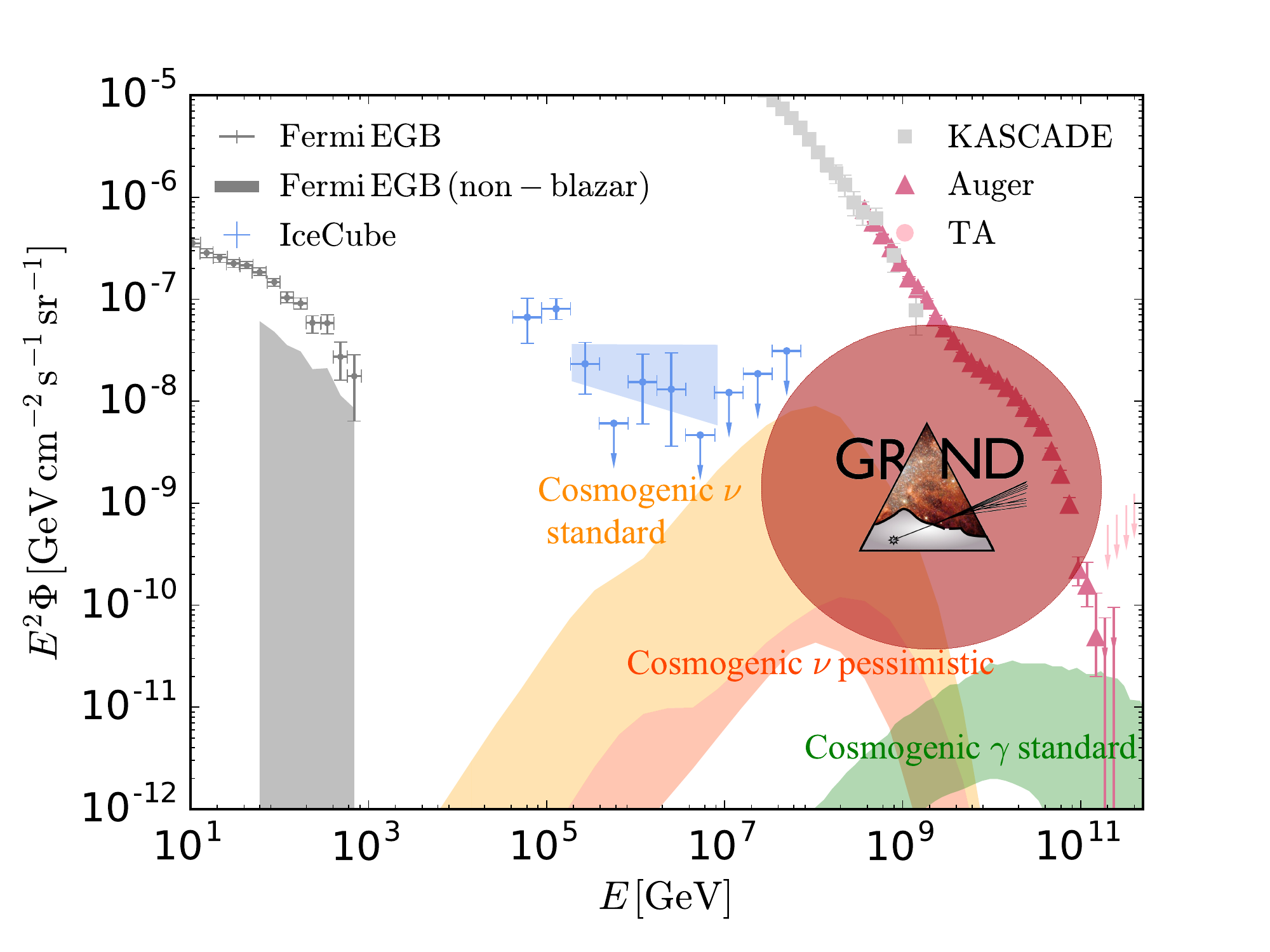}
 \caption{GRAND target zone overlaid on the energy spectra of astrophysical and cosmogenic messengers.  For gamma rays, we show the extragalactic gamma-ray background (EGB) measured by {\it Fermi}-LAT\ \cite{Ackermann:2014usa, TheFermi-LAT:2015ykq} and, shaded, the contribution to the EGB due to unresolved, non-blazar sources.  For neutrinos, we show the all-flavor 6-year measurements by IceCube of High Energy Starting Events (HESE)\ \cite{Aartsen:2017mau} and through-going muons\ \cite{Aartsen:2016xlq}.  For cosmic rays, we show measurements by KASCADE-Grande\ \cite{Apel:2013uni}, Auger\ \cite{Aab:2015bza}, and the Telescope Array (TA), including the Telescope Array Low Energy (TALE) extension\ \cite{Jui:2016amg}.
 We show the predicted conservative and standard ranges of cosmogenic neutrino fluxes from \Ref\ \cite{AlvesBatista:2018zui} (see the main text for details) and the predicted range of cosmogenic gamma rays from \Ref\ \cite{Decerprit:2011qe} bracketing light to pure-iron UHECR models.}
 \label{fig:spectrum_all}
\end{figure}

Figure \ref{fig:spectrum_all} shows the ranges of energy and flux that GRAND targets, overlaid on the observed spectra of cosmic rays, gamma rays, and astrophysical neutrinos.  It also shows the spread in predicted fluxes of the undiscovered cosmogenic gamma rays\ \cite{Decerprit:2011qe} and cosmogenic neutrinos\ \cite{AlvesBatista:2018zui}.  UHECR unknowns broaden the spread of predictions.

Below, we refer to intermediate construction stages of GRAND: the 300-antenna array GRANDProto300; the 10\,000-antenna array GRAND10k; and the 200\,000-antenna array GRAND200k, made up of several replicas of GRAND10k built at separate geographical locations.  See Section\ \ref{section:construction_stages} for details on the construction stages.

\subsection{Ultra-high-energy neutrinos}
\label{section:uhe_neutrinos}

\subsubsection{Cosmogenic neutrinos}

\begin{figure}[t!]
 \centering
 \includegraphics[width=\columnwidth]{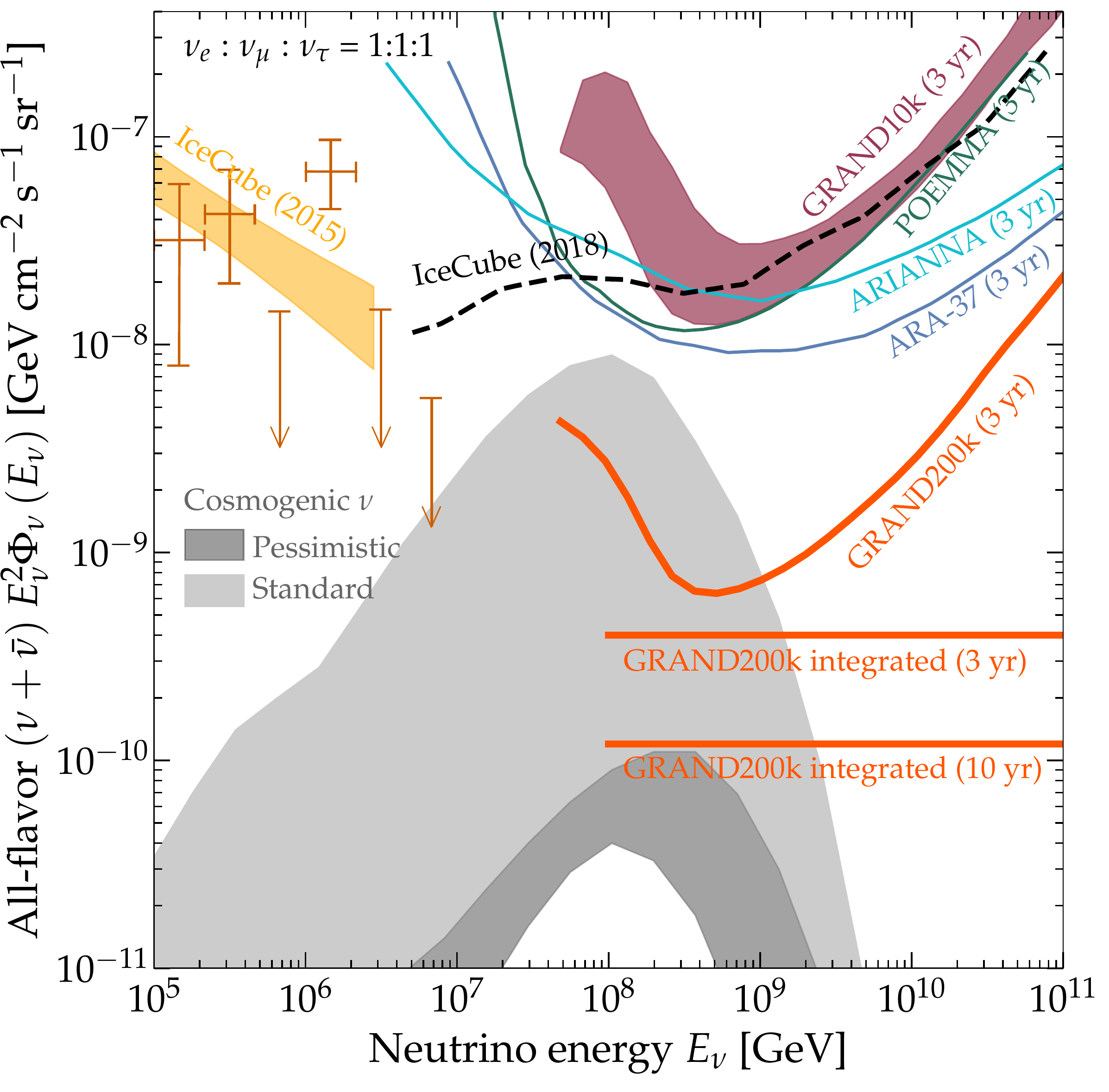}
 \caption{Predicted cosmogenic neutrino flux, compared to experimental upper limits and sensitivities.  Gray-shaded regions are generated by fitting UHECR simulations to Auger spectral and mass-composition data\ \cite{AlvesBatista:2018zui}.  See the main text for details.  The astrophysical neutrino signal below 3~PeV was reported by IceCube\ \cite{Aartsen:2015knd}.  We show the most restrictive upper limit (90\%~C.L.), from IceCube\ \cite{Aartsen:2018vtx}; limits from Auger\ \cite{Aab:2015kma} and ANITA\ \cite{Allison:2018cxu} are less restrictive.  Projected 3-year sensitivities of planned instruments are for ARA-37\ \cite{Allison:2015eky} (trigger level), ARIANNA\ \cite{PhDPersichilli} (``optimal wind'' sensitivity), POEMMA\ \cite{POEMMA_UHECR2018:Talk} (assuming full-sky coverage), the 10\,000-antenna array GRAND10k, and the full 200\,000-antenna array GRAND200k.  The GRAND10k band is spanned by the choice of antenna detection voltage threshold, from a conservative threshold at the top of the band to an aggressive one at the bottom of it; see Section\ \ref{section:performances} for details.}
 \label{fig:cosmo_neut}
\end{figure}

A diffuse flux of cosmogenic neutrinos\ \cite{Beresinsky:1969qj} is \emph{guaranteed} to exist, produced in interactions of UHECRs with cosmic background photon fields such as the cosmic microwave background (CMB) and the extragalactic background light (EBL).  The neutrino flux depends on properties of UHECRs and their sources: the distribution of sources with redshift, the neutrino source emissivity, the injected UHECR spectrum --- typically assumed to be a power law suppressed above a maximum energy --- and the mass composition of the injected cosmic rays.  Because these parameters are uncertainly known, there is a large spread in the predictions of the cosmogenic neutrino flux.  

Because cosmogenic neutrinos are produced in photo-pion interactions of UHECRs, the relative number of $\nu + \bar{\nu}$ of different flavors is $(N_{\nu_e}:N_{\nu_\mu}:N_{\nu_\tau}) = 1:2:0$.  This holds to within 1\%, regardless of whether the mass composition of UHECRs is light or heavy\ \cite{Moller:2018isk}.  Neutrinos oscillations re-distribute the flavors, so that at Earth the relative number of all flavors should be about the same, \ie, $1:1:1$.  Even if oscillation parameters are allowed to vary within uncertainties, $\nu_\tau$ make up no less than 15\% of the flux\ \cite{Bustamante:2015waa}.

Figure \ref{fig:cosmo_neut} shows the range of cosmogenic neutrino flux predictions, at 90\% confidence level, resulting from fitting the UHECR spectrum and mass composition simulated with CRPropa\ \cite{Batista:2016yrx} to those measured by the Pierre Auger Observatory\ \cite{Valino:2015zdi,Aab:2014kda}, as computed in \Ref\ \cite{AlvesBatista:2018zui}.  The predictions assume that the sources are uniformly distributed up to redshift $z=6$ and that they emit UHECRs with the same luminosity and spectra.  The prominent bump around $10^{8}$~GeV is due to photohadronic interactions of the most energetic UHECRs with the peak of the CMB spectrum.  In \Ref\ \cite{AlvesBatista:2018zui}, the redshift integration conservatively stopped at $z=1$, but we have continued it to $z=6$, which is why the cosmogenic neutrino fluxes shown in \figu{cosmo_neut} are larger.

The conservative range in \figu{cosmo_neut} is obtained using a generic form for the evolution of the source emissivity $\propto (1 + z)^m$; the fit favors negative source evolution, \ie, $m < 0$\ \cite{Taylor:2015rla}.  The standard range in \figu{cosmo_neut} is spanned by the fluxes generated with all other choices of source emissivity: star formation rate, gamma-ray bursts (GRB), and active galactic nuclei (AGN).  The simulations neglect the effect of extragalactic magnetic fields and inhomogeneities in the source distribution, which, in reality, would increase the flux past the standard range.

Figure \ref{fig:cosmo_neut} shows that the estimated differential sensitivity of GRAND200k covers a large part of the standard flux range. 
GRAND10k, ARA-37, ARIANNA, and POEMMA have comparable sensitivities, of about $10^{-8}$~GeV~cm$^{-2}$~s$^{-1}$~sr$^{-1}$.  Other experiments in planning, not shown, could reach sensitivities of this order.  These include Ashra-NTA\ \cite{Sasaki:2017zwd}, Trinity\ \cite{Otte:2018uxj}, and CTA\ \cite{Gora:2016mmy} --- that look for Cherenkov and fluorescence light emission by tau-initiated showers --- and TAROGE\ \cite{Liu:2018hux} and BEACON\ \cite{BEACON_TeVPA2017:Talk} --- that look for radio emission from elevated sites.  However, only GRAND200k is planned with the goal of reaching sensitivities of a few times $10^{-10}$~GeV~cm$^{-2}$~s$^{-1}$~sr$^{-1}$.

GRAND could potentially detect a large number of cosmogenic neutrinos.  Within the standard range of fluxes, the projected event rate in the $10^8$--$10^{11}$~GeV neutrino energy range is 1--18 events per year in GRAND200k versus less than one event per year in the planned full-sized configurations of the ARA-37 and ARIANNA in-ice radio detectors.  For the latter two, rates were calculated using the effective area from \Ref\ \cite{Allison:2015eky} for ARA-37 and the effective volume from \Ref\ \cite{Barwick:2014pca} for ARIANNA, re-scaled to 300 stations. 

Detecting a large number of cosmogenic neutrinos can indirectly measure the energy at which the cosmic-ray spectrum cuts off, contingent on the neutrino energy resolution, which is currently under study; see Section\ \ref{section:GRANDDesignPerf-Performance-Reconstruction}.  This 
would accurately locate the peak energy of cosmogenic neutrinos, overcome most parameter degeneracies, and constrain the major cosmic-ray injection and source properties.  Conversely, if the flux of cosmogenic neutrinos is low, it would not add to the background for searches of UHE neutrinos produced directly at the source environment, easing their detection; we explore this below.

Similarly, searches for UHE neutrinos constrain the fraction of UHE protons arriving at Earth could be constrained by observing EeV neutrinos\ \cite{vanVliet:2017obm, vanVliet:2019nse}.  With a projected sensitivity of about $10^{-10}$~GeV~cm$^{-2}$~s$^{-1}$~sr$^{-1}$, GRAND would constrain the proton fraction to less than 30\% for $m=-6$, and to 5\% for $m \geq 0$, thereby providing a complementary approach to infer the composition of UHECRs without being dominated by uncertainties inherent to the modeling of hadronic interactions in the atmosphere.


\subsubsection{Neutrinos from astrophysical sources}

\paragraph{Diffuse neutrino fluxes} 

\smallskip

\begin{figure}[t!]
 \centering
 \includegraphics[width=\columnwidth, trim = 0 0 0 1.6cm, clip=true]{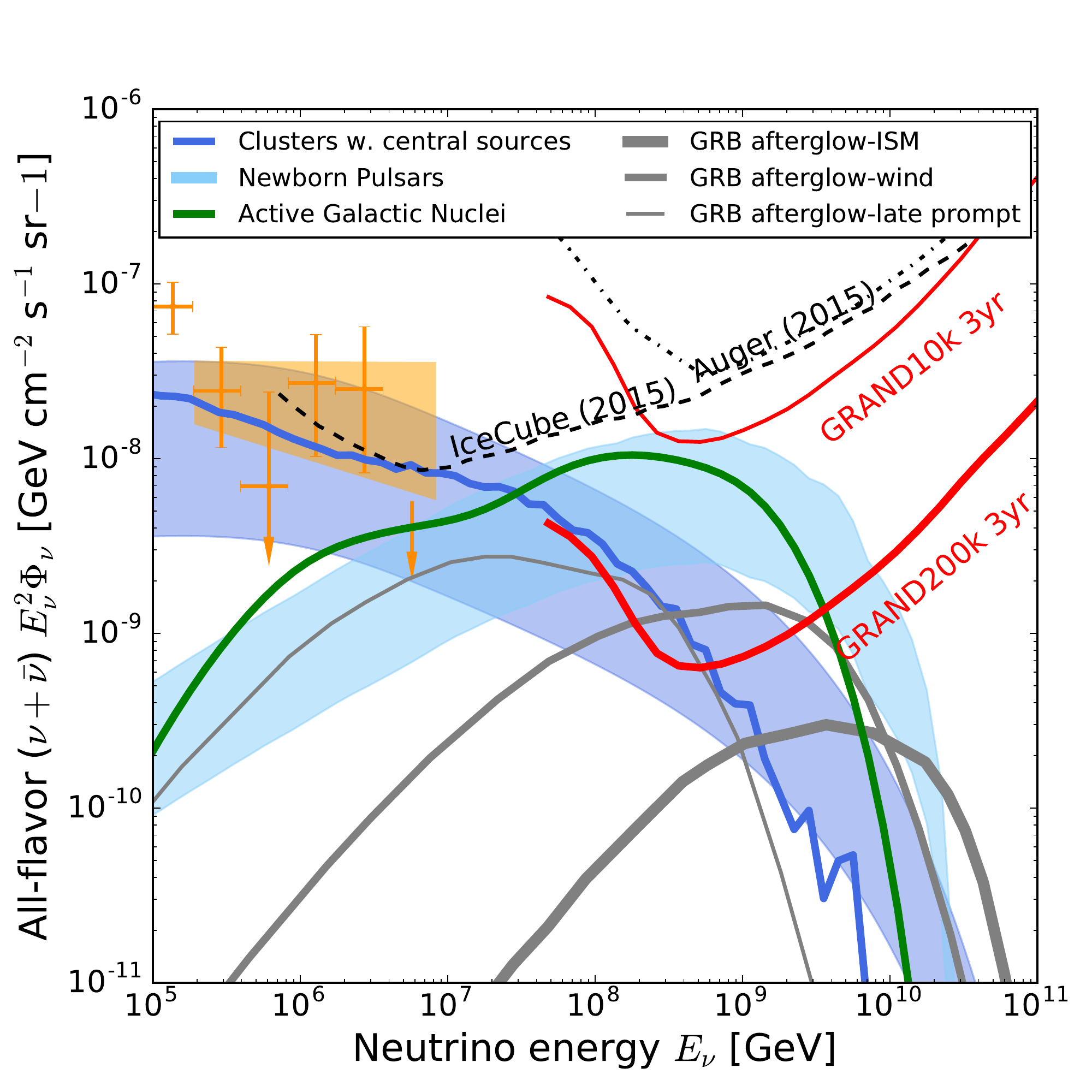}
 \caption{\label{fig:flux}Predicted neutrino flux from different classes of astrophysical sources, compared to upper limits on UHE neutrinos from IceCube\ \cite{Aartsen:2016ngq} and Auger\ \cite{Aab:2015kma}, and projected 3-year sensitivity of GRAND10k and GRAND200k (Sections \ref{section:GRANDStages-GRAND10k} and \ref{section:GRANDStages-GRAND200k}).  Several source classes can account for the observed UHECR spectrum: galaxy clusters with central sources\ \cite{2008ApJ...689L.105M, 2017arXiv170400015F}, fast-spinning newborn pulsars\ \cite{2014PhRvD..90j3005F}, active galactic nuclei\ \cite{2015arXiv151101590M}, and afterglows of gamma-ray bursts\ \cite{Murase:2007yt}.}
\end{figure}

EeV neutrinos are not only produced when UHECRs interact with extragalactic background photons during propagation from their sources to the Earth, but also when UHECRs interact with photons and hadrons inside the sources themselves.  

\begin{figure}[t!]
 \centering
 \includegraphics[width=\columnwidth, trim = 5.0cm 3.2cm 4.0cm 3.4cm, clip=true]{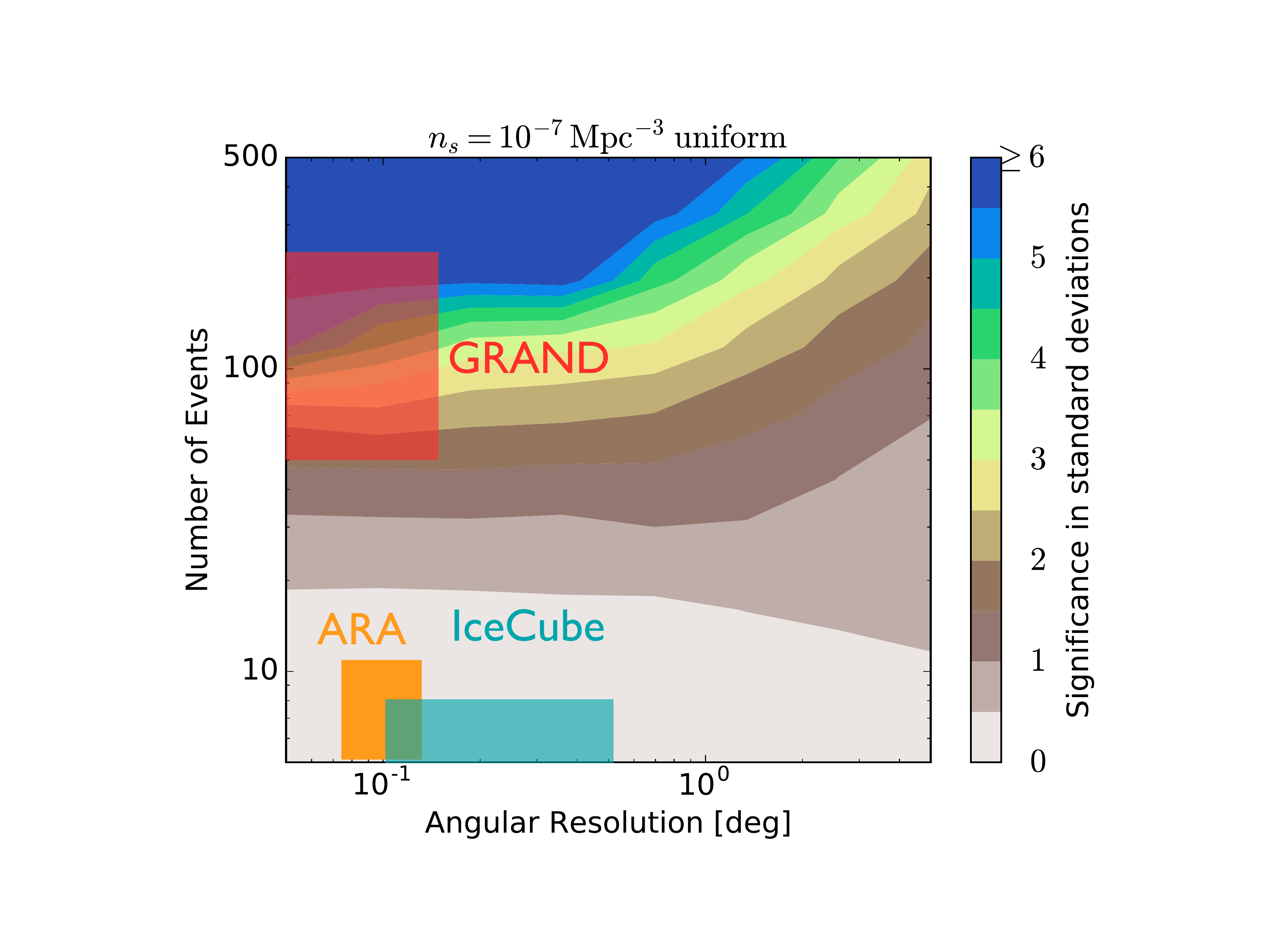}
 \caption{\label{fig:sourceSearch}Significance of detection of point sources of UHE neutrinos by experiments with various angular resolutions and numbers of detected events.  The source density is assumed to be $n_s = 10^{-7}\,\rm Mpc^{-3}$ up to 2~Gpc.  Each shaded box represents uncertainties in the source spectrum and detector angular resolution.  Exposure times are 15 years for IceCube, 3 years for ARA, and 3 years for GRAND.  Figure adapted from \Ref\ \cite{Fang:2016hop}.}
\end{figure}

Because different classes of astrophysical sources would produce UHE neutrinos on-site on different time scales and under different production conditions, the integrated neutrino fluxes from different source classes may have different spectra.  Thus, the diffuse neutrino spectrum contains important information about the dominant source class.

\begin{figure*}[t!]
 \centering
 \includegraphics[width=\columnwidth, trim = 1.6cm 2.5cm 1cm 2.5cm, clip=true]{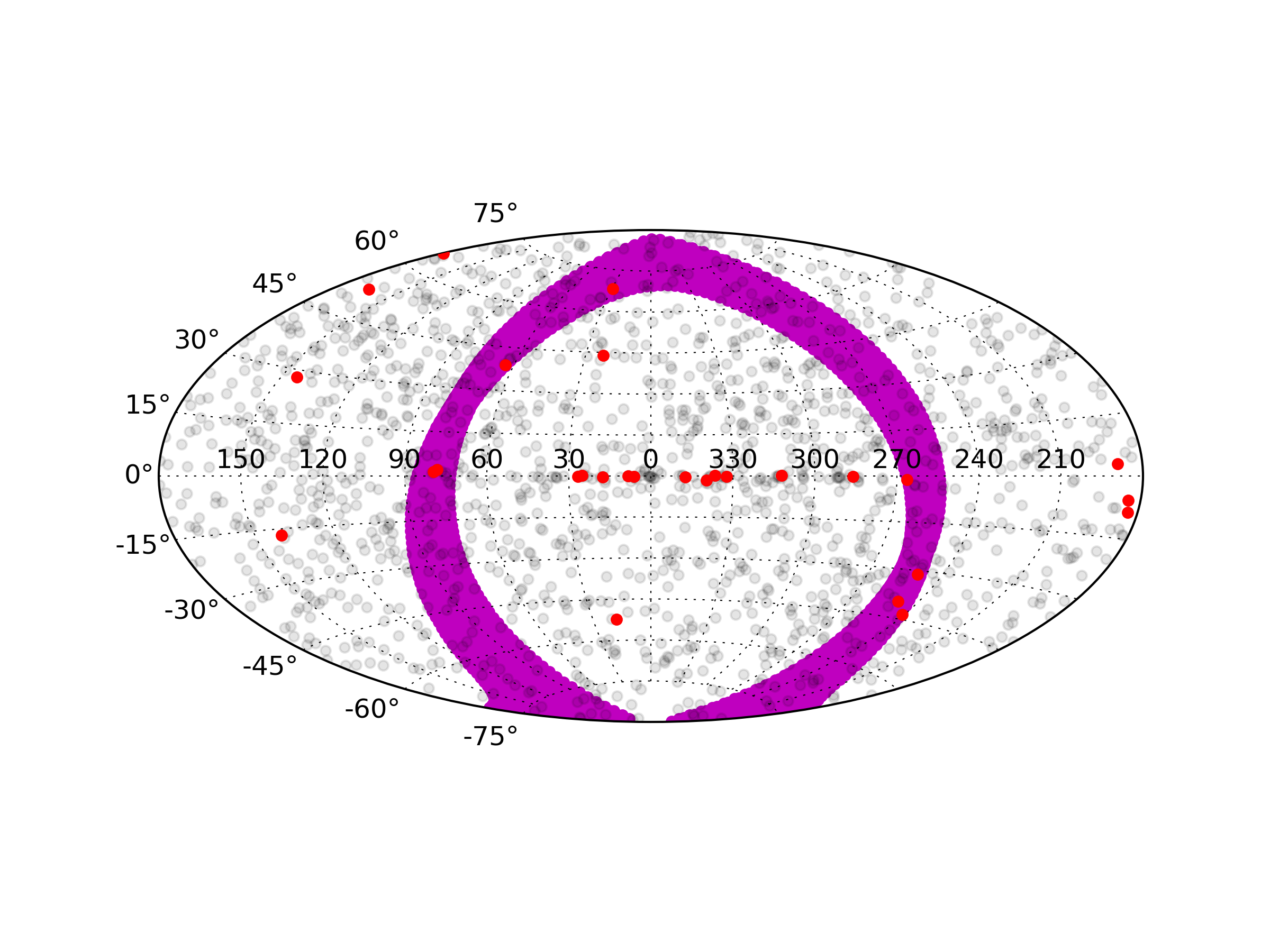}
 \includegraphics[width=0.93\columnwidth]{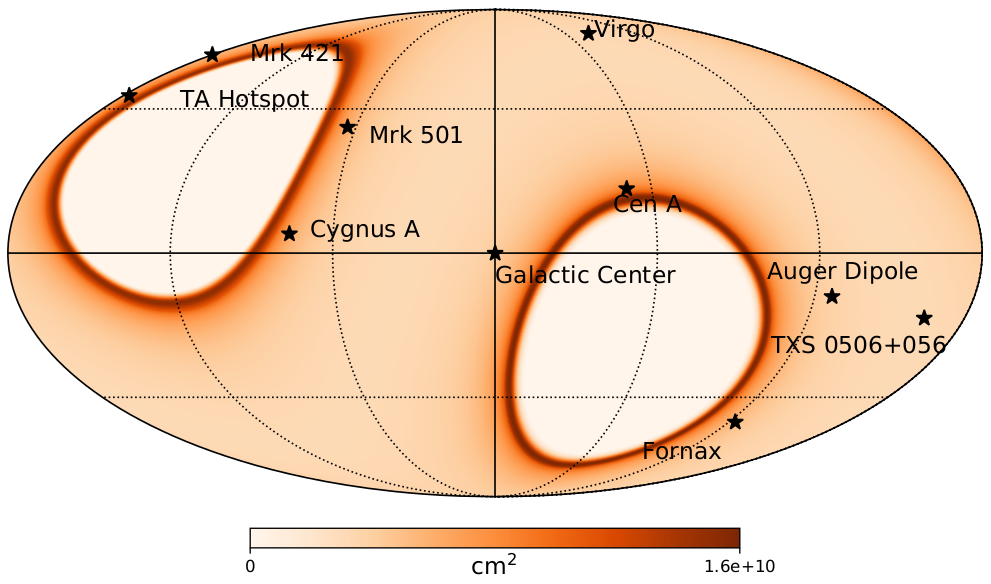}
 \caption{\label{fig:FOV}{\it Left:}  Field of view of GRAND200k for a 1-hour exposure, shaded purple, in Galactic coordinates.  Overlaid dots mark the positions of sources from the {\it Fermi} 3FHL catalog\ \cite{TheFermi-LAT:2017pvy}; red dots indicate the most significant (test statistic TS$\,>25$), brightest {\it Fermi} sources at energies $>\,50\,$GeV with spectral index $<2.5$, and maximum photon energy $>100$~GeV.  {\it Right:} Effective area of GRAND200k as a function of position in the sky in Galactic coordinates for UHE neutrinos with energy $3 \times 10^9$~GeV.  Stars represent the coordinates of interesting astrophysical sources in the field of view of GRAND.}
\end{figure*}

Figure\ \ref{fig:flux} summarizes predictions of the diffuse fluxes of EeV neutrinos from astrophysical sources, including AGN\ \cite{1991PhRvL..66.2697S, 2014PhRvD..90b3007M, 2015arXiv151101590M}, GRBs\ \cite{1999PhRvD..59b3002W, 2012ApJ...752...29H, 2006PhRvD..73f3002M, 2015NatCo...6E6783B, 2017ApJ...837...33B, 2006ApJ...651L...5M, 2008PhRvD..78b3005M}, galaxy clusters\ \cite{2008ApJ...689L.105M, 2009ApJ...707..370K, 2016ApJ...828...37F, 2017arXiv170400015F}, and pulsars and magnetars\ \cite{2009PhRvD..79j3001M, 2014PhRvD..90j3005F}. 
GRAND200k will detect the flux from most of these source models within 3 years of operation and characterize their spectrum.

\medskip

\paragraph{Point-source neutrino fluxes}

\smallskip

Figure\ \ref{fig:sourceSearch} shows the required angular resolution and number of events in order to resolve individual point sources\  \cite{Fang:2016hop}.  The color coding represents the confidence level at which an isotropic background can be rejected, using the statistical method from \Ref\ \cite{2016ApJ...826..102F}, assuming that all of the sources have the same luminosity, and that the sources follow a uniform distribution with number density $10^{-7}$~Mpc$^{-3}$ up to 2~Gpc.  We assume full sky coverage; fewer events are required in the field of view in case of a smaller sky coverage. We contrast that to the angular resolution and event rates to be reached by IceCube with 15 years of operation, and ARA and GRAND with 3 years of operation, assuming that the integrated flux of point sources is comparable to $10^{-8}$~ GeV~cm$^{-2}$~s$^{-1}$~sr$^{-1}$ with an $E_\nu^{-2}$ spectrum around EeV\ \cite{2016ApJ...826..102F}.  The event rate of each detector is computed for the energy range $10^{9}$--$10^{10}$~GeV, with each color box representing uncertainties in the neutrino spectrum and the detector angular resolution.  Figure\ \ref{fig:sourceSearch} shows that GRAND could discover the first sources of UHE neutrinos at a significance of $5\sigma$.

\medskip

\paragraph{Transient EeV neutrino astronomy}

\smallskip

A promising way to identify EeV neutrino sources is to detect transient neutrino emission in coincidence with electromagnetic emission.  GRAND makes this possible, due to its excellent angular resolution and large sky coverage; see Section \ref{section:GRANDDesignPerf-Performance-Reconstruction}.

Figure\ \ref{fig:FOV} shows that the instantaneous field of view of GRAND is a band between zenith angles $85^\circ \leq \theta_z \leq 95^\circ$, corresponding to $<5\%$ of the sky.  For simplicity, the figure assumes that the 200\,000 antennas envisioned for the final configuration of GRAND are grouped in a single array covering 200\,000~km$^2$. Since all azimuth angles are observed at any instant, about 80\% of the sky is observed every day.
Transients lasting less than a day have a low probability of being spotted, but for longer 
transients --- blazar flares, tidal disruption events, superluminous supernovae, \etc\ --- offline analysis at the location of existing transients and stacking searches can be performed.  Depending on the background discrimination efficiency, GRAND could send alerts to other experiments; see Section\ \ref{section:multi_messenger}.  In reality, because GRAND200k will consist of several sub-arrays placed at different geographical locations (see Section \ref{section:array_layout}), the instantaneous field of view will be larger.

\begin{figure}[t!]
 \centering
 \includegraphics[trim = 0.15cm 0.3cm 0.3cm 0.3cm, clip=true, width=\columnwidth]{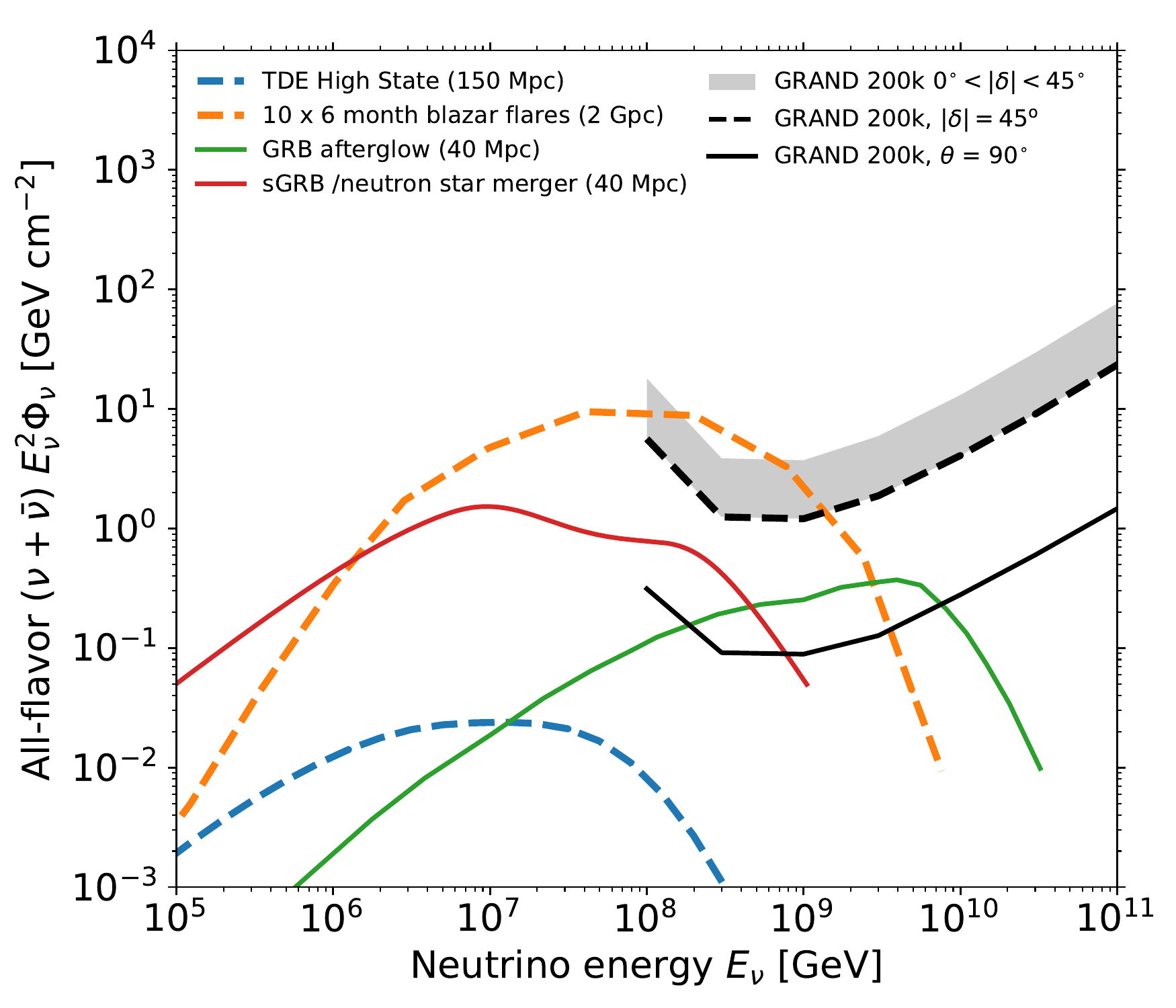}
 \caption{Neutrino fluence from transient sources.  Short-duration transients --- a short-duration GRB (sGRB) and a GRB afterglow --- are compared to the GRAND200k instantaneous sensitivity at zenith angle $\theta_z = 90^\circ$ (solid black line).  A long-duration transient --- a TDE --- is compared to the GRAND200k declination-averaged sensitivity (gray-shaded band).  The stacked fluence from 10 six-month-long blazar flares in the declination range $40^\circ< |\delta| < 45^\circ$ is compared to the GRAND200k sensitivity for a fixed $\delta=45^\circ$ (dashed black line).  See the main text for details.   The sources were assumed to lie at distances such to allow for a conservative rate of $\sim$1 event per century, using population rates inferred from \Ref\ \cite{Zhang:2017lpb} for short-duration GRBs and associated neutron-star mergers, \Ref\ \cite{Murase:2007yt} for GRB afterglows, and \Ref \cite{Guepin:2017abw} for TDEs. The sensitivity is the Feldman-Cousins upper limit per decade in energy at 90\% C.L., assuming a power-law neutrino spectrum $\propto E_{\nu}^{-2}$, for no candidate events and null background.}
 \label{fig:point_sources}
\end{figure}

Figure\ \ref{fig:point_sources} shows the neutrino fluence predicted from candidate classes of astrophysical transient sources, of short and long duration, compared to the GRAND200k point-source sensitivity. 
The predictions for short-duration, sub-hour transients in the instantaneous GRAND field of view are compared to the instantaneous sensitivity at $\theta = 90^\circ$.  These are a short-duration GRB (sGRB) possibly associated with a double neutron-star merger\ \cite{Kimura:2017kan} at 40~Mpc and a GRB afterglow\ \cite{Murase:2007yt} at 40~Mpc.  The prediction for a longer transient --- a TDE at 150~Mpc\ \cite{Guepin:2017abw} --- is compared to the declination-averaged sensitivity.  The stacked fluence of 10 blazar flares in the declination range $40^\circ< |\delta| < 45^\circ$ --- calculated using as template a 6-month long flare of the blazar 3C66A at 2~Gpc\ \cite{Murase:2014foa} --- is compared to the sensitivity for a fixed declination $\delta=45^\circ$. 

Short GRBs and GRB afterglows that occur within the field of view of GRAND will be readily detectable as neutrino point sources by GRAND if they take place at the specified distances.  Identification of individual blazars, magnetars, and TDEs is more challenging, but GRAND will be able to identify the stacked neutrino signal from such sources, as they are individually less than one or two orders of magnitude below the point source sensitivity.  The performance of GRAND in detecting neutrinos from a given class of transient sources can be estimated using the criterion for detection of neutrino flares described in \Ref\ \cite{Guepin:2017dfi}.

\subsection{Fundamental neutrino physics} 

Astrophysical and cosmogenic neutrinos provide a chance to test fundamental physics in new regimes.
Numerous new-physics models have effects whose intensities are proportional to some power of the neutrino energy $E_\nu$ and to the source-detector baseline $L$, {\it i.e.}, $\sim \kappa_n E_\nu^n L$, where the energy dependence $n$ and the proportionality constant $\kappa_n$ are model-dependent\ \cite{Colladay:1998fq, Kostelecky:2003xn, Kostelecky:2003cr, Diaz:2011ia, Kostelecky:2011gq}.  For instance, for neutrino decay, $n = -1$; for CPT-odd Lorentz violation or coupling to a torsion field, $n = 0$; and for CPT-even Lorentz violation or violation of the equivalence principle, $n=1$.  If GRAND were to detect neutrinos of energy $E_\nu$ coming from sources located at a distance $L$ then, nominally, it could probe new physics with exquisite sensitivities of
$\kappa_n \sim 4 \cdot 10^{-50} ( E_\nu / \text{EeV} )^{-n} ( L / \text{Gpc} )^{-1}\ \text{EeV}^{1-n}$.   This is an enormous improvement over current limits of $\kappa_0~\lesssim~10^{-32}$~EeV and $\kappa_1 \lesssim 10^{-33}$, obtained with atmospheric and solar neutrinos\ \cite{Abbasi:2010kx,Abe:2014wla}.
This holds even if the diffuse neutrino flux is used instead, since most of the contributing sources are expected to be at distances of Gpc.

New physics could affect any of the following observables:
\begin{itemize}
 \item
  {\bf Spectral shape:}  Neutrino energy spectra are expected to be power laws.  New physics could introduce additional spectral features, like peaks, troughs, and varying slopes.  New physics models include neutrino decay\ \cite{Baerwald:2012kc,Shoemaker:2015qul,Bustamante:2016ciw}, secret neutrino interactions \cite{Ioka:2014kca,Ng:2014pca,Blum:2014ewa,DiFranzo:2015qea,Araki:2015mya}, and scattering off dark matter\ \cite{Kopp:2015bfa,Davis:2015rza,Reynoso:2016hjr}.
  
  In GRAND, detection of EeV neutrinos with large statistics and sufficient energy resolution (see Section \ref{section:GRANDDesignPerf-Performance-Reconstruction}) would allow to infer their energy spectrum and potentially identify sub-dominant features introduced by new physics, including their energy dependence\ \cite{Kashti:2005qa,Lipari:2007su,Shoemaker:2015qul,Bustamante:2015waa,Bustamante:2016ciw}.
 \item
  {\bf Angular distribution:}  When neutrinos travel inside the Earth, the neutrino-nucleon cross section imprints itself on the distribution of their arrival directions.  This has allowed to measure the cross section up to PeV energies in IceCube\ \cite{Aartsen:2017kpd, Bustamante:2017xuy}.  EeV neutrinos could extend the measurement.  Further, we can look for deviations due to enhanced neutrino-nucleon interactions\ \cite{Blennow:2009rp,Connolly:2011vc,Gonzalez-Garcia:2016gpq} and interactions with high-density regions of dark matter\ \cite{Davis:2015rza,Arguelles:2017atb}.  The sub-degree pointing accuracy of GRAND would precisely reconstruct the distribution of arrival directions.
 \item
  {\bf Flavor composition:}  Flavor ratios --- the proportion of each neutrino flavor in the incoming flux --- are free from uncertainties on the flux normalization and so could provide clean signals of new physics\  
  \cite{Pakvasa:1981ci, Barenboim:2003jm, Xing:2006uk, Lipari:2007su, Pakvasa:2007dc, Esmaili:2009dz, Lai:2009ke, Choubey:2009jq, Bustamante:2010nq, Mehta:2011qb, Pakvasa:2012db, Dorame:2013lka, Mena:2014sja, Xu:2014via,
  Aeikens:2014yga, Fu:2014isa, Palladino:2015zua, Palomares-Ruiz:2015mka, Aartsen:2015ivb, Palladino:2015vna, Arguelles:2015dca, Bustamante:2015waa, Shoemaker:2015qul, Nunokawa:2016pop, Vincent:2016nut, 
  Gonzalez-Garcia:2016gpq, Brdar:2016thq, Bustamante:2018mzu, Ahlers:2018yom}. Possibilities include neutrino decay\ 
  \cite{Pakvasa:1981ci, Lindner:2001fx, Beacom:2002cb, Beacom:2002vi, Barenboim:2003jm, Beacom:2003nh, Beacom:2003zg, Meloni:2006gv, Maltoni:2008jr, Bustamante:2010nq, Baerwald:2012kc, Pakvasa:2012db, Dorame:2013lka, Pagliaroli:2015rca, Bustamante:2015waa, Huang:2015flc, Shoemaker:2015qul, Bustamante:2016ciw}, Lorentz-invariance violation (LIV)\ \cite{Colladay:1998fq, Kostelecky:2003xn, Barenboim:2003jm, Hooper:2005jp, Ellis:2011ek, Bustamante:2010nq}, coupling to a torsion field\ \cite{DeSabbata:1981ek}, active-sterile neutrino mixing\ \cite{Brdar:2016thq}, pseudo-Dirac neutrinos\ \cite{Pakvasa:2012db, Shoemaker:2015qul, Ahn:2016hhq}, renormalization-group running of mixing parameters\ \cite{Bustamante:2010bf}, and interaction with dark matter\ \cite{deSalas:2016svi, Reynoso:2016hjr} or dark energy\ \cite{Klop:2017dim}.
  
  GRAND will be sensitive to $\nu_\tau$.  Other EeV-neutrino experiments --- ARA, ARIANNA, ANITA --- are sensitive to neutrinos of all flavors, though they are unable to distinguish between them; however, see \Refs\ \cite{Supanitsky:2012pp, Wang:2013bvz}.  Comparing GRAND $\nu_\tau$ data with all-flavor data from other experiments could yield the tau flavor ratio.  Alternatively, this could be done with GRAND data alone, by comparing showers initiated by neutrinos of all flavors interacting in the atmosphere to showers initiated by $\nu_\tau$ interacting underground; see Section\ \ref{section:detection_principle_sub_detection_strategy}.
\end{itemize}

Ultimately, the ability of GRAND to probe fundamental physics at the EeV scale will depend on the level of the cosmogenic neutrino flux.  If the flux is low, probing new physics will be challenging.  On the other hand, with a flux high enough to yield tens of events, we could probe fundamental physics in a completely novel regime.


\subsection{Ultra-high-energy gamma rays}
\label{section:uhe_photons}

Like cosmogenic neutrinos, cosmogenic UHE gamma rays are a guaranteed by-product of photo-pion interactions of UHECRs with the CMB.  They can also be generated through inverse-Compton scattering of CMB photons by electrons or positrons produced by UHECRs scattering off the CMB.  Like for neutrinos, higher fluxes are expected if UHECRs are dominated by protons than if they are dominated by heavy nuclei or have a mixed mass composition.  To date, UHE gamma rays have not been detected.

\begin{figure}[t!]
 \begin{center}
 \includegraphics[width=\columnwidth]{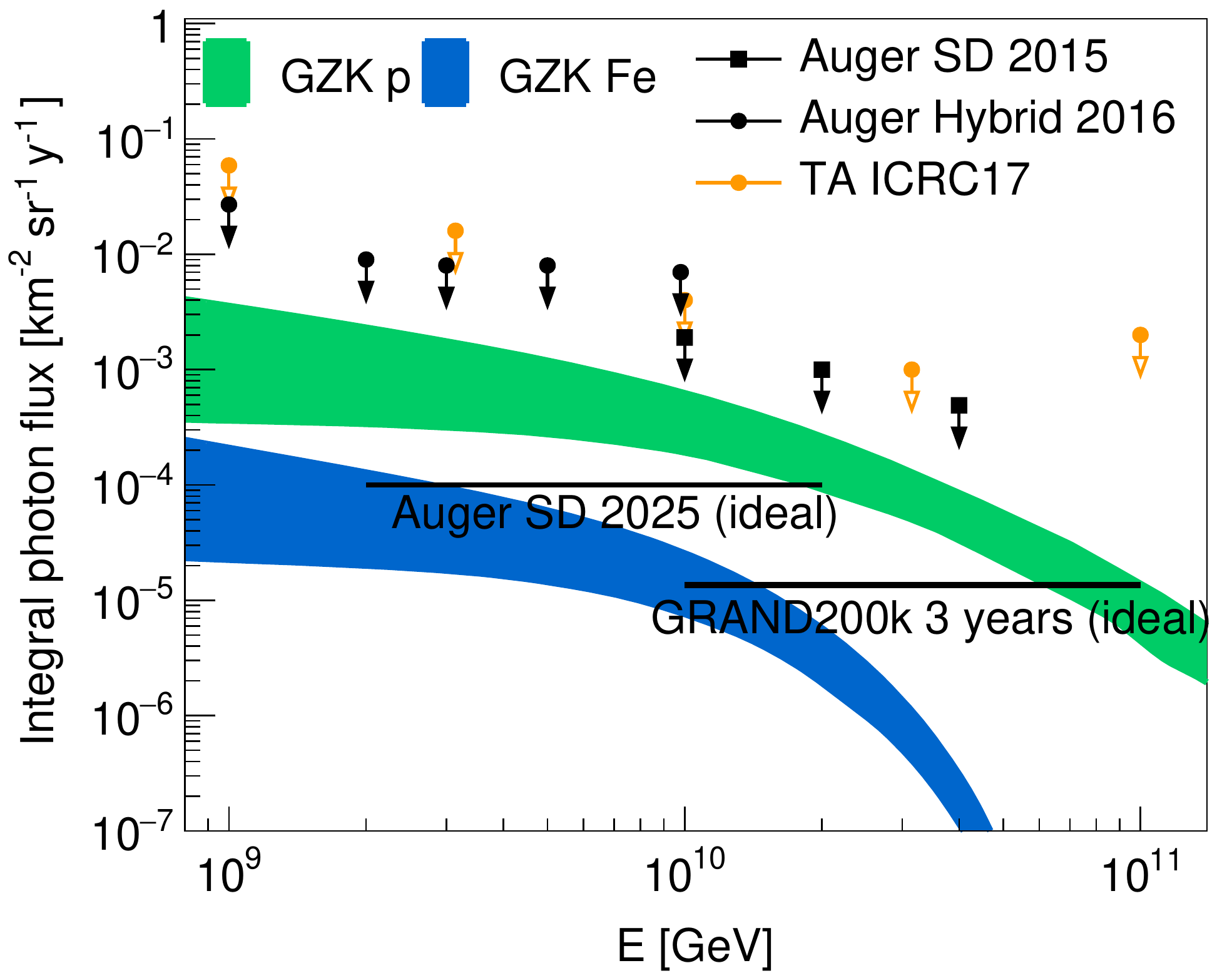}
 \end{center}
 \caption{\label{fig:phA}Predicted cosmogenic UHE photon flux from pure-proton and pure-iron UHECRs, as estimated in \Ref\ \cite{Sarkar11}.  For comparison, we include the existing upper limits from Auger and the Telescope Array (TA)\ \cite{Aab:2016agp, Niechciol:2017vqf, Rubtsov17}, the projected reach of Auger by 2025, and of GRAND after 3 years of operation.}
\end{figure}

Figure \ref{fig:phA} shows the current state of the art in searches for UHE gamma rays and the preliminary sensitivity of GRAND, in terms of the fraction of air showers initiated by gamma rays; see Section~\ref{section:GRANDDesignPerf-Performance-PhotonsCR}.  The most stringent upper limits to date come from Auger\ \cite{Aab:2016agp, Niechciol:2017vqf}.  With its present sensitivity, Auger constrains the photon fraction to be $\leq 0.1\%$ of their total event rate, and thus rules out some of the region of photon fluxes predicted in astrophysical scenarios for a proton-dominated mass composition\ \cite{Aab:2016agp, Aab:2016bpi, Niechciol:2017vqf}.  The Telescope Array (TA) provides complementary limits in the same energy range in the Northern Hemisphere\ \cite{Rubtsov17}.  Auger will continue to lower the upper limits on the flux of UHE gamma rays, or discover them, until 2025.  By then, \Fig\ \ref{fig:phA} shows that Auger will have reached sensitivity to even conservative predictions of the flux, assuming proton-dominated UHECRs\ \cite{Aab:2016vlz}.  A few years later, GRAND200k will be sensitive to cosmogenic gamma-ray fluxes, even for iron-dominated UHECRs.

Searches for UHE gamma rays with GRAND will contribute to several science goals.  The primary objective, with a guaranteed scientific return, is measuring the flux of cosmogenic gamma rays above $10^{10}$~GeV, or strongly constraining it.  A first detection of UHE gamma rays is in close reach if a fraction of UHECR primaries are protons.  Figure \ref{fig:phA} shows that GRAND will be able to detect or disfavor proton-dominated UHECR models within 3 years of operation, even for models with the lowest predictions of UHE photons.  A non-detection, on the other hand, would be evidence of a heavier UHECR composition.

In addition, because gamma rays produced inside astrophysical sources point back at them, GRAND could detect nearby sources of UHE gamma rays, \ie, sources that lie within the mean free path of EeV gamma rays on the CMB, of about 10~Mpc.  This is particularly attractive for searches of transient multi-messenger sources; see Section\ \ref{section:multi_messenger}. 

The detection of UHE gamma rays would probe the little-known diffuse cosmic radio background (CRB)\ \cite{Protheroe:1996si, Fixsen:1998kq}.  While the increasingly stringent constraints on the EBL come from the steadily increasing quantity and quality of very-high-energy observations with imaging air Cherenkov telescopes\ \cite{Biteau:2015xpa}, GRAND could be the first experiment to put such indirect constraints on the CRB.  The energy range from $10^{10}$ to $10^{11}$~GeV, where GRAND will reach full efficiency for photon detection, is optimal to constrain the impact of the CRB on UHE photon propagation. 

Further, as with neutrinos, UHE photons could be used to probe open questions in fundamental physics, such as the existence of axion-like particles\ \cite{DeAngelis:2007dqd, SanchezConde:2009wu} and LIV\ \cite{AmelinoCamelia:2002dx, Stecker:2003pw, Stecker:2011ps, Vasileiou:2013vra, Ellis:2018lca}.  Since models of LIV predict energy-dependent delays in photon arrival times that are linear or quadratic in the photon energy, LIV studies would benefit significantly from the detection of UHE gamma rays.


\subsection{Ultra-high-energy cosmic rays}
\label{section:uhe_messengers_uhecr}

\begin{figure*}[t!]
 \centering
 \includegraphics[width=0.44\textwidth]{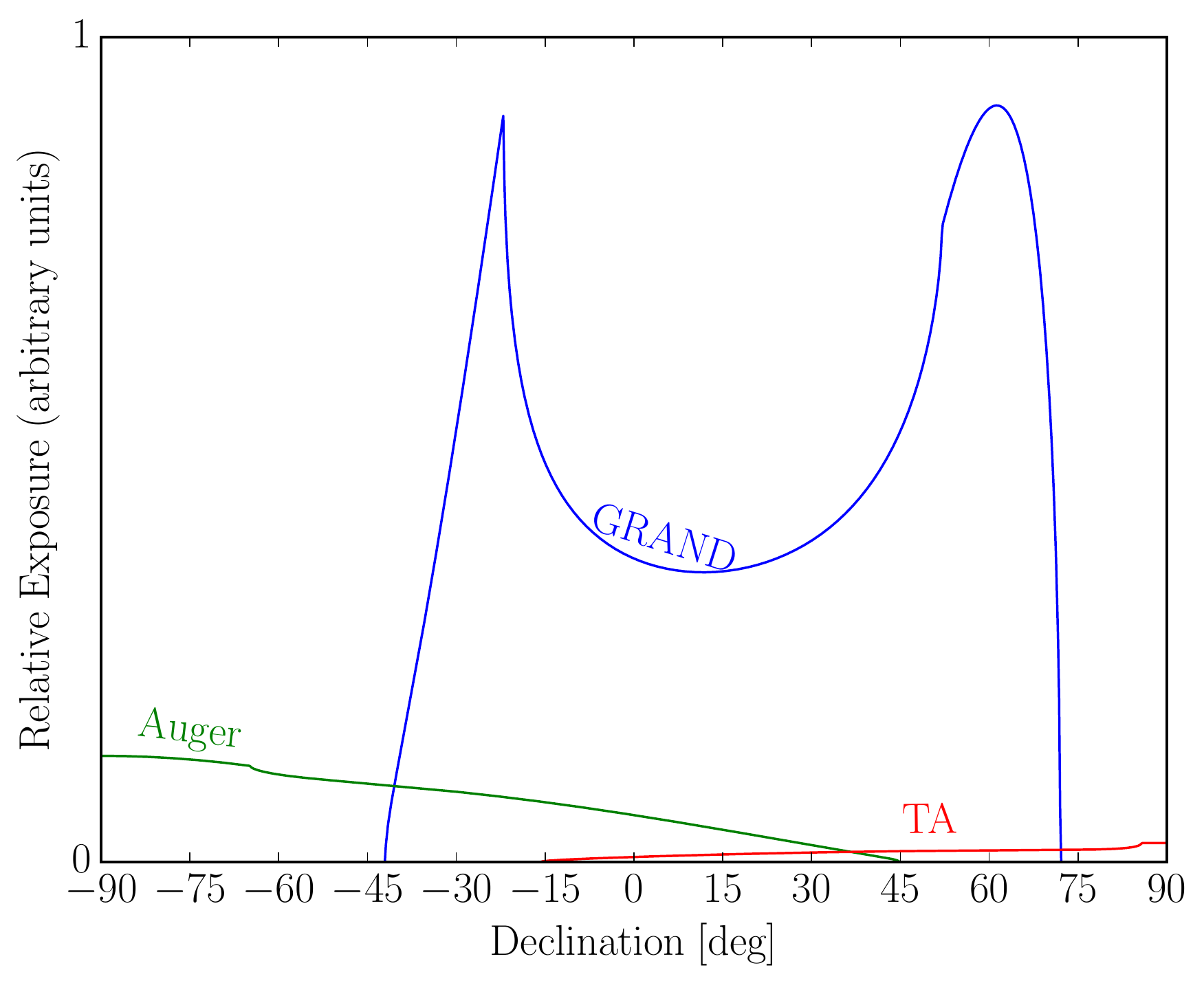}
 \includegraphics[width=0.55\textwidth, trim=0 -1cm 0 0, clip=true]{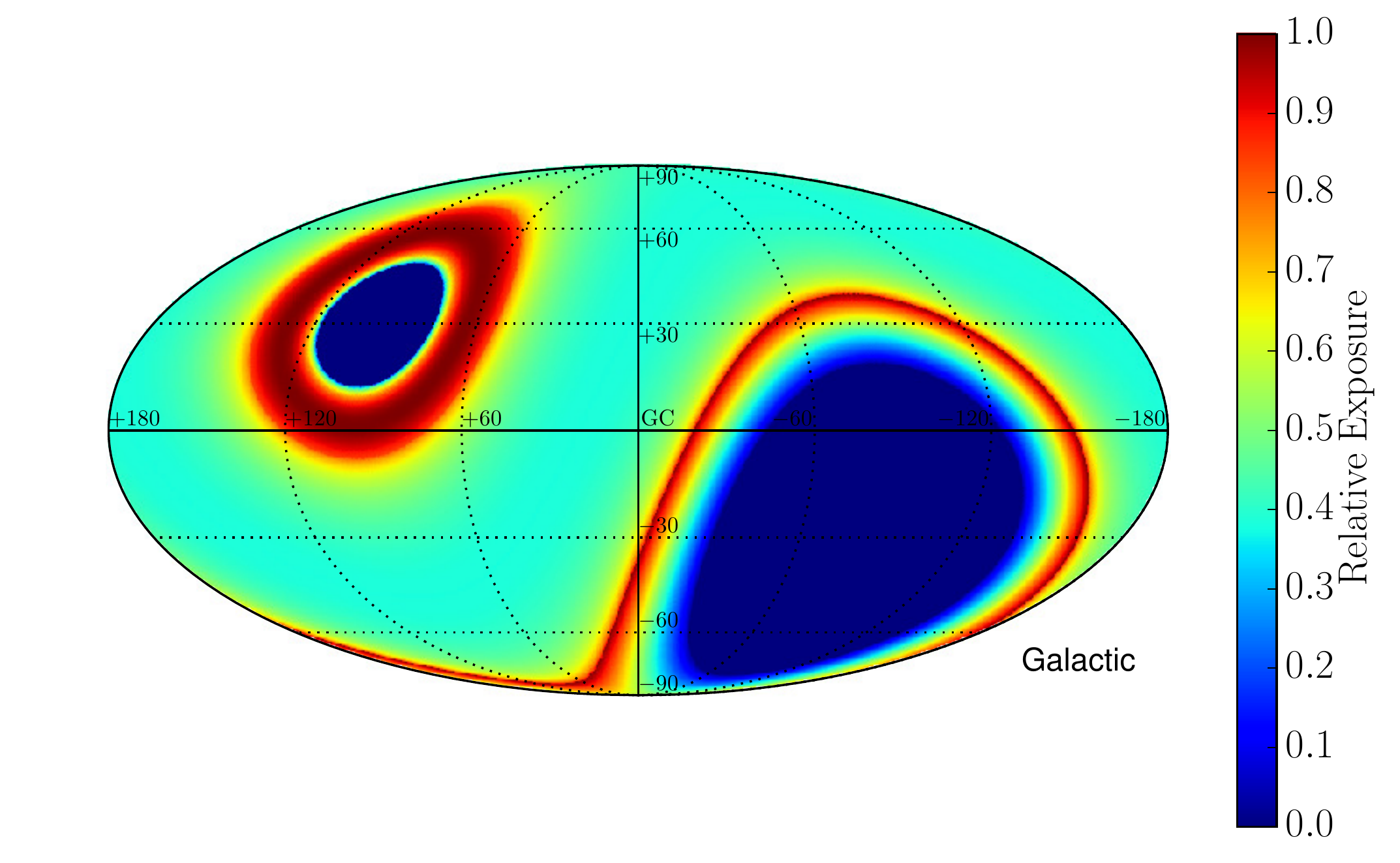}
 \caption{{\it Left:} The relative annual geometric exposure to UHECRs of GRAND, Auger, and TA.
 At high energies these detectors are fully efficient so the geometric exposure approximates well the true exposure.
 {\it Right}: 3-year integrated exposure to UHECRs of GRAND in Galactic coordinates.}
 \label{fig:exposure}
\end{figure*}

{\it Energy spectrum and mass composition.---}  Presently, the most precise results on UHECRs come from the two largest cosmic-ray detectors, Auger\ \cite{Aab:2017njo} and TA\ \cite{Ikeda:2017dci}.  Statistics are needed to confirm and refine features seen in the UHECR energy spectrum at the extreme end, above the observed cut-off energy of $4 \cdot 10^{10}$~GeV, where events are scarce.  For instance, finding the precise shape of the cut-off would reveal whether it is due to sources running out of power or to the GZK process; see, \eg, \Ref\ \cite{Kotera2011}.  GRAND has the potential to provide the required statistics.

Since GRAND will be fully efficient above $10^{10}$~GeV and sensitive to cosmic rays in a zenith-angle range of $65^\circ$--$85^\circ$, it will have an aperture of $107\,000$~km$^2$~sr, leading to an exposure of $535\,000$~km$^2$~sr~yr after 5 years of live-time.  In comparison, the current aperture of Auger is $\sim$5\,400~km$^2$~sr and its cumulative 9-year exposure is $\sim$48\,000~km$^2$~sr~yr\ \cite{Aab:2016ban}.  Below, to calculate event rates, we use the flux fitted to the Auger spectrum by \Ref\ \cite{Fenu:2017hlc}, though it slightly overestimates data points at the highest energies.

The resulting expected UHECR event rate in GRAND is 20 times higher than in Auger.  In 1 year, GRAND should detect 6\,400 events above $10^{10.5}$~GeV, versus 320 in Auger; and 150 events above $10^{11}$~GeV, versus 8 in Auger.  In 5 years, GRAND should detect 32\,000 UHECRs above $10^{10.5}$~GeV, far exceeding the combined $\sim$750 collected events of Auger and TA to date\ \cite{Fenu:2017hlc, Tsunesada:2017ICRC}. 

\medskip

{\it Mass composition.---}  The mass composition of UHECRs is a key ingredient to understand the transition between Galactic and extragalactic source populations around the ``ankle'', at 1~EeV.  At the highest energies, it provides insight on the mechanisms that accelerate UHECRs.

The best understood observable in determining the mass composition of cosmic rays is the column depth $X_{\rm max}$ at which the electromagnetic particle content of the air shower --- electrons and photons --- is maximum.  GRAND could measure $X_{\rm max}$ with a precision of 20~g~cm$^{-2}$, sufficient to distinguish between different nuclei; see Section\ \ref{section:GRANDDesignPerf-Performance-PhotonsCR}.  Already GRANDProto300 --- the early 300-antenna construction stage of GRAND --- should separate showers initiated by light and heavy primaries, and study their energy spectrum and arrival directions separately; see Section\ \ref{section:construction_stages}.

\medskip

{\it Arrival directions.---}  With the number of detected UHECRs exceeding thousands of events, GRAND will be able to detect small-scale clustering of arrival directions in the sky and probe source population distributions \cite{2015JCAP...01..030O}.

Measuring large-scale anisotropies in the arrival directions is significantly harder with partial sky exposure than full sky coverage\ \cite{Denton:2015bga}. Currently, there is a $\sim$20\% difference in the measured fluxes of Auger and TA, complicating the combination of the data sets.  This combination requires the addition of a fudge factor determined by the overlap region which is dominated by the low statistics of highly inclined showers\ \cite{Array:2013dra}.  GRAND would have high exposure in the field of view of both Auger and TA, allowing for a consistent measurement of the normalization of the flux and for more accurate anisotropy studies.

TA has detected a relative excess in the number of detected UHECRs above 50~EeV --- a {\it hotspot} --- coming from declination $43.2^\circ$, with significance of $3.2\sigma$\ \cite{Abbasi:2014lda, Abbasi:2017vru}.
This hotspot is near the peak of the exposure of GRAND in the Northern Hemisphere, so GRAND could study the nature of this excess with higher statistics.  In addition, Auger has reported evidence for a dipolar excess at declination $-24^\circ$\ \cite{Aab:2017tyv}, within the field of view of GRAND, which we will be able to confirm within a year of operation.

Figure \ref{fig:exposure} shows, in the left panel, the relative annual UHECR exposure of Auger, TA, and GRAND as a function of declination.  Due to highly inclined showers, the exposure of GRAND is bimodal with peaks in both hemispheres.  The right panel of \Fig\ \ref{fig:exposure} shows the 3-year integrated exposure of GRAND, in Galactic coordinates, assuming that the zenith angle acceptance is $65^\circ \le \theta_z \le 85^\circ$.  For this computation, as illustration, we assumed that the final configuration of GRAND is a single array of 200\,000 antennas located at the site of the TREND array\ \cite{Charrier:2018fle}, at latitude $43^\circ$ North.  GRAND will sweep out a region on the sky covering the declination band $-43^\circ < \delta < 63^\circ$.

\medskip

{\it Measuring the proton-air cross section.---}  Large UHECR statistics and precise $X_{\rm max}$ resolution will allow GRAND to measure the proton-air cross section up to center-of-mass energies of $\sqrt{s} \sim 2 \cdot 10^{5}$~GeV, thus extending the world data to slightly higher energies; see \Fig\ 2 in \Ref\ \cite{Collaboration:2012wt}. 

Since protons penetrate the atmosphere deeper than heavier nuclei, they dominate the tail of the $X_{\rm max}$ distribution of a sample of collected showers.  Therefore, we can extract the proton-air cross section from the shape of the event distribution, following a method similar to the one used by Auger in \Ref\ \cite{Collaboration:2012wt}.  Assuming that protons make up 10\% of the UHECR composition above $10^{10.5}$~GeV, GRAND will accumulate 3\,200 proton events in 5 years, comparable to what Auger used.   

\medskip

{\it UHECR science in GRANDProto300.---}  The intermediate, 300-antenna stage GRANDProto300 has two important UHECR science goals.  First, it will study the transition of UHECRs from having a Galactic origin to having an extragalactic origin --- which is believed to occur around the ankle of the cosmic-ray spectrum, $10^{7.5}$--$10^9$~GeV --- with high statistics.  Second, it will use a co-located ground particle array to study the existing discrepancy in the muon content of air showers\ \cite{Aab:2014pza}.  See Section\ \ref{section:construction_stages} for details.


\subsection{Multi-messenger studies}
\label{section:multi_messenger}

The era of multi-messenger astrophysics has begun, signaled by the detection of gravitational waves from a binary-neutron-star inspiral by LIGO\ \cite{TheLIGOScientific:2017qsa}, in coincidence with electromagnetic counterparts including a gamma-ray burst\ \cite{GBM:2017lvd} and possibly the first high-energy cosmic neutrino source associated with a blazar flare\ \cite{IceCube:2018cha, IceCube:2018dnn}. 
By the time of construction of the later stages of GRAND, time-domain astroparticle physics will be even more developed, encompassing all the different messengers --- neutrinos, cosmic rays, photons, and gravitational waves. 

Future gravitational-wave detectors Advanced LIGO Plus\ \cite{ALIGOPlus:Talk}, Einstein Telescope\ \cite{Sathyaprakash:2012jk}, and LIGO Cosmic Explorer\ \cite{LIGOCosmicExplorer:Talk} will be able to observe mergers at cosmological distances. 
In the event of a transient, multiple instruments will cover a large field of view quickly across the electromagnetic spectrum.  
GeV--TeV gamma-ray counterparts will be detectable not only by current telescopes, such as {\it Fermi}-GBM\ \cite{Meegan:2009qu}, {\it Fermi}-LAT\ \cite{Atwood:2009ez}, H.E.S.S.\ \cite{Abdalla:2017brm}, MAGIC\ \cite{Ahnen:2017ite}, VERITAS\ \cite{Abeysekara:2015rna}, and HAWC\ \cite{Abeysekara:2017yqc}, but also by the planned Cherenkov Telescope Array (CTA)\ \cite{Acharya:2017ttl} and LHAASO\ \cite{DiSciascio:2016rgi}.  The optical band will be covered by the future Large Synoptic Survey Telescope\ \cite{2009arXiv0912.0201L} and the Tomo-e Gozen Camera\ \cite{2016SPIE.9908E..3PS} at the Kiso Observatory.  These instruments will make frequent combined observations of transient phenomena. 

Detecting UHE neutrinos from transient point sources, in coincidence with electromagnetic observations, would be a vital step towards revealing the sources of UHECRs.  GRAND is ideally positioned to do this; see Section\ \ref{section:uhe_neutrinos}.
GRBs, blazars, and TDEs are known to be powerful non-thermal emitters that can be seen at multiple wavelengths, whereas supernovae are observed as bright optical transients.  The angular resolution of GRAND is expected to be a fraction of a degree, within which a number of galaxies exist.  To pinpoint the source of the neutrino transient among them, GRAND will use timing information from their electromagnetic counterparts.  

UHE transients should produce not only neutrinos, but also gamma rays.  Depending on the spectral energy distribution of the sources, UHE gamma rays can also escape, and UHE gamma-ray flares can be detected up to a distance of 10--100~Mpc\ \cite{Murase:2009ah}.  As noted above, GRAND will be sensitive to UHE gamma rays; see Section\ \ref{section:uhe_photons}.  A fraction of the UHE gamma rays should induce electromagnetic cascades by interacting with the large scale structure in which the sources are embedded, and the associated GeV--TeV synchrotron emission\ \cite{Murase:2011yw} should be detectable by CTA as transients with longer time scales. 

Promising UHE gamma-ray source candidates include short GRBs and their off-axis counterparts, which are accompanied by gravitational wave signals.  By the time GRAND is completed, third-generation gravitational-wave detectors will be able to see compact-object mergers at $z \approx$ 2--6. Given that the fluence sensitivity is around $0.1~{\rm GeV}~{\rm cm}^{-2}$, late emission of short GRBs\ \cite{Kimura:2017kan} or mergers leaving a magnetar remnant\ \cite{Fang:2017tla} could be detected by GRAND from distances of 50--100~Mpc.  The signals are expected to arrive hours or days later than the gravitational wave signal, and would give us critical insight into the fate of neutron-star mergers and the physics of outflows.

Further, due to its unprecedented UHE neutrino sensitivity, GRAND will be a crucial triggering and follow-up partner in multi-messenger programs.  Currently, initiatives are bilateral between individual experiments or coordinated by multi-messenger networks such as the Astrophysical Multi-messenger Observatory Network (AMON)\ \cite{Smith:2012eu}. 

As a triggering partner, the design of GRAND will make it possible to reconstruct the arrival direction and issue an initial alert of an incoming neutrino-initiated air-shower from close to the horizon, owing to its distinctive polarization pattern, with sub-degree accuracy away from known sources of noise and with sub-minute latency.  This will make GRAND an important triggering partner in multi-messenger networks.  It will allow follow-up partner experiments to swiftly slew to the direction of the alert and search for electromagnetic counterparts.  Simulations of such ``real-time'' alerts are underway. 

As a follow-up partner, GRAND will be crucial in particular when alerts are issued by experiments without good angular resolution, such as neutrino-induced cascades from IceCube\ \cite{Aartsen:2013bka} and IceCube-Gen2\ \cite{Aartsen:2014njl}, and gravitational wave detectors.  If the alert from these experiments happens to be in the instantaneous field of view of GRAND, it will be possible to place a limit on the UHE neutrino emission from the observed transient, or detect the UHE neutrino counterpart, with sub-degree angular resolution, aiding to direct other follow-up instruments and possibly identifying the transient source.



\section{Radioastronomy and cosmology}
\label{section:cosmology_radioastronomy}

\mybox{{\bf At a glance}}{Tan!70}{grand_brown!20}
{
 \begin{itemize}[leftmargin=*]
  \item
   The wide field of view, frequency band, and size of GRAND will probe millisecond astrophysical transients: fast radio bursts and giant radio pulses
  \item
   By mapping the sky temperature with mK precision, GRAND could measure the global signature of the epoch of reionization and study the Cosmic Dawn 
  \item
   These measurements will be feasible already during the intermediate construction stages GRANDProto300 and GRAND10k
 \end{itemize}
}


\subsection{Fast radio bursts}
\label{section:FRB}

Fast radio bursts (FRBs) are a recently discovered class of
astrophysical transient events. They are short radio pulses, typically lasting a few ms,
emitted in a broad frequency band, and heavily dispersed in arrival times\ \cite{Masui:2015kmb}.  The temporal dispersion
is due to the presence of free electrons along the line of sight. The
delay in arrival time is
$\delta t \propto \mathrm{DM} \times f^{-2}$, where $f$ is the 
observing frequency, and the dispersion measure ${\rm DM} \sim \int n_e dl$ is the column depth of free electrons.
The brevity and large dispersion measure of the pulses
suggest that FRBs are made by extragalactic compact sources with sizes of a few thousand kilometers.

The first FRB was reported in 2007\ \cite{Lorimer:2007qn}; since then, about 30 FRBs have been
detected\footnote{FRBCAT: \href{http://frbcat.org/}{http://frbcat.org/}} \cite{Petroff:2016tcr}. 
Only two have been found to repeat\ \cite{Spitler:2016dmz, Amiri:2019bjk}.
Extrapolations from present-day small-number statistics
suggest that a few thousand FRBs occur every day.
Their origin remains unexplained, though several possible explanations exist\ \cite{Mottez:2014awa, Platts:2018hiy}.

GRAND could detect FRBs by incoherently adding the signals from individual antennas.  This method allows to infer the dispersion measure, though it does not locate the FRB. The resulting sensitivity is proportional to the square root of the number of antennas and the field of view is as large as for a single antenna.  These features are unmatched by other instruments that search for FRBs by adding signals coherently.  If the FRB is detected by more than one GRAND sub-array, we could use the difference in their detection times to estimate the position of the FRB.  

With potentially orders of magnitude more FRBs detected than the currently available sample, GRAND could discover different categories of FRB-like events, with unique, repeating, chaotic, or regular signatures, nearby or at cosmological distances.  The large statistics will help to answer
key questions, including what is the space density of FRBs in the
local Universe, how their radio spectra evolve at low frequencies,
whether the spectra at low frequencies are as dispersed as at high frequencies,
whether there is a low-frequency cut-off in the spectrum, and
how common FRB repeaters are.

\begin{figure}[t!]
 \centering
 \includegraphics[width=\columnwidth]{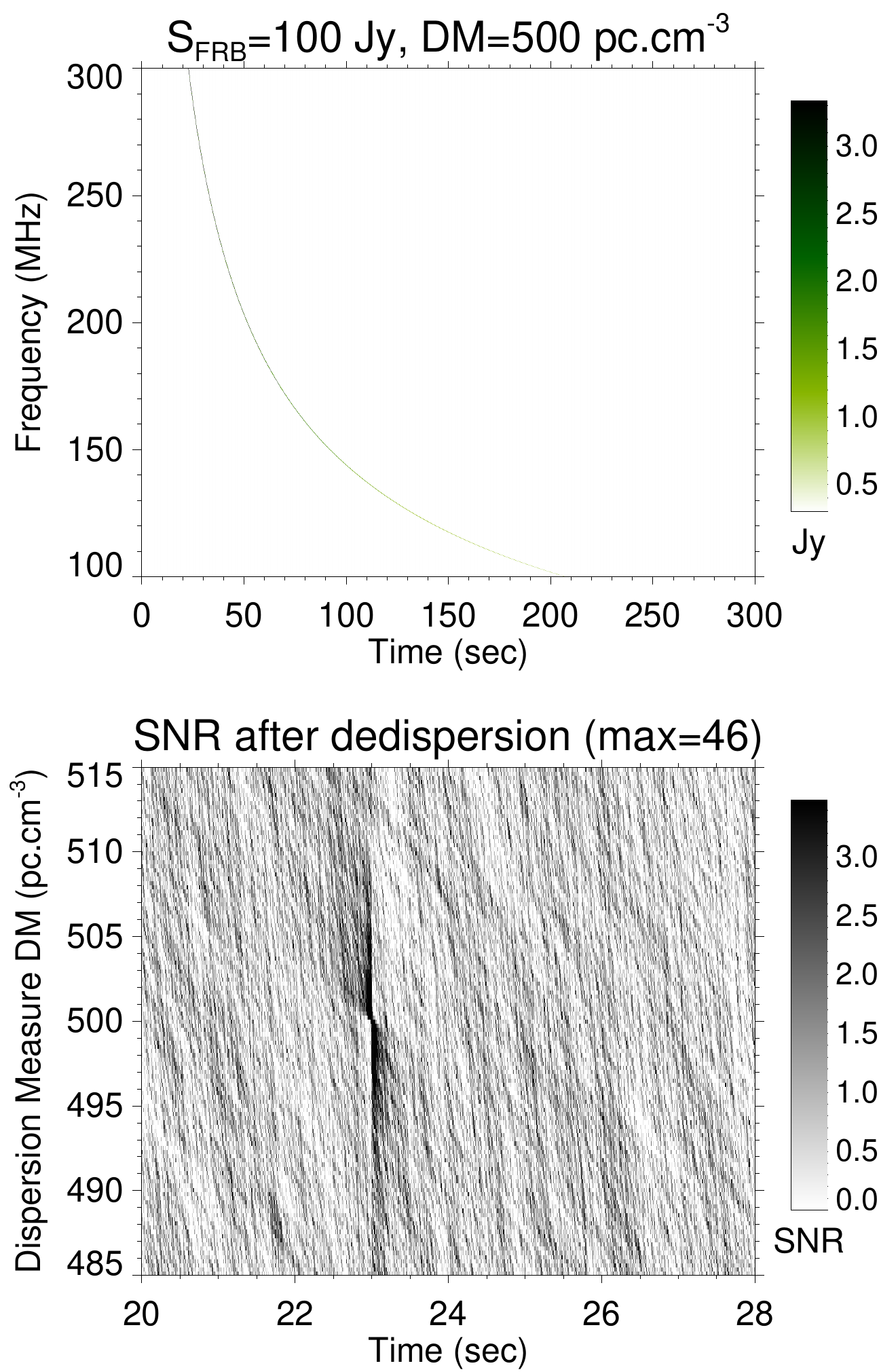}
 \caption{\label{fig:FRB}{\it Top:} Simulated FRB with a flat spectrum of 100 Jy, intrinsic duration of 5~ms, and dispersion measure of 500~pc~cm$^{-3}$, after being dispersed and scattered by propagation and detected with a resolution of 10~ms and 25~kHz.  The dominant Galactic background noise is not shown.  The dispersive drift starts at time $t = 23$ s in our simulation. {\it Bottom:} Result of a blind search for FRBs using GRAND. For each trial DM value, the dynamic spectrum is de-dispersed and integrated in frequency, and the resulting intensity profile is normalized by its standard deviation after subtracting its mean, \ie, it is displayed as a signal-to-noise ratio (SNR).  The SNR is small except near $t =23$ s, where it increases to reach a maximum value of $\sim$46 at 500~pc~cm$^{-3}$.}
\end{figure}

So far, verified FRBs have been observed between 580~MHz and $\sim$2~GHz\ \cite{Petroff:2016tcr}.
GRAND could test whether the FRB spectrum extends down to 50--200~MHz, complementing the search by CHIME, which operates between 400--800~MHz\ \cite{Ng:2017pyl, 2018ATel11901....1B}.
Below, we estimate the sensitivity to FRBs in GRAND.  In our simulations, following \Ref\ \cite{2016sf2a.conf..347Z}, in order to limit the data transfer
rate to $\lesssim2$ MB s$^{-1}$ 
we consider a dynamic spectrum domain of 
100--200$\ \mathrm{MHz}\times 300\ \mathrm{s}$ with a
resolution of $25\ \mathrm{kHz}\times 10\ \mathrm{ms}$.

We have simulated a wide variety of FRBs, taking into account Galactic noise,
the dispersion measure of the signal during propagation, and its scattering off electrons along the line of sight. Since electrons are inhomogeneously distributed, they affect differently the arrival time of different parts of the pulse front.  
We have not considered radio-frequency interference (RFI), since it should be mitigated before detection.

Figure \ref{fig:FRB} shows a representative simulated FRB spectrum based on values measured for a real FRB\ \cite{Petroff:2016tcr} with de-dispersed ${\rm DM} =500$~pc~cm$^{-3}$.  Using the full array of 200\,000 antennas --- \ie, GRAND200k, made up of 10\,000-antenna sub-arrays (see Section \ref{section:GRANDStages-GRAND200k}) --- in single-polarization mode and resolution of 10~ms $\times$ 25 kHz in the 100--200 MHz~band, we could detect a flat spectrum of 30~Jy at $\sim$$10\sigma$.  High detection significance is possible for values of DM of up to 1000~pc~cm$^{-3}$.

\begin{figure}[H]
 \includegraphics[width=0.45\textwidth]{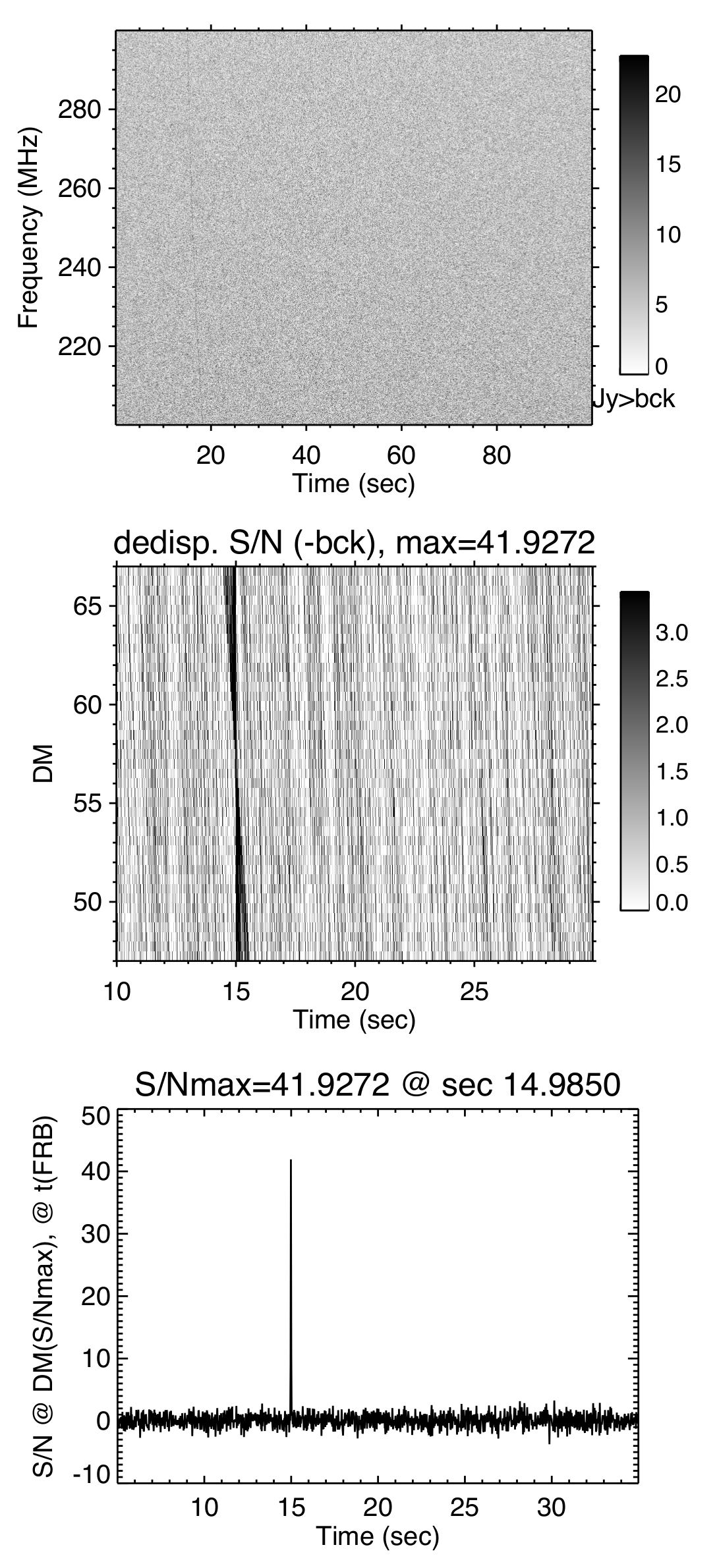}\\
 \caption{\label{fig:GP}{\it Top:} Simulated giant radio pulse (GP) with intrinsic duration of 1 ms, and dispersion measure of 57~pc~cm$^{-3}$, after being dispersed and scattered by propagation and detected with a resolution of 20~ms and 50~kHz.  The pulse is superimposed onto Galactic noise fluctuations, the average value of which is subtracted at each frequency.  The flux is given in Jy above the sky background. {\it Middle:}  Result of a blind search for GPs using GRAND.  The signal-to-noise ratio (SNR) is saturated at 3.5, at best-fit values 57~pc~cm$^{-3}$ and 15~s.  {\it Bottom:} Fixing the dispersion measure to its best-fit value, the SNR reaches 42.}
\end{figure}

We take the daily all-sky event rate to be $N(f,S)=N_0(f_0,S_0) S^{-3/2} f^{-\alpha}$, at a frequency $f$ for a
sensitivity $S$, where $\alpha$ is the FRB spectral index.  The dependence $S^{-3/2}$ is compatible with a cosmological origin and with FRB as standard candles\ \cite{Rowlinson:2016zjw}.  Though the expression we adopt for $N(f,S)$ is speculative and simplified, it is enough to make estimates.
Using Parkes Observatory data, \Ref\ \cite{Rane:2015sxa} derived $N_0(f_0 = 1.4~\mbox{GHz}, S_0 = 1~\mbox{Jy})=3000$~day$^{-1}$.
For $S=30$~Jy, and $f=300$~MHz, we obtain $N=100$~day$^{-1}$ for a flat
spectrum ($\alpha=0$) and $N=460$~day$^{-1}$ for $\alpha=1$.  
These are attractive numbers, compared to the tens of FRBs detected in total so far.  

Thus, a large number of FRBs should be detectable in GRAND at intensity levels
comparable to the prototypical Lorimer burst\ \cite{Lorimer:2007qn}. 
The largest uncertainty is whether the FRB spectrum extends to low frequencies and with what shape.  In the best-case scenario, FRBs could be detected at a rate of a few thousand per day.



\subsection{Giant radio pulses}
\label{section:Giant_radio_pulses}

Like FRBs, giant radio pulses (GP) are transient astrophysical events: coherent, ms-long, intense, intermittent radio pulses associated to Galactic pulsars, like the Crab\ \cite{Soglasnov:2007mm, Eftekhari:2016fyo} and PRS B1937+21\ \cite{2005JRASC..99Q.139K}.  They are 1000 times shorter and brighter than FRBs, and have smaller dispersion measures.  They are seen at frequencies from tens of MHz to a few GHz --- containing the GRAND frequency band --- with fluxes up to 500~kJy.  Their rate is high: from the Crab, they are detected on the scale of minutes.  

Contrary to the high-energy emission from pulsars, which is due to incoherent synchrotron and curvature emission, the coherent radio emission from pulsars is poorly understood\ \cite{Beskin:2015ria}.  Giant radio pulses add a degree of complexity. 
For instance, \Ref\ \cite{2004A&A...424..227P} proposed that giant pulses could result from radio emission that is Compton-scattered by the electron-positron plasma of the pulsar\ \cite{Sturrock:1971zc, Voisin:2017uqb}.

GPs have lower dispersion measures than FRBs.  For GPs, the dispersion is dominantly attributed to the pulsars themselves, whose values of DM are known.  This can be used to identify GPs in GRAND: if a transient is detected when a pulsar is within view, with a compatible DM value, then it is likely to be a GP from that pulsar.  Otherwise, the DM and the pulse broadening could differentiate between an FRB and a newly discovered GP.  Observation of GPs in the 100--200~MHz in GRAND would improve low-frequency statistics in this range, which is presently poor.  Further, simultaneous observation of GPs by GRAND and other instruments would help to calibrate GRAND.

Continuous coverage of half or more of the sky by GRAND --- {\it vs.}\ a few square degrees for radio-telescopes --- could discover new classes of sources of GPs that have not yet been identified because their rate is too low to be detected by pulsar surveys, \eg, $< 1$~hr$^{-1}$ or $< 1$~day$^{-1}$.  
With half-sky coverage, GRAND may detect GPs from the Crab at 200 MHz above $5 \cdot 10^3$ Jy at a rate of $200$ per day. 
GRAND could also address whether GPs have a high-intensity cut-off: the most intense GP recorded is well above $10^6$~Jy, but it is unknown if this is an upper limit.  It has been proposed that FRBs could be super-giant pulses in galaxies at cosmological distances\ \cite{Pen:2015ema}.  If such a GP occurred in the Crab, it would have a flux of $10^9$~Jy.  Detection of such an event could help understand both GPs and FRBs.  

For GPs, the detection principle is the same as for FRBs, though with larger signal amplitudes, lower DM, and less pulse broadening.  We have simulated the detection of GPs in GRAND following a procedure similar to the FRB simulations.  To produce the signal, based on low-frequency observational data\ \cite{Meyers2017}, we model the GP flux as $S=1~{\rm kJy} (f/100~{\rm MHz})^{-0.7}$ and fix the fluence at 1~Jy~s.  We disperse the signal using ${\rm DM} = 57$~pc cm$^{-3}$ --- the Crab dispersion measure --- and a characteristic scattering time of 10~ms at 100~MHz\ \cite{Meyers2017}.  As for FRBs, we take into account Galactic radio noise, but not RFI.  To simulate detection, the signals received by all of the GRAND200k antennas are added incoherently.  For the results shown here, we used the 200--300~MHz band; however, simulations show that our conclusions hold for the 100--200~MHz band.  

Figure \ref{fig:GP} shows results for one of the GP simulations run for GRAND200k, where the pulse starts at around 18~s in the simulation; see the top panel.   After de-dispersion, GRAND200k would detect the simulated GP with SNR = 58.  An array of 20\,000 antennas --- or two GRAND10k sub-arrays --- would detect it with SNR = 16--19.  Even a smaller array of 300 antennas --- GRANDProto300 --- would detect it with SNR of 6--8, provided the fluence is three times higher and the integration time is reduced to 5~ms.  Thus, prospects for detecting GPs in GRAND are good, even in intermediate construction stages.


\subsection{Cosmology: epoch of reionization}
\label{section:EoR}

The Universe remains largely unknown between redshifts $z = 1100$ --- the surface of last scattering --- and $z = 6$ --- the end of the epoch of reionization (EoR).  At redshifts $z > 20$ --- the Dark Ages --- the only light in the Universe was from photons emitted earlier by the formation of neutral hydrogen --- which makes up the CMB today --- and by the hyperfine transition of the 21-cm spin state of neutral hydrogen.  At $z = 20$ --- the Cosmic Dawn, the beginning of the EoR --- the first generation of stars appeared.  Through $z = 6$, stars ionized neutral hydrogen, which then emitted via the same 21-cm hyperfine transition.  The 21-cm line from the Cosmic Dawn imprinted itself onto the cosmic radiation background as a line-like absorption feature, redshifted today to frequencies between 10 and 200~MHz.

Thus, measuring the 21-cm signal would reveal how the Universe transitioned from a dark phase to a bright phase, how the growth of large-scale structure changed from the linear to the non-linear regime, and how baryonic matter became pre-eminent in the formation and evolution of cosmic structures\ \cite{2012RPPh...75h6901P, 2013ASSL..396...45Z, Ferrara:2014sda}.  Direct imaging of ionized regions\ \cite{2015aska.confE...1K} and statistical studies of brightness fluctuations in the 21-cm spectrum using interferometry\ \cite{2011ApJ...734L..34J, 2012JAI.....150004T, Paciga:2013fj, 2013A&A...556A...2V, 2013PASA...30...31B, 2014MNRAS.445.1084Z, DeBoer:2016tnn,Zheng:2016} could measure the 21-cm cosmic signal with high sensitivity, though they are regarded as challenging observations.

GRAND will have access to an alternative, more direct method.  By measuring the temperature of the sky with mK precision, as a function of frequency, it will reconstruct the global EoR signature and identify the absorption feature due to reionization below 100~MHz.  Unlike other observables, the global EoR signature can reveal the cosmic history of neutral hydrogen.  With this method, using one single-polarization antenna, EDGES\ \cite{Bowman:2018yin} recently found a 500~mK-deep absorption feature centered at 78~MHz, the first claimed observation of the Cosmic Dawn.  However, the depth and shape of the feature --- a relatively flat plateau --- differ from theoretical predictions, which has prompted possible explanations involving charged dark matter \cite{Fraser:2018acy, Cheung:2018vww, Wang:2018azy}.  Improved measurements in GRAND will help decide between competing hypotheses.
 
Below, we compute the sensitivity of GRAND to the global EoR signature.  The sensitivity of a single GRAND antenna, in units of temperature, is
$T_b = T_{\rm sys}/{\sqrt{\triangle\nu\triangle t}}$, where $T_{\rm sys}$ is the equivalent system temperature, $\triangle\nu$ is the frequency bandwidth, and $\triangle t$ is the observation time. 
For the GRAND antennas, we assume $\triangle\nu=1$~MHz and a conservative value of 50~K for the receiver noise, so that $T_{\rm sys} = (50 + T_{\rm sky})$~K, with $T_{\rm sky}=60~(\nu/300~{\rm MHz})^{-2.55}$~K\ \cite{Chapman:2012yj}.  This yields a resolution of $T_b$ = 6.1~mK at 80~MHz to measure the global EoR signature with a single antenna in 24~h.  Adding the signals from $N$ identical antennas improves the resolution by a factor of $1/\sqrt{N}$, potentially achieving 1~mK using just 30 antennas. 

For the method to work, the antennas must be calibrated at the 1~mK level.  This can be achieved, \eg, using the methods in \Refs\ \cite{2012RaSc...47.0K06R,Monsalve:2016xbk}.  The large number of antennas in GRAND will help to control systematics in the calibration, identify and correct unstable behavior in the system, or exclude misbehaving units. 

Since 30 antennas are already enough to search for the global EoR signal, we will explore the potential to carry out this measurement already during GRANDProto300 construction stage (see Section \ref{section:GRANDStages-GRANDProto300}).

\newpage



\section{Detector design and performance}
\label{section:detector_design}

\mybox{{\bf At a glance}}{Tan!70}{grand_brown!20}
{
 \begin{itemize}[leftmargin=*]
  \item
   We have designed the GRAND \HA\ to be specially sensitive to horizontal air showers
  \item
   Simulations predict the 3-year sensitivity to neutrinos to be $4 \cdot 10^{-10}$~GeV~cm$^{-2}$~s$^{-1}$~sr$^{-1}$ around $10^9$~GeV
  \item
   GRAND will be fully efficient for UHECRs and gamma rays above $10^{10}$~GeV and zenith angles $> 65^\circ$
  \item
   100 UHECRs detected per day above $10^{10}$~GeV
  \item
   Several techniques could reduce the background from steady and transient radio sources
  \item
   Shower angular resolution is $< 0.5^\circ$ and should improve
  \item
   Shower energy resolution is 15\%
  \item
   Targeted $X_{\rm max}$ resolution is 20--40~g~cm$^{-2}$
 \end{itemize}
}

Today, radio detection of extensive air showers (EAS) is a mature technique that rivals traditional ones, as demonstrated by experiments such as AERA, CODALEMA, LOFAR, and Tunka-Rex; see \Refs \cite{Connolly:2016pqr, Schroder:2016hrv, Huege:2017khw} for recent reviews.  EAS initiated by UHE particles emit radio signals that are coherent, broadband, and impulsive, and that are only weakly attenuated in the atmosphere, leading to radio footprints of up to 100~km$^2$ on the ground, tens of kilometers away from the shower source.  Because radio antennas are relatively inexpensive and robust, they are suitable to build giant arrays to detect even tiny fluxes of UHE particles.

GRAND will build on the substantial technological, theoretical, and computational progress experienced by the field of radio-detection.  It will extend the field by demonstrating the radio-detection of inclined showers initiated by UHE neutrinos.  Below, we detail the detection principle, design, layout, and expected performance of GRAND.


\subsection{Detection principle}
\label{section:detection_principle}

\begin{figure}[t!]
 \centering
 \includegraphics[trim = 0.38cm 0.2cm 0.2cm 0.1cm, clip=true, width=\columnwidth]{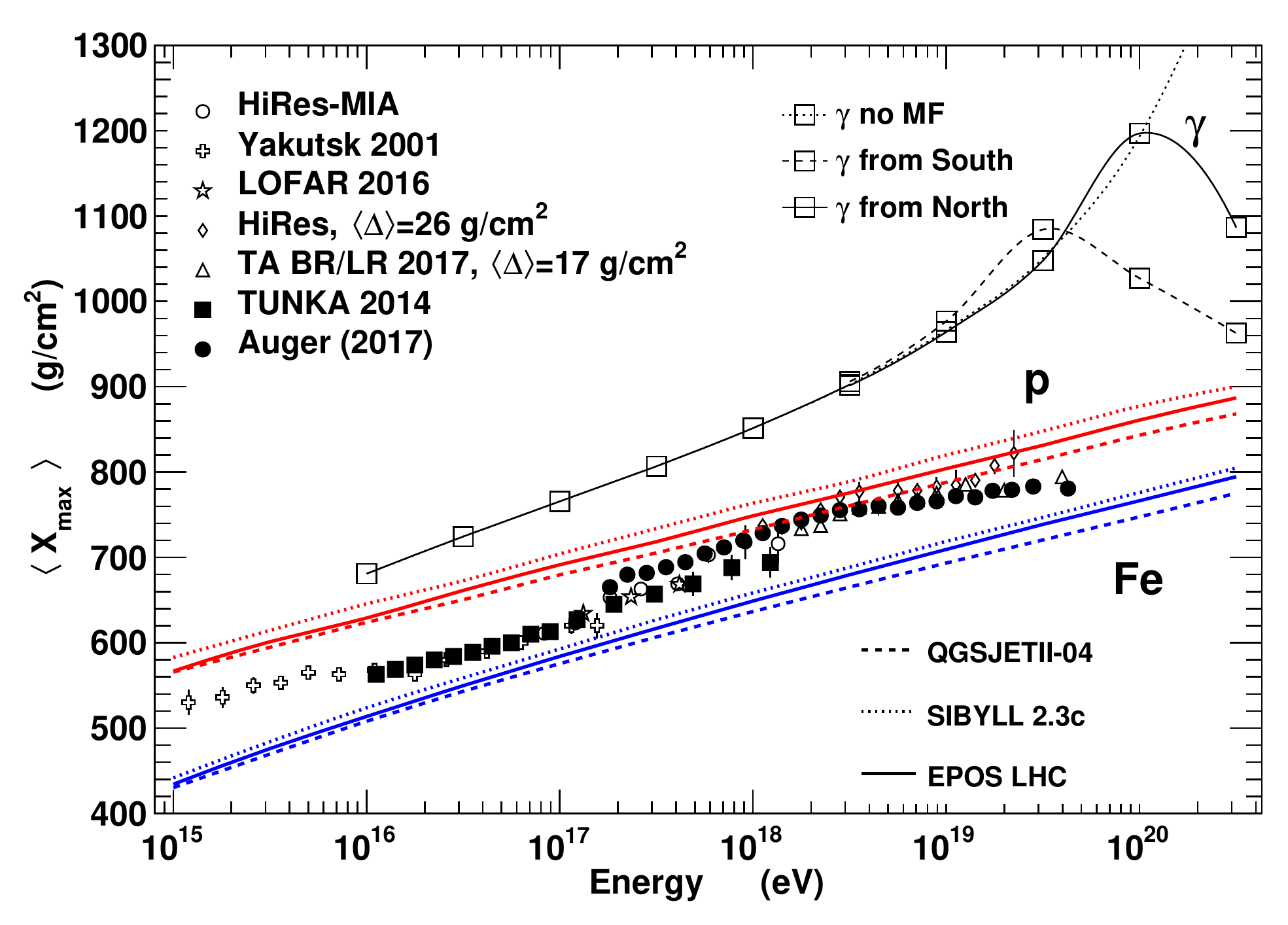}
 \caption{\label{fig:elongrate}Measurements of $\langle X_{\rm max} \rangle$ for air showers initiated by UHECRs by non-imaging Cherenkov detectors  --- Yakutsk\ \cite{2011ASTRA...7..251K, Knurenko:2011zz}, Tunka\ \cite{Prosin:2016jev} --- fluorescence detectors --- HiRes-MIA\ \cite{AbuZayyad:2000ay}, HiRes\ \cite{Abbasi:2009nf}, Telescope Array (TA)\ \cite{Belz:2015cvi}, Auger\ \cite{Bellido:2017cgf, Sanchez-Lucas:2017nhr} --- and a radio detector --- LOFAR\ \cite{Buitink:2016nkf} --- compared to simulations performed using hadronic interaction models QGSJETII-04, Sybill 2.3c, and EPOS LHC, assuming a pure-proton or pure-iron composition.  HiRes and TA data have been corrected for detector effects by shifting them by an amount $\langle \Delta \rangle$, to allow comparison with the unbiased Auger data.  Gamma-ray-initiated air showers are denoted by open squares.  The effect of the geomagnetic field, taken here at the Auger site, depends on the direction of the shower\ \cite{Homola:2006kx}.}
\end{figure}

\subsubsection{Radio emission from extensive air showers}

When a high-energy particle --- a charged particle, a gamma ray, or a neutrino --- interacts with an atom in the atmosphere, the ensuing chain of reactions produces an EAS, containing up to millions of particles, including gamma rays, electrons, positrons, muons, and hadrons.  As the shower propagates toward the ground, the number of particles in it grows until it reaches a maximum number, proportional to the energy of the primary particle that triggered the shower.  Henceforth, particles in the shower are gradually absorbed in the atmosphere.  See, \eg, \Ref\ \cite{Gaisser:2016uoy} for a review of EAS.

The atmospheric column depth at which the shower reaches its maximum particle content, $X_{\rm max}$, is statistically related to the identity of the primary particle.  On average, protons travel a longer distance in the atmosphere than heavier nuclei of the same energy before interacting and triggering an EAS.  In the atmosphere, neutrino-initiated showers are expected to be rare, due to the long mean free path of neutrinos.

Figure\ \ref{fig:elongrate} shows that the average $\langle X_{\rm max} \rangle$ is about 100~g~cm$^{-2}$ deeper for a proton primary than for an iron primary of the same energy.  Gamma ray-initiated showers reach their maximum even deeper.  In addition, in cosmic ray-initiated showers, the spread in $\langle X_{\rm max} \rangle$ is larger for lighter primaries because shower-to-shower fluctuations are larger, which complicates inferring the identity of the primary.  Uncertainties in the hadronic interaction models used to simulate EAS further complicate the issue.
 
As the shower develops in the atmosphere, the geomagnetic field separates positive and negative charges.  This creates a time-varying electric current that induces geomagnetic radio emission\ \cite{1966RSPSA.289..206K}.  Additionally, an excess of negative charge builds up in the shower during propagation due to Compton scattering, which also induces radio emission.  The process is known as the Askaryan effect\ \cite{Askaryan1, Askaryan2}.  In dense media, like ice, the Askaryan effect dominates, while in air the geomagnetic effect dominates.  

Both types of emission are coherent for wavelengths longer than the size of the particle shower, so the signal amplitude scales linearly with the energy of the electromagnetic component of the shower for frequencies up to 100~MHz.  Each type of emission has a different polarization pattern.  Because the emitting particles are relativistic, the emission is beamed in the forward direction, inside a narrow cone of half-width equal to the Cherenkov angle.

\begin{figure}[t!]
 \includegraphics[width=\columnwidth, keepaspectratio, trim = 0 0 0 0.6cm, clip=true]{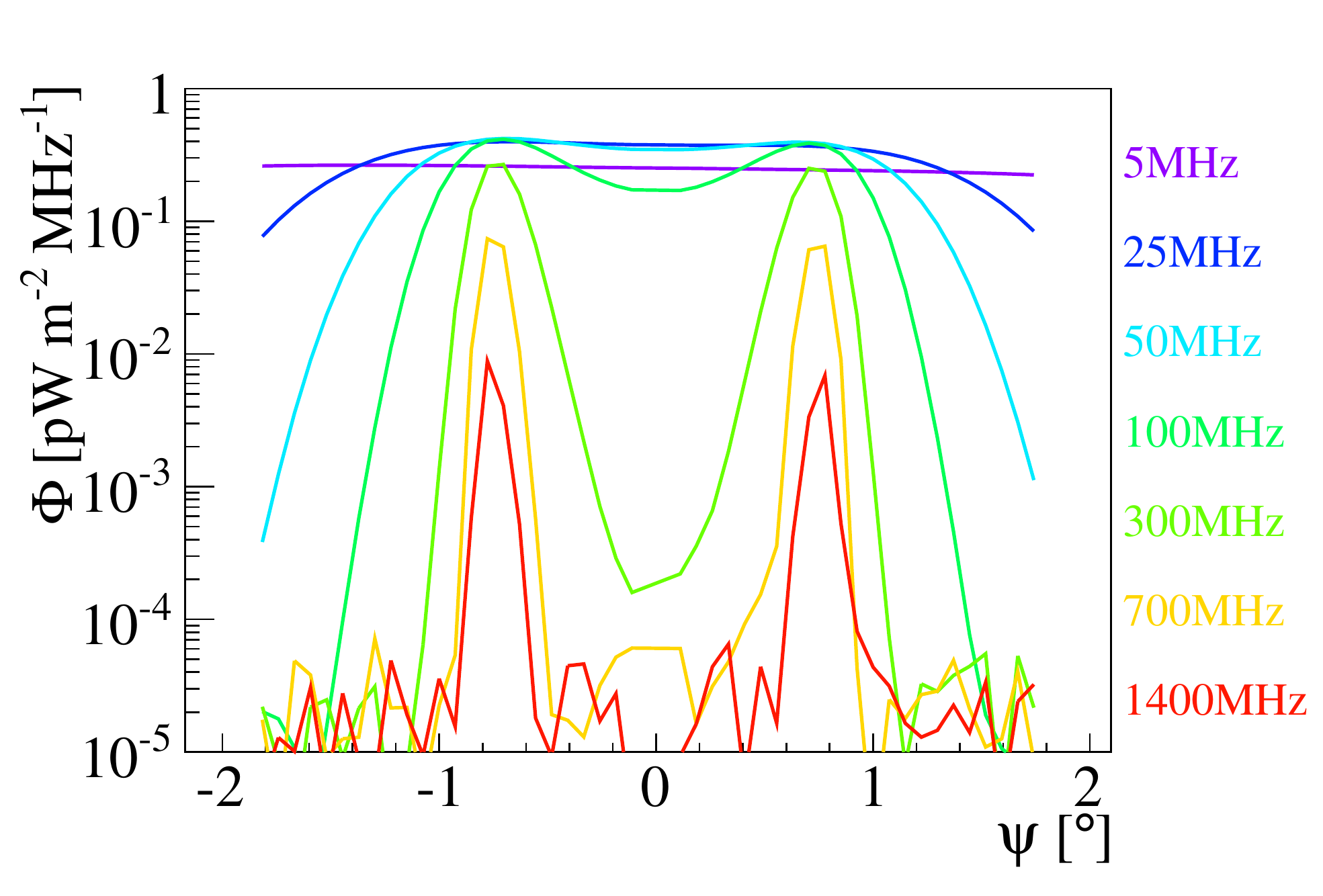}
 \caption{Flux density $\Phi$ as a function of frequency and off-axis angle $\psi$ for an air shower with zenith angle $\theta_z=71^{\circ}$ and energy of $10^{8.8}$~GeV, simulated with ZHAireS\ \cite{Zhaires:2012}.  At each frequency, the flux density
 is computed as the power spectrum averaged over a period of 10~ns.  Figure taken from \Ref\ \cite{Alvarez-Muniz:2014dza}.}
 \label{fig:fig4_freqcone}
\end{figure}

\medskip

{\it Cherenkov effects.---}  In a medium with constant index of refraction $n$, the Cherenkov angle is given by
$\theta_{\rm Ch}~=~\arccos\left[ (n\beta)^{-1} \right]$,
where $\beta \equiv v/c \approx 1$ is the speed of the relativistic particle shower.  The index of refraction in air is close to unity --- at ground level, $n\approx 1.0003$ --- and the corresponding Cherenkov angle is 1--2$^\circ$ (in ice, where $n \approx 1.78$, it is $56^\circ$).  For an observer located at this specific angle to the shower, the radio signals emitted from all points along the shower arrive simultaneously, boosting the signal along a ``Cherenkov ring'', up to GHz frequencies.  

Figure \ref{fig:fig4_freqcone} shows the flux density of a simulated EAS as a function of off-axis angle, for different frequencies\ \cite{Alvarez-Muniz:2014dza}.  The Cherenkov ring is increasingly more visible the higher the frequency.  LOFAR detected a Cherenkov ring prominently in the 120--200~MHz band and less prominently in the 30--80~MHz band\ \cite{Nelles:2014dja}.  When moving away from the Cherenkov ring, coherence is lost at the highest frequencies.

For cosmic ray-initiated showers with small zenith angles and $X_{\rm max}$ at a height of 4~km above sea level, the ``Cherenkov distance'' is roughly 100~m.  This is the distance between the shower core and the position of the Cherenkov ring in the plane perpendicular to the shower propagation axis.  For more inclined showers, like the ones initiated by Earth-skimming neutrinos and targeted by GRAND, the peak emission comes from farther away, leading to larger Cherenkov distances.  A detailed description of the Cherenkov effect in air showers is given in \Refs\ \cite{PalomaresRuiz:2005xw, AlvarezMuniz:2012sa, deVries:2013dia}.

\medskip

{\it Shower geometry.---}  The different radio emission mechanisms from air showers, combined with the Cherenkov effect, result in complex emission patterns in terms of amplitude, frequency, and polarization.  They are well understood and can be modeled in great detail if the shower geometry is known.  Conversely, the radio-detection of an air shower by a large number of antennas at different locations can be used to reconstruct the shower geometry and infer the properties of the primary particle; see Section \ref{section:GRANDDesignPerf-Performance-Reconstruction}.  

\begin{figure}[t!]
 \centering
 \includegraphics[width=\columnwidth, keepaspectratio]{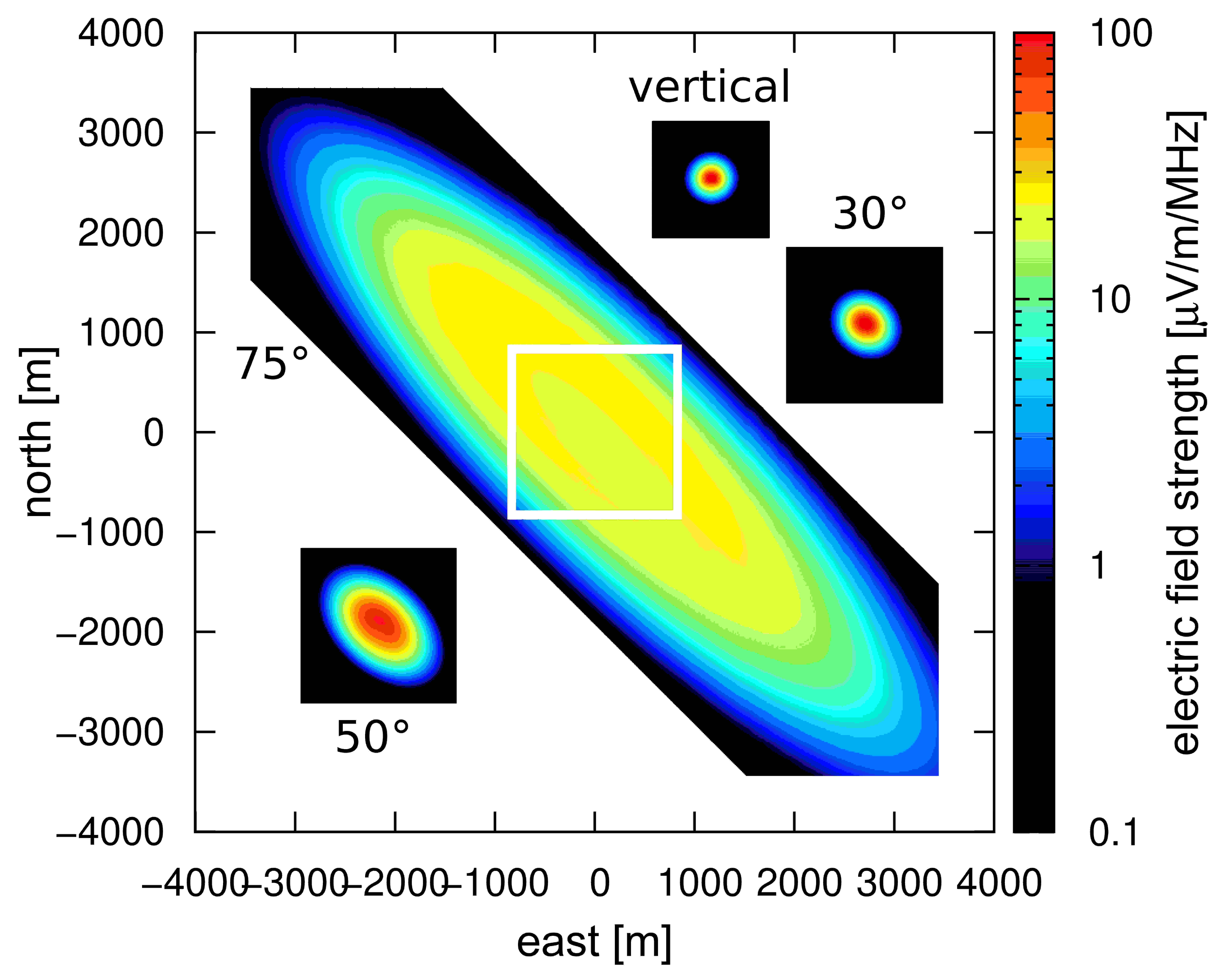}
 \caption{\label{fig:fig7_kdv}Radio footprint for various shower inclinations, from CoREAS simulations\ \cite{Huege:2013vt}. Figure taken from \Ref\ \cite{Huege:2016veh}.}
\end{figure}

Figure \ref{fig:fig7_kdv} shows that, for vertical showers, the radio footprint is small and concentrated; dense antenna arrays, with spacings smaller than 100~m, are needed to sample it.  For showers inclined by $75^\circ$, the elongated footprint spans an area of several tens of km$^2$, and sparser antenna arrays are sufficient to sample it, as shown by AERA\ \cite{Aab:2018ytv} and as planned for GRAND.

\medskip

{\it Neutrino-initiated air-showers.---}  
At EeV energies, the neutrino-nucleon deep-inelastic-scattering cross section increases roughly $\propto E^{0.363}$\ \cite{Gandhi:1998ri}.  
Thus, the interaction length inside Earth decreases from 6\,000~km at PeV to a few hundred km at EeV, making the Earth opaque to even Earth-skimming neutrinos.  If the neutrino interaction is charged-current, it produces an outgoing high-energy charged lepton of the same flavor as the interacting neutrino, whose decay or interaction products might be observable by GRAND, depending on the flavor and on whether the charged lepton is able to exit into the atmosphere.

If the interacting neutrino is a $\nu_e$, the outgoing electron initiates an electromagnetic shower that is promptly damped by radiative energy losses.  Typically, the shower is short-lived, concentrated close to the neutrino interaction vertex, and undetectable in radio.  At ultra-high energies, if the $\nu_e$ interacts close to the Earth surface, a fraction of the ensuing shower could emerge into the atmosphere and produce coherent transition radiation in the MHz--GHz range\ \cite{deVries:2015oda, Motloch:2015wca}, detectable by GRAND.  The Landau-Pomeranchuk-Migdal effect\ \cite{Landau:1953um, Landau:1953gr, Migdal:1956tc} would suppress radiative losses and elongate the shower, further improving its chances of reaching the atmosphere\ \cite{Tartare:2012zz}.  However, at 1~EeV, the maximum column depth that the shower can traverse is 3000--4000~g~cm$^{-2}$\ \cite{AlvarezMuniz:1997sh}, corresponding to only 30--40~m underground.  Therefore, the effective volume for this detection channel is small.  Accordingly, in the present study we neglect the contribution of $\nu_e$ interactions in rock.

\begin{figure*}[!htb]
 \centering
 \includegraphics[width=0.9\textwidth]{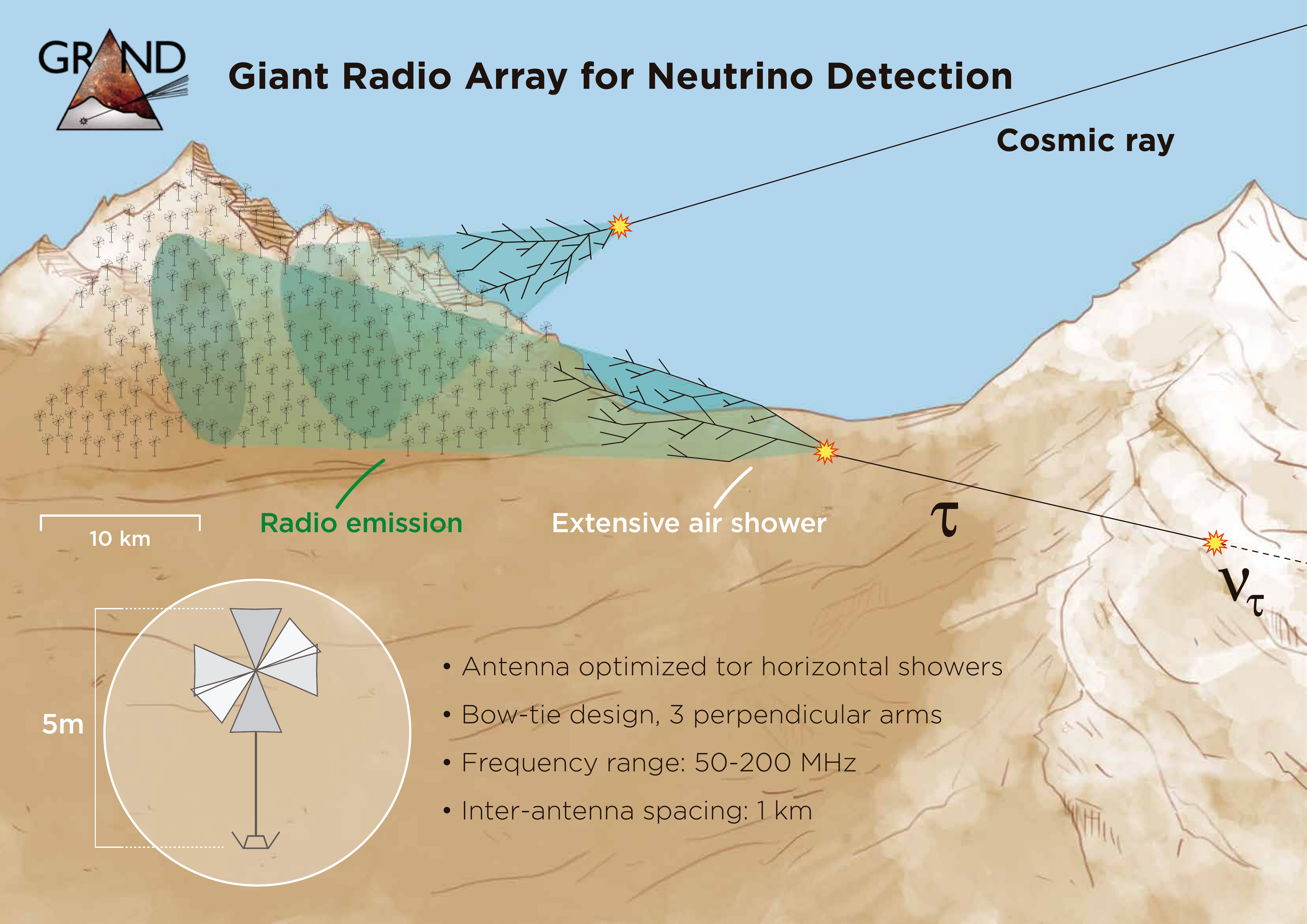}
 \caption{\label{fig:grand_det_principle}GRAND detection principle, illustrated for one of the 10\,000-antenna GRAND10k arrays located at a hotspot.  See main text for details.  Ultra-high-energy cosmic rays and gamma rays (not shown) interact in the atmosphere, while ultra-high-energy $\nu_\tau$ interact underground and create a high-energy tau that exits into the atmosphere and decays.  The ensuing extensive air showers emit a radio signal that is detected by the antennas.  The inset shows a sketch of the \HA~designed for GRAND.}
\end{figure*}

If the interacting neutrino is a $\nu_\mu$, the outgoing high-energy muon can travel underground from its point of creation and exit into the atmosphere, since radiative losses are suppressed due to the larger mass of the muon\ \cite{Groom:2001kq}.  
Because the range of the muon in the atmosphere is of several kilometers, the probability that it decays above the radio array and generates a detectable air shower is negligible.  Therefore, for all practical purposes, GRAND will not be sensitive to Earth-skimming $\nu_\mu$.

If the interacting neutrino is a $\nu_\tau$, it is possible to detect the outgoing tau in GRAND.  On the one hand, the tau is heavier than the muon, so radiative losses are suppressed even further.  On the other hand, the short lifetime of the tau (0.29~ps) gives it a range of 50 meters per PeV of energy before decaying.  As a result, a tau born from an UHE $\nu_\tau$ underground could exit the rock and decay above the antenna array, triggering a particle shower that emits a radio signal detectable by the antennas; later, we detail how.  Because the branching ratio of tau into hadrons --- mostly pions --- is about 65\%, and nearly 20\% to electrons, the chance of the tau decay creating an air shower is high: only when the tau decays into a muon (17\% of cases) does the shower not emit a radio signal, as explained above.

In addition, since the decay of a tau makes a new $\nu_\tau$, the attenuation of the neutrino flux due to matter interactions as it propagates inside the Earth is partially balanced by the regeneration of $\nu_\tau$.  However, since the outgoing $\nu_\tau$ receives only 30\% of the parent tau energy, multiple regenerations shift the flux to lower energies\ \cite{Palomares-Ruiz:2015mka, Vincent:2017svp}.


\subsubsection{Detection strategy}
\label{section:detection_principle_sub_detection_strategy}

The detection strategy of GRAND is built upon the fundamental principles introduced above.  Below, we focus on the detection of Earth-skimming $\nu_\tau$ that interact underground and give preliminary remarks about the possibility of detecting neutrino interactions in the atmosphere.

Figure\ \ref{fig:grand_det_principle} sketches the detection principle of GRAND.  Below, we explain it.

\medskip

\paragraph{Earth-skimming underground events}  

\smallskip

The dominant neutrino detection channel in GRAND, first proposed in \Ref\ \cite{Fargion:1999se}, is described by:
\begin{enumerate}
 \item
  A $\nu_\tau$ makes a tau by interacting underground or inside a mountain
 \item
  The tau exits the rock into the atmosphere and decays in-flight 
 \item
  The decay produces an EAS whose radio signal is detected in GRAND
\end{enumerate}
The high density of rock helps the chances of neutrinos interacting in it.  Further, instrumenting the surface with radio antennas is arguably easier than instrumenting a dense medium.  The strategy described above is efficient only at ultra-high energies, where the tau range is of several km, and for incoming neutrinos with Earth-skimming trajectories, for which the distance traveled by the neutrino is comparable to its interaction length.

\begin{figure}[t!]
 \centering
 \includegraphics[trim=0.8cm 0.1cm 0.2cm 0.4cm, clip=true, width=\columnwidth]{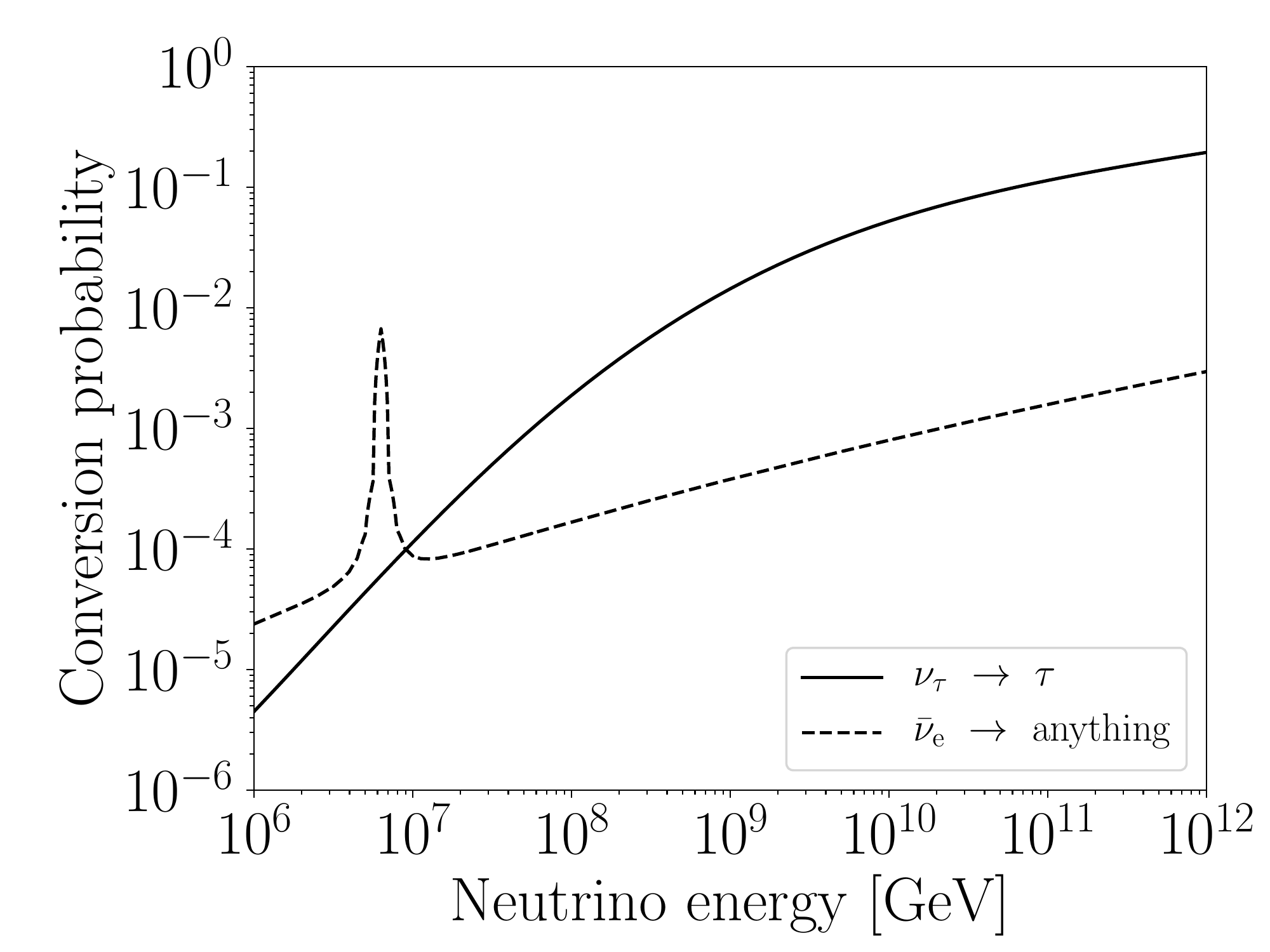}
 \caption{\label{fig:nu_tau_efficiency} Conversion probability
 of a $\nu_\tau$ into a tau emerging with $1^\circ$ of elevation from a flat
 Earth made up of standard rock, with density $2.65$~g~cm$^{-3}$,
 and probability of interaction in the atmosphere 
 of a downward-going $\overline{\nu}_e$ with $1^\circ$ of
 elevation. The latter assumes the U.S.\ standard atmosphere density profile, a spherical
 Earth, and detection at sea level.  The peak around 6.3~PeV is due to the $\bar{\nu}_e$ Glashow resonant cross section\ \cite{Glashow:1960zz},
 not including Doppler broadening from the motion of atomic electrons~\cite{Loewy:2014zva}.
 The probabilities were computed using NuTauSim\ \cite{Alvarez-Muniz:2017mpk}.}
\end{figure}

In our simulations, we compute steps 1 and 2 using a custom-made numerical program that calculates the probability for a $\nu_\tau$ to generate a tau that emerges from the ground into the atmosphere; see Section \ref{section:GRANDDesignPerf-Performance-NuDetection} for details.  

Figure\ \ref{fig:nu_tau_efficiency} shows the $\nu_\tau \to \tau$ conversion probability as a function of neutrino energy.  Between 1 and 10~EeV, the probability is significant, at a few percent.  Several factors may modify this number\ \cite{Alvarez-Muniz:2017mpk}.  In particular, the conversion probability strongly depends on the thickness of the surface layer: it can be significantly enhanced by placing the detector in mountainous terrain.  Figure\ \ref{fig:nu_tau_efficiency} does not take into account the detection efficiency, which may differ significantly for Earth-skimming underground and atmospheric events, as explained below.

\medskip

\paragraph{Earth-skimming atmospheric events}  

\smallskip

GRAND may also detect Earth-skimming neutrinos that interact in the atmosphere.  For neutrinos propagating in the atmosphere, coming from slightly above the horizon, the traversed column depth is equivalent to about 100~m of rock.  Hence, the probability of interaction in the atmosphere is non-negligible, though still significantly smaller than for Earth-skimming neutrinos interacting in rock.  However, in atmospheric events, all neutrino flavors could be detected via the hadronic showers that they trigger in charged-current and neutral-current interactions.  

Figure\ \ref{fig:nu_tau_efficiency} shows the interaction probability of a $\bar{\nu}_e$ in the atmosphere.  Preliminary estimates indicate that the rate of atmospheric events might be one-tenth that of underground events.  Detailed simulations will improve this estimate.  At present, we neglect the contribution of atmospheric events to the total event rate.


\subsection{Antenna design}
\label{section:antenna_design}

\begin{figure}[t!]
 \centering
 \includegraphics[width=\columnwidth]{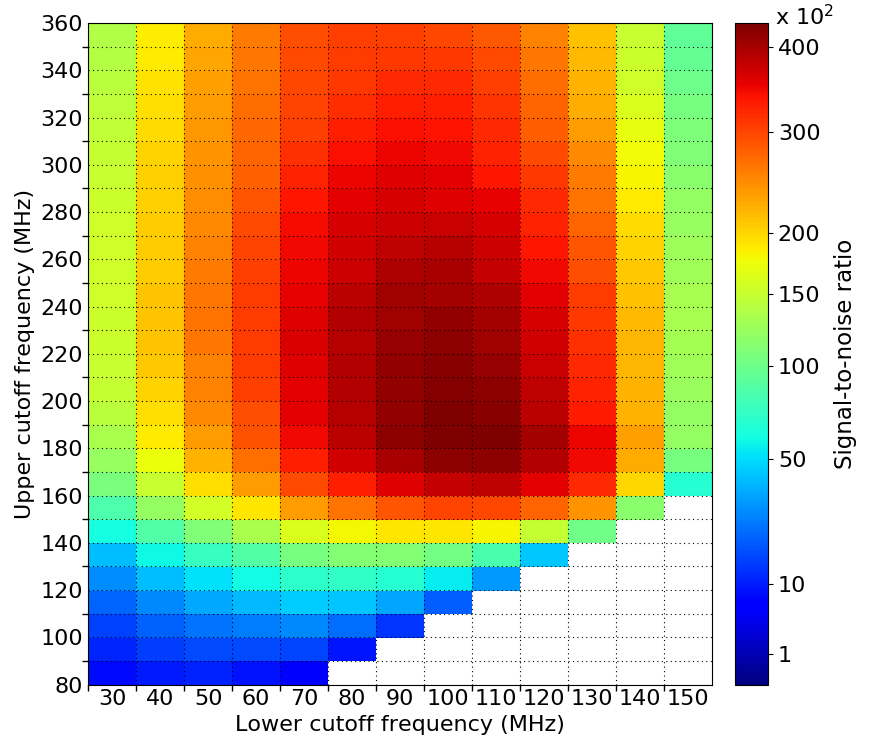}
 \caption{\label{fig:optfreqband}Simulation of the signal-to-noise (SNR) ratio seen in a typical GRAND \HA\ located on the Cherenkov ring made by a slightly up-going neutrino-initiated air shower of energy 0.5~EeV.  The lower and upper cut-off in frequency were varied to maximize the SNR and optimize the frequency band.}
\end{figure}

In GRAND, radio signals from air showers initiated by Earth-skimming neutrinos will arrive with zenith angles close to 90$^{\circ}$ and a polarization that is mostly horizontal.  This introduces a serious challenge for radio-detection, as the diffraction of radio waves off the ground severely alters the antenna response.

\begin{figure*}[t!]
 \centering
 \includegraphics[width=\columnwidth]{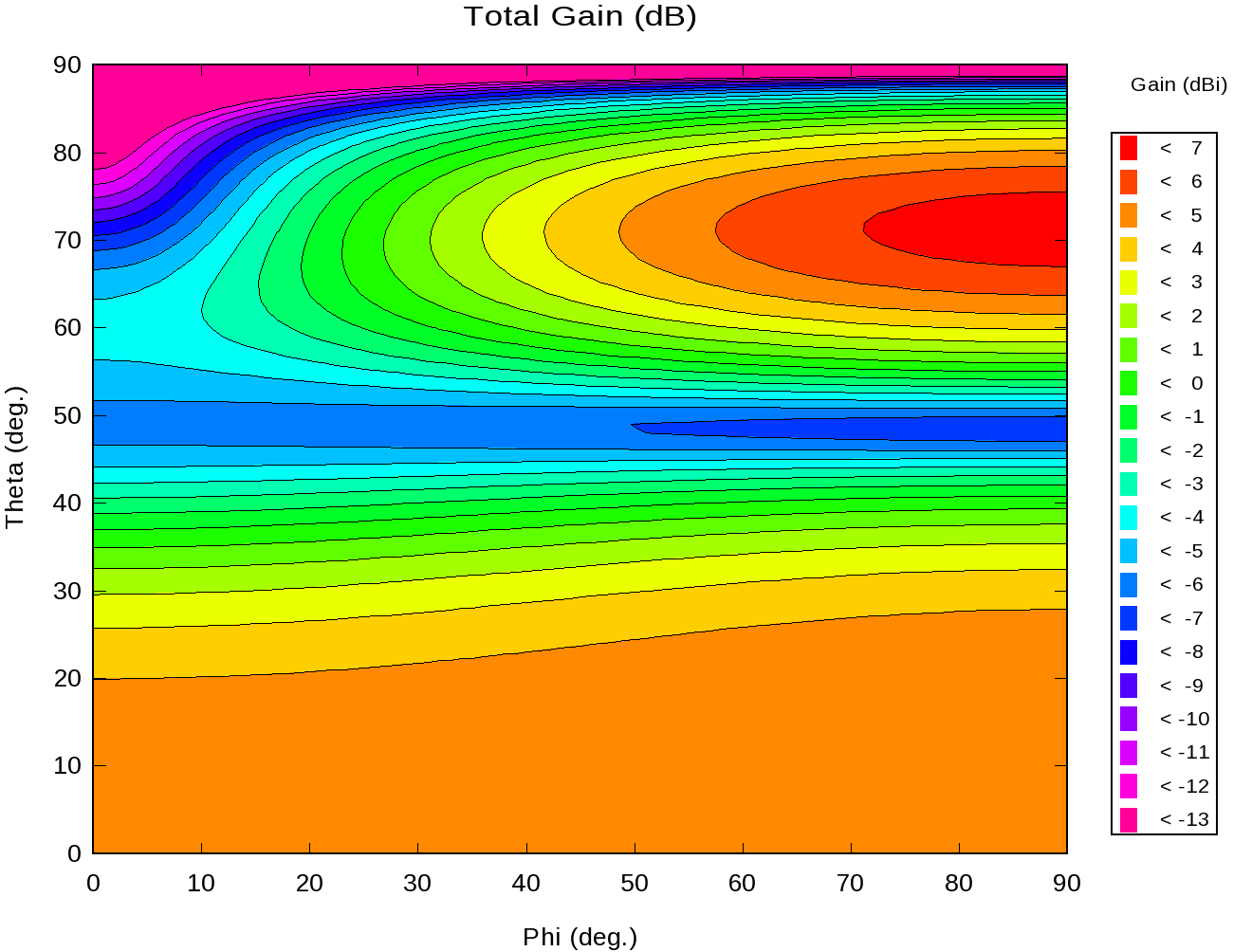}
 \includegraphics[width=0.8\columnwidth]{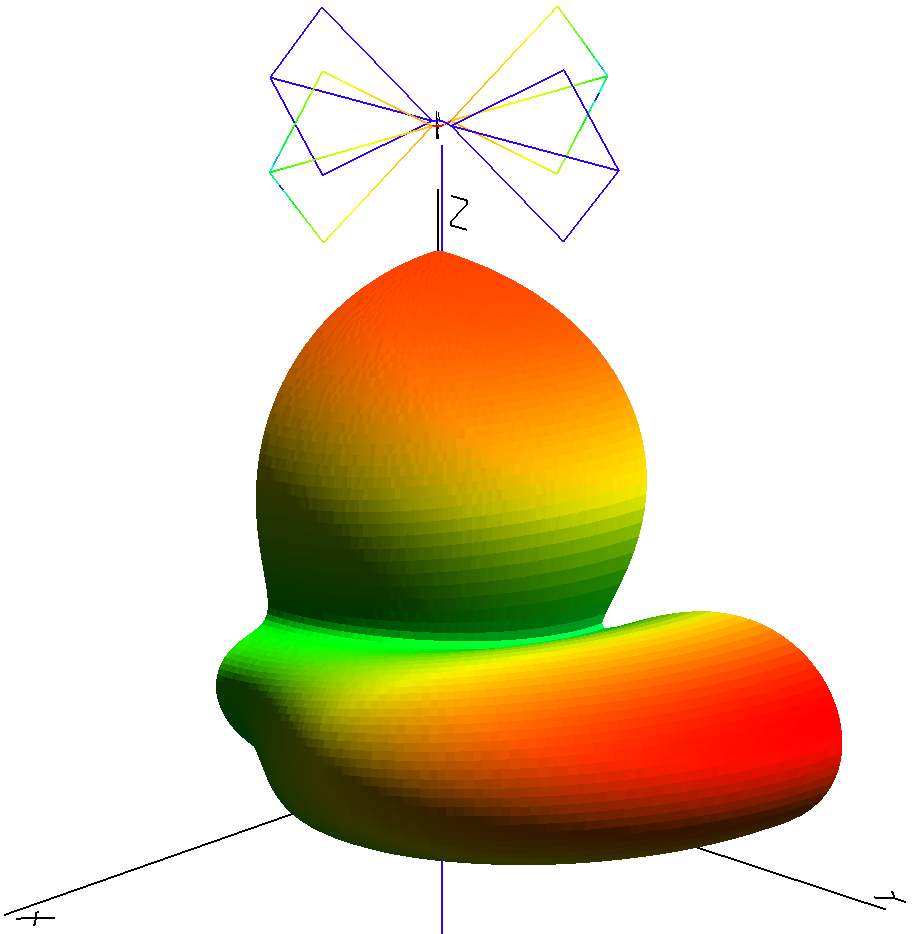}
 \hspace*{0.7cm}
 \includegraphics[width=\columnwidth]{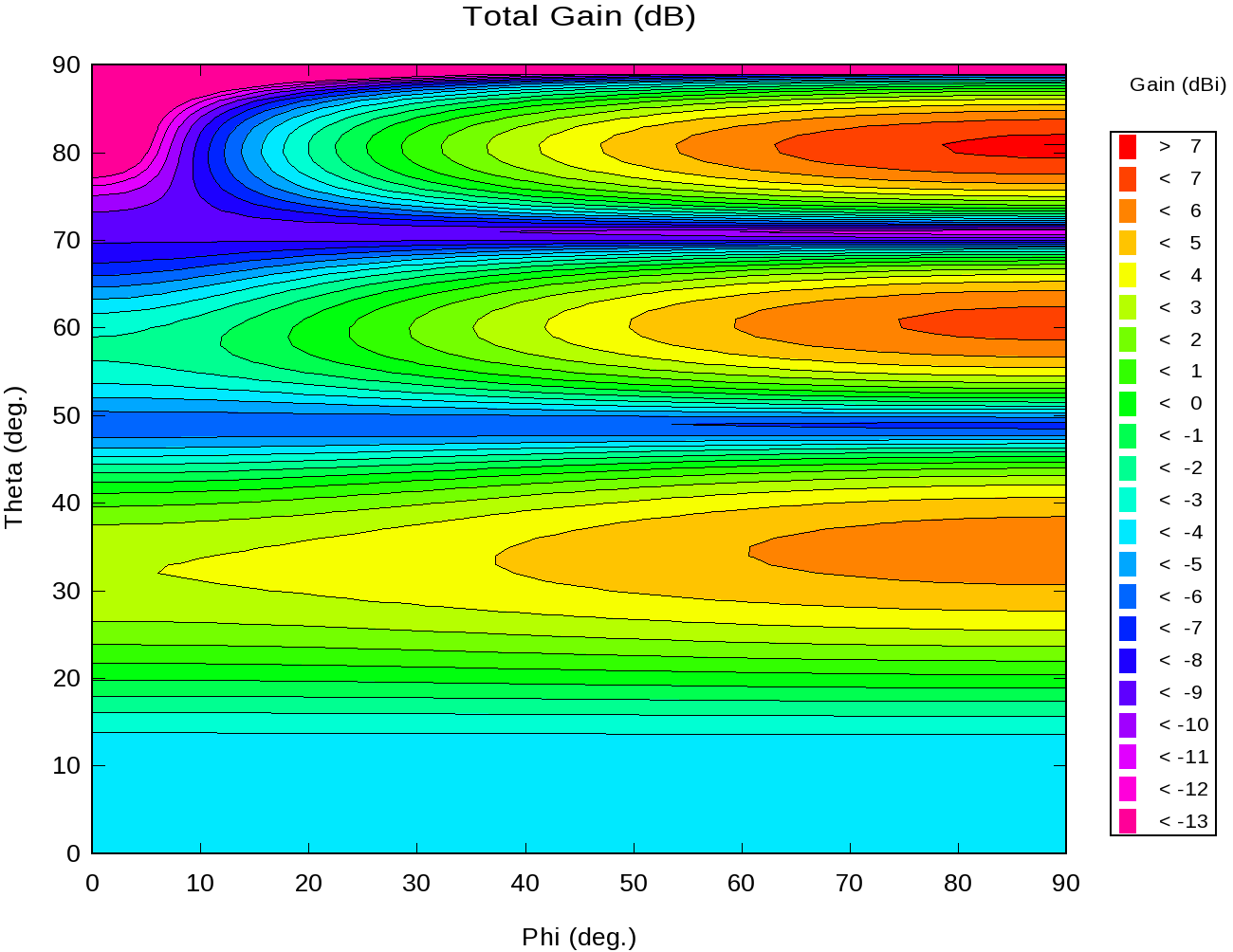}
 \includegraphics[width=0.9\columnwidth]{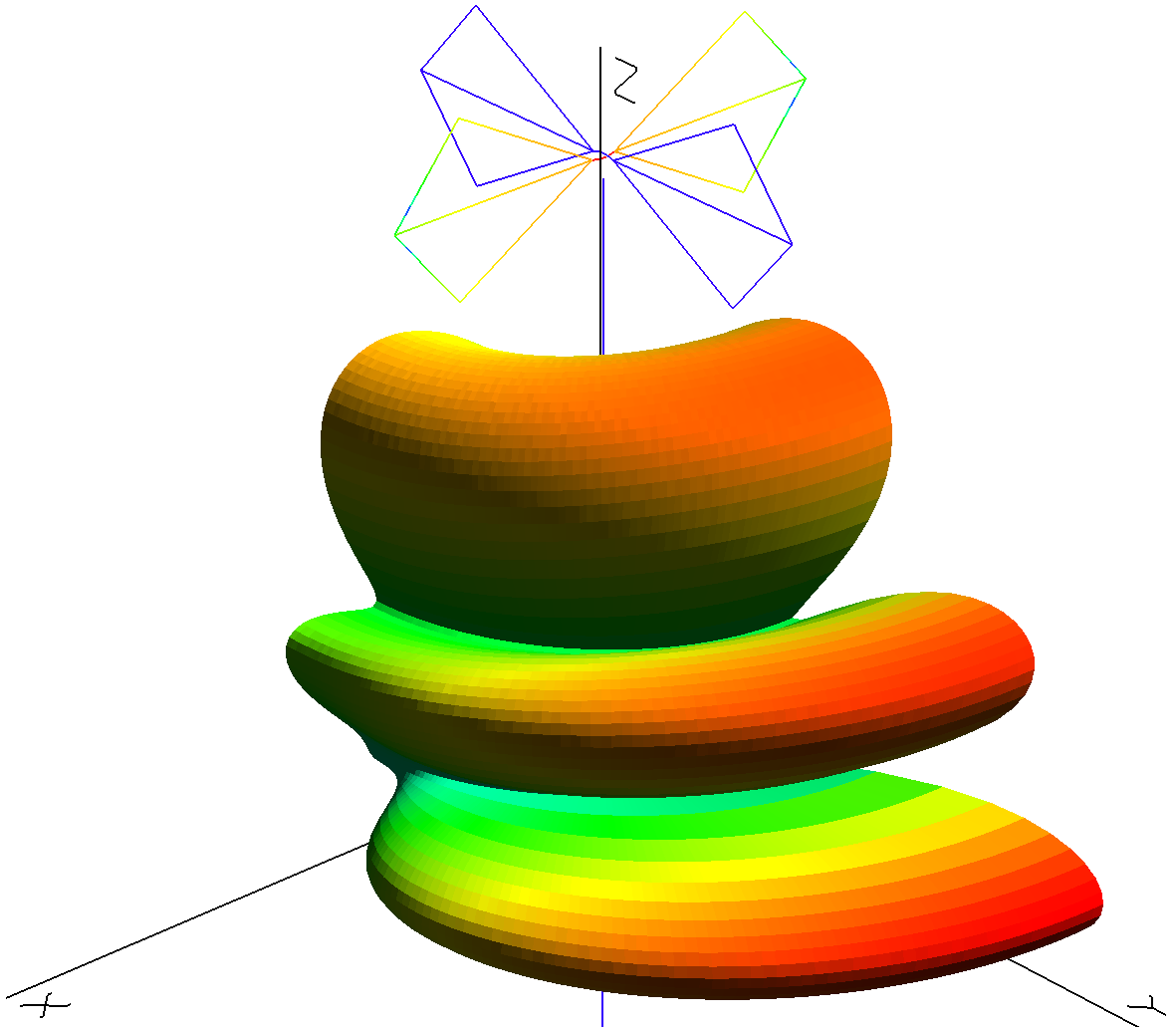}
 \caption{\label{fig:ha_gain}Two-dimensional ({\it left column}) and three-dimensional ({\it right column}) total gain of the X-arm of the GRAND \HA\ as a function of direction. {\it Top row:} At 50~MHz.  {\it Bottom row:} At 100~MHz.}
\end{figure*}

To address this problem, we have designed the GRAND antennas to have a high detection efficiency along the horizon --- we call the design \HA.  Because the effect of ground reflection decreases with $h/\lambda$, where $h$ is the detector height above ground and $\lambda$ is the radio wavelength, we place the \HA\ at $h=5$~m --- atop a wooden pole --- and the frequency range to $f>50$~MHz ($\lambda < 6$~m).  Because we would like to detect radio Cherenkov rings --- which could help background rejection and signal reconstruction (see Section \ref{section:GRANDDesignPerf-Performance-Reconstruction}) --- we set the upper limit of the frequency range to 200~MHz, instead of the 80~MHz or 100~MHz used in most existing arrays.  This is aided by the radio background dropping significantly above 100~MHz; see Section\ \ref{section:background}.  Further, recent studies made for other air-shower arrays confirm that extending the frequency band to 200~MHz significantly improves the signal-to-noise ratio and lowers the detection threshold\ \cite{V.:2017kbm}. To confirm the validity of this result for horizontal showers, we found the optimal frequency band for GRAND by following a procedure similar to the one in \Ref\ \cite{V.:2017kbm}, using the response of a dipole antenna.  We based it on ZHAireS simulations of horizontal showers, using the physical conditions at the GRANDProto35 location; see Section\ \ref{section:GRANDStages-GRANDProto35}.

Figure\ \ref{fig:optfreqband} shows results from one of our simulated showers.  The determination of the signal-to-noise ratio (SNR) in different frequency bands is based on the signals of the North-South and East-West polarization.  For the radio noise, we assumed the average Galactic background plus additional thermal noise of 300~K.  We found the optimal frequency band for a GRAND array to be 100--180~MHz, consistent with the results obtained in \Ref\ \cite{V.:2017kbm}.

The \HA\ is an active bow-tie antenna with a relatively flat response as a function of azimuthal direction and frequency.
Its design is inspired by the ``butterfly antenna''\ \cite{Charrier:2012zz} developed for CODALEMA, and later used in AERA\ \cite{Abreu:2012pi}.  It has 3 perpendicular arms (X, Y, Z) oriented along two horizontal directions and a vertical one.  The \HA\ uses the same low-noise amplifier, but its radiating element is half the size of that in CODALEMA and AERA, in order to increase the sensitivity to the 50--200~MHz range.

Figure \ref{fig:ha_gain} shows the two- and three-dimensional total gain of the \HA\ as a function of direction, at 50~MHz and 100~MHz, computed with the NEC4 simulation code\ \cite{NEC4}.  The antenna gain $G(\theta_z,\phi)$ is defined, in emission mode, as the ratio of the power $U$ radiated in the direction $(\theta_z,\phi)$ to the mean radiated power, \ie, $G(\theta_z,\phi) = U(\theta_z,\phi)/\langle U \rangle_{4\pi}$.  By emission-reception reciprocity, this gain also determines the antenna sensitivity.  At all frequencies, the \HA\ has an optimized response down to a few degrees above the horizon.  

Figure \ref{fig:ha_gain} shows that the gain varies strongly with zenith angle. 
For $\theta_z \approx 90^{\circ}$, this is due to simplified simulations settings in NEC4, \ie, using a radio source placed at infinite distance away from the antenna and perfectly flat ground.  In reality, the variation may be milder, especially for the shortest wavelengths. 
Away from the horizon, the strong variation of the lobes depend strongly on the incoming wave frequency.  Therefore, they are smoothed when receiving broadband waves, such as those emitted by air showers.  Careful experimental verification of the \HA\, response as a function of direction remains to be performed.

A prototype of the \HA\ was successfully tested in 2018 during the site survey for GRANDProto300; see Section \ref{section:GRANDStages-GRANDProto300}.


\subsection{Array layout}
\label{section:array_layout}

The large size of the radio footprint for very inclined showers makes it possible to instrument a large area using a sparse array.  Below, we show that a convenient strategy for deploying GRAND is to make it modular, \ie, to divide it into 10--20 geographically separate and independent GRAND10k sub-arrays, each containing about $10^4$ antennas.  Even individually, each GRAND10k sub-array will have a rich science program; see Section\ \ref{section:GRANDStages-GRAND10k}.  The modular strategy will allow GRAND to build up sensitivity to progressively smaller fluxes of UHE particles, while distributing construction efforts.

\begin{figure*}[t!]
 \centering
 \includegraphics[width=\textwidth]{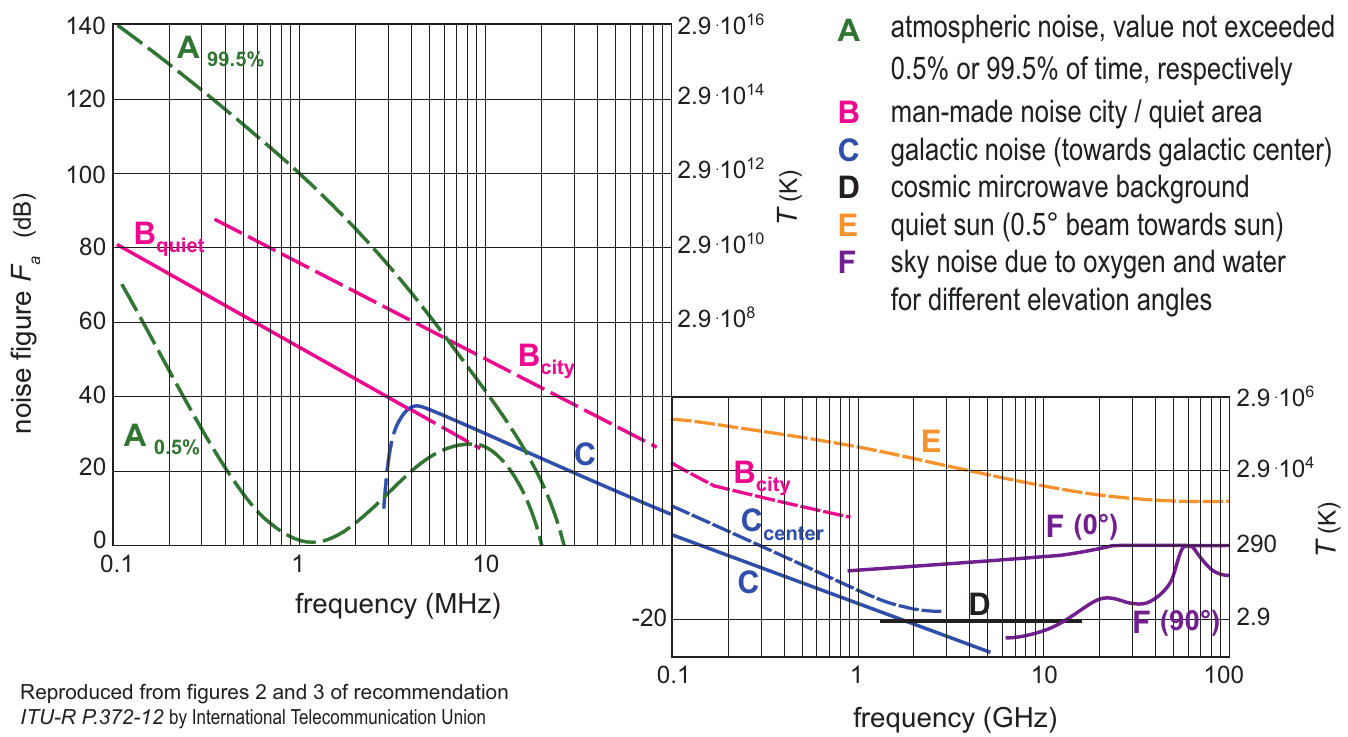}
 \caption{Sources of external radio noise as a function of frequency, expressed as temperature or noise figure $F = 10 \log_{10}(1+T/T_{\rm amb})$.  The blackbody radiation emitted by the ground corresponds to a straight line at $T = T_{\rm amb} = 290$~K. Figure taken from \Ref\ \cite{Schroder:2016hrv}, adapted from \Ref\ \cite{ITU_R}.}
 \label{fig:radio_noise}
\end{figure*}

GRAND10k arrays will be deployed on sites with topographies favorable to neutrino detection.  These sites are found on mountain slopes, which feature the following appealing characteristics:
\begin{itemize}
 \item
  {\bf Collection of a larger fraction of the radio emission compared to a flat site:}  Geometric considerations show that a mountain slope with an elevation of 1000--2000~m acts as an efficient projection screen for the forward-beamed radio signal of a neutrino-initiated shower that emerges from underground, while antennas lying in a valley would fail to detect the signal\ \cite{Montanet:2011zz}
 \item
  {\bf Increased antenna sensitivity:}  Geometric considerations show that the response of a dipole antenna placed on a 10$^{\circ}$ mountain slope to a horizontal shower propagating towards this mountain should be the same as the response to a $\theta_z =80^{\circ}$ downward-going shower of an identical antenna placed on a flat ground.  This is confirmed with good precision by NEC4 simulations.  This is a significant effect, since antenna response drops when the waves come from the horizon; see \Fig\ \ref{fig:ha_gain}. 
 \item
  {\bf Improved reconstruction:} The difference in antenna altitudes on a slope provides a good handle on the reconstruction of the zenith angle for very inclined showers; see Section \ref{section:GRANDDesignPerf-Performance-Reconstruction} for details.
\end{itemize}

Ideally, the antenna array deployed on a mountain slope should face another mountain that could serve as an additional target for interaction of downward-going neutrinos, which would trigger downward-going showers.  The opposing mountain should be distant enough --- a few tens of kilometers --- for the shower to develop and the radio emission cone to enlarge before hitting the antennas.   Simulations indicate that, for specific topographies, these downward-going ``mountain events'' could be as frequent as the upward-going ``underground events''; see Section \ref{section:GRANDDesignPerf-Performance-NuDetection}.  The opposing mountain also provides a means of background rejection, since it acts as a natural screen to stop very inclined UHECRs, simplifying their discrimination from neutrinos; see Section\ \ref{section:background}.

GRAND will be deployed in radio-quiet areas.  Our criteria for radio-quietness are that the stationary noise level $\sigma_{\rm noise}$ is close to the Galactic radio background in the 50--200~MHz band and that the rate of transient signals with peak amplitude larger than $5 \sigma_{\rm noise}$ is below 1~kHz in that band under standard operating conditions.

Logistics is also an important aspect.  Access to the detection units must be reasonably easy.  A large power supply to run the DAQ system and broadband Internet connection should be available.  Weather conditions at the site should allow for stable operation of the electronics, in particular of solar panels; see Section \ref{section:GRANDStages-GRAND200k}. 

Our surveys have found that several sites in China fulfill the above requirements, making it the leading candidates to host GRAND.  This is where the prototype stages of GRAND will be deployed; see Section \ref{section:construction_stages}.  Construction of multiple, separate GRAND10k sub-arrays will increase the chances of finding ideal sites.  Further, having arrays at different locations would enlarge the instantaneous field of view of GRAND, which could improve the rate of detection of transient events; see 
(see Sections \ref{section:uhe_neutrinos} and \ref{section:multi_messenger}) and help to reconstruct the direction of origin of FRBs (see Section\ \ref{section:FRB}), provided they are detected by multiple arrays.


\subsection{Background rejection}
\label{section:background}

Natural and man-made radio sources, stationary and transient, are the background in the search for EAS.  Below, we show that GRAND will have multiple strategies to deal with them.  Nevertheless, the background at a given geographical location is too diverse in nature and intensity to be accurately modeled.  A complete understanding of the background and its rejection requires performing prolonged tests on-site.  This is one of the main goals of GRANDProto300, the pathfinder of GRAND; see Section \ref{section:GRANDStages-GRANDProto300}.

\medskip

{\it Stationary noise.---}  Figure \ref{fig:radio_noise} shows the main sources of radio background as a function of frequency.  In the range 50--200~MHz of the GRAND antennas, there are two irreducible sources of stationary noise, both of natural origin: emission from the sky --- dominated by synchrotron radiation from the Galactic plane --- and thermal emission from the ground.  The equivalent temperature of the sky decreases from $T_{\rm sky} \sim 5\,000$~K at 50~MHz to $\sim$100~K at 200~MHz.  Above 150~MHz, the blackbody radiation of the ground at ambient temperature $T_{\rm ground} \sim 290$~K becomes the dominant source of stationary noise.

\begin{figure}[t!]
 \centering
 \includegraphics[width=\columnwidth, trim=0.6cm 0 1cm 1.5cm, clip=true]{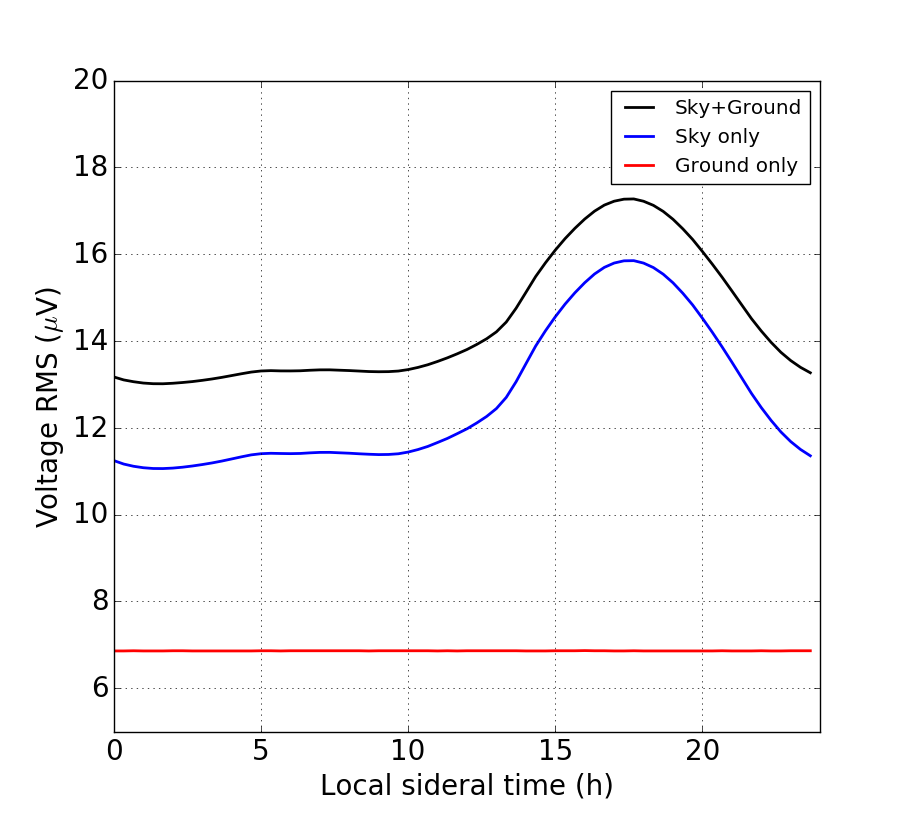}
 \caption{Total stationary noise level $V_{\rm rms}$ expected in one arm of a GRAND \HA\ oriented along the East-West direction, as a function of local sidereal time.}
 \label{fig:antennaLST}
\end{figure}

Figure \ref{fig:antennaLST} shows the response of the \HA, simulated using NEC4\ \cite{NEC4}.  The radio noise from the sky plus the ground, with mean value of 15~$\mu$V, is the minimum noise level in one antenna arm; other sources of steady noise may add to that.
The periodic variation with time is due to the transit of the Galactic plane across the field of view of the antenna, which can be used for calibration\ \cite{Lamblin:2007dya, LeCoz:2017ICRC}. 

Looking for causal coincidences among antenna triggers provides an efficient way to discriminate between antenna triggers induced by transient radio waves and random coincidences due to stationary noise fluctuations.  However, because trigger algorithms are usually based on detecting a signal excess above the stationary noise, this noise sets the detection energy threshold and the extent of the measurable radio footprint in GRAND.

\medskip

{\it Transient sources.---}  In any terrestrial location outside the polar regions, there is a wide variety of background sources --- thunderstorms, high-voltage power lines, airplanes, \etc --- that emit transient electromagnetic signals in a wide frequency range, including  the GRAND band.  The associated rate of detected background events in a radio array strongly depends on the local environment.  Typically, it ranges from tens of Hz per antenna in the most remote areas to kHz or more.  This rate is always higher by several orders of magnitude than the rate of detectable cosmic particles.  Section \ref{section:construction_stages} discusses the technical challenge of handling the event rate without affecting the DAQ live-time. Here we focus on another challenge, related to data analysis: how can this enormous background rate be discriminated from the true EAS events?  Fortunately, their radio signatures differ significantly:
\begin{itemize}
 \item 
  Background events often cluster in time, in position, or both, because sources are either static (\eg, high-voltage power lines) or follow a trajectory (\eg, an airplane in flight), and often emit radio signals in bursts.  For example, ANITA-2 filtered out 99\% of their total events using a cluster analysis\ \cite{Mottram:2012jsa}.  TREND reached a similar performance, at the cost of $\sim$10\% loss in EAS detection efficiency\ \cite{Charrier:2018fle}.
 \item
  Background time-traces are often longer than shower signals and their shapes less regular, allowing for efficient background rejection based on templates\ \cite{Barwick:2016mxm, Charrier:2018fle}
 \item
  Background point sources emit isotropically, while in EAS the shower fronts are, to first order, plane waves and are contained within a narrow Cherenkov cone ($\theta_{\rm Ch} \sim 1^{\circ}$).  Thus, the shape of their wavefronts and signal amplitude patterns on the ground differ radically.  This is especially true when the background sources lie inside the detector array, so that their positions can be determined accurately.  Besides, emission from EAS is enhanced along the Cherenkov ring, more so at higher frequencies; see \figu{fig4_freqcone}.  Detection of this feature will be a smoking gun to establish that a detected event is due to an EAS.
 \item
  The characteristic polarization pattern of the EAS signal can be distinguished from the polarization of the background\ \cite{Schellart:2014oaa, Carduner:2017mqf}
\end{itemize}

{\it UHECRs vs.\ UHE neutrinos.---}  In the search for UHE neutrinos, atmospheric neutrinos are a negligible background, due to their soft energy spectrum.  However, UHECRs with inclined trajectories may generate radio signals comparable to those made by UHE neutrinos, and at a much higher rate.  Preliminary simulations show that, in its final configuration (see Section\ \ref{section:GRANDStages-GRAND200k}), GRAND could detect as many as $3 \cdot 10^8$ cosmic rays per year with energies larger than $10^{8}$~GeV and zenith angles larger than $85^{\circ}$.

Nevertheless, UHECR-initiated showers can be discriminated efficiently using two criteria:
\begin{itemize}
 \item
  Using the excellent angular resolution of GRAND, we can reject events coming from above the horizon; see Section\ \ref{section:GRANDDesignPerf-Performance-Reconstruction}.  In the mountainous regions where GRAND10k arrays will be deployed, we can use a more aggressive angular cut --- \eg, 0.5$^{\circ}$ below the horizon --- without affecting significantly the sensitivity to neutrinos, since such trajectories correspond to small column densities underground. 
 \item
  Neutrino-initiated showers start deeper in the atmosphere than UHECR-initiated showers.  For instance, for a $10^9$~GeV proton with $\theta_z = 85^\circ$, its average shower maximum of $\langle X_{\rm max} \rangle_{\rm CR} = 750$~g~cm$^{-2}$ corresponds to less than one tenth of the total atmospheric depth, and thus translates into distances to the ground larger than 100~km.  Under the conservative assumption that GRAND has a resolution of $\sigma_{X_{\rm max}}=40$~g~cm$^{-2}$ (see Section \ref{section:GRANDDesignPerf-Performance-Reconstruction}), selecting showers with $X_{\rm max} > \langle X_{\rm max} \rangle_{\rm CR} + 10~\sigma_{X_{\rm max}}$ could discriminate between old showers from cosmic rays and young showers from neutrinos.  The efficiency of this cut will be determined via simulations.
\end{itemize}

Preliminary estimates indicate that these cuts could reduce the number of UHECRs mis-identified as neutrinos to less than 0.1 per year in GRAND200k, with a combined area of 200\,000~km$^2$; see Section\ \ref{section:GRANDStages-GRAND200k}.


\subsection{Detector performance}
\label{section:performances}

\subsubsection{Sensitivity to neutrinos}
\label{section:GRANDDesignPerf-Performance-NuDetection}

To determine the sensitivity of GRAND to neutrinos, we implemented the detection principles presented in Section\ \ref{section:detection_principle} in a custom-made end-to-end numerical simulation. The simulation takes into account the unique features of the problem, like the local topography of the array site --- which complicates the geometry of the simulation --- and the large instrumented area --- which increases the computational resources needed.  
We included all relevant physical processes while optimizing the use of computing resources. To validate it, we successfully tested its different parts against existing codes. Thus, the simulation results that we present below are reliable.

The simulation chain is divided into four independent parts.  First, for each simulated $\nu_\tau$, we propagate it before and after its interaction underground, carrying over daughter neutrinos, down to the point where the tau decays.  Second, we make the tau decay and evaluate the radio signal from the ensuing shower at the positions of the GRAND antennas.  Third, we simulate the response of the \HA\, to the radio signal.  And, fourth, we run a trigger algorithm to determine if the shower is detected by the array, following pre-defined detection criteria.  Below, we present the simulation chain in detail.

\bigskip

\paragraph{Simulation chain} 
\label{sensitivity}

\medskip

{\it From neutrino trajectory to tau decay: DANTON.---} The simulation from the primary neutrino to the tau decay is performed with a custom-made Monte-Carlo code, DANTON\ \cite{Niess:2018opy, DANTON:GitHub}.  It is a detailed simulation of neutrino and tau interactions with matter, including stochastic effects like transverse scattering.  The differential neutrino deep-inelastic-scattering cross sections are computed from the CT14NLO parton distribution functions\ \cite{Dulat:2015mca}. Tau transport is performed by PUMAS\ \cite{PUMAS:GitHub}, a specialized muon and tau transport code.  Photonuclear interactions are modeled following \Ref\ \cite{Dutta:2001}.  Tau decays are simulated with TAUOLA\ \cite{Jadach:1993hs}.  In addition, DANTON also allows to define a detailed topography, via the TURTLE package\ \cite{TURTLE:GitHub}: we use data from the NASA Shuttle Radar Topography Mission v3 Global 1-arcsecond dataset (SRTMGL1)\ \cite{SRTMGL1}. 

We have tested that the flux of emerging taus simulated by DANTON and NuTauSim\ \cite{Alvarez-Muniz:2017mpk} agree to within 10\%, compatible with uncertainties in neutrino cross sections. 
DANTON uses Monte-Carlo backtracking\ \cite{Niess:2017rlv}: starting from the position of the tau just prior to its decay, it backtracks its trajectory, interactions, and energy losses up to the primary neutrino.  Backtracking reduces the running time of simulations by several orders of magnitude below $10^9$~GeV and by 1--2 orders of magnitude above that energy.  Further, it allows us to exclusively sample the flux that yields a certain final tau state.

Since DANTON was built as a general-purpose tool for $\nu_\tau$ and tau propagation, we developed a simulation framework tailored to GRAND, called RETRO\ \cite{RETRO:GitHub}.  RETRO
generates candidate tau decays over the instrumented area and backward-samples their flux using DANTON.  Following that, ray-tracing is used to pre-select antenna positions inside a $3^\circ$ half-angle cone with vertex at the tau decay point. Antennas masked by the topography are rejected from the selection.

DANTON\ \cite{DANTON:GitHub, DANTON:GitHub}, PUMAS\ \cite{PUMAS:GitHub}, TURTLE\ \cite{TURTLE:GitHub}, and RETRO\ \cite{RETRO:GitHub} are available online.
    
\medskip

{\it Electromagnetic radiation simulation: radio morphing.---}
Radio emission by air showers is well understood and simulated \cite{Zhaires:2012,Huege:2013vt,EVA:2013}, but publicly available simulation codes often require intense CPU use, \eg, several hours to compute an electric field at 100 positions.  This prevents them from being used for a detector as large as GRAND.  Therefore, we developed an innovative semi-analytic approach, called {\it radio morphing}\ \cite{ZillesARENA:2018}, based on concepts similar to shower universality\ \cite{showerUniversality}.  Below, we present it briefly.

We start by simulating the electromagnetic radiation emitted by one reference shower $\mathcal{A}$, in 3D, using ZHAireS\ \cite{Zhaires:2012}.  In the simulation, we compute the electric field at approximately 3\,000 locations $x_i$ inside a conical volume oriented along the shower axis with an opening half-angle of $\sim 3^{\circ}$ and an extension of 120~km from the point of first interaction of the shower.  This volume contains the region where the electromagnetic emission is expected to be significant.  The high-granularity 3D mesh provides a detailed mapping of the radio signal in space and time. 

Based on the reference simulation, we can use radio morphing to compute the electric field generated by a different shower $\mathcal{B}$, at any target position $y$, without having to run a dedicated simulation for it.  First, at each of the original locations $x_i$, we re-scale the amplitude of the electric field of the reference shower by applying coefficients that depend on the energy, direction, and injection height of shower $\mathcal{B}$.  The orientation and size of the 3D mesh is also transformed via an isometry --- changing $x_i \rightarrow x_i^\prime$ --- depending on the geometry and height of maximum development of shower $\mathcal{B}$.  The result is the electric field induced by shower $\mathcal{B}$ at specific locations $x'_i$.  Finally, we interpolate the mesh to compute the electric field at the requested location $y$. 

Radio morphing yields electric-field amplitudes within 20\% of those calculated with ZHAireS for positions inside the Cherenkov cone, with a computation time typically 100 times faster.  This means that, based on a single full simulation, radio signals can be computed at any antenna location and for any simulated tau decay, within a few seconds and with high precision, thus making it possible to run GRAND neutrino sensitivity simulations in reasonable times.
 
\medskip

{\it Antenna response simulation.---}  To simulate the response of the \HA\ to the transient radio signal, we calculate the voltage at the input of the electric circuit on the antenna arm oriented along axis $\vec{u}$, \ie, $V_u(t) = \int E_u(\nu) \times l_{\rm eq}(u,\theta,\phi,\nu)e^{-2i\pi \nu t}d\nu$, where $\vec{u}$ = North-South or East-West, since, for now, the vertical arm of the antenna is not yet simulated. Here, $E_u$ is the Fourier transform of the electric-field waveform $\vec{E}(t)$ computed at one antenna position and projected along the $\vec{u}$ axis, and $l_{\rm eq}$ is the projection of the antenna equivalent length\ \cite{Balanis} along $\vec{u}$, for a frequency $\nu$ and an incoming wave direction $(\theta, \phi)$, computed with respect to the normal to the ground at the antenna position. 

\begin{figure}[t!]
 \centering
 \includegraphics[trim = 0.5cm 0.5cm 0.2cm 0.2cm, clip=true, width=\columnwidth]{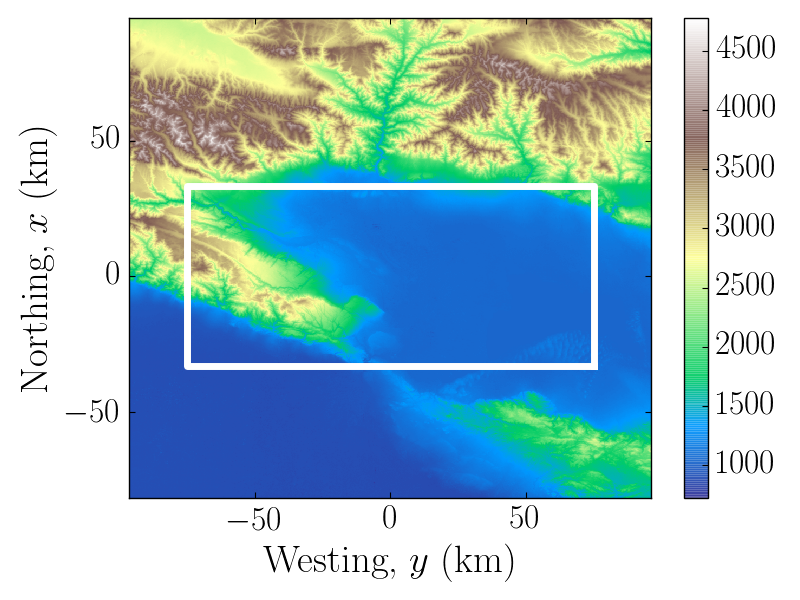}
 \caption{Elevation map of the GRAND10k simulation area, showing UTM altitude a.s.l.  The white area encloses HotSpot1, where 10\,000 simulated antennas are deployed over 10\,000~km$^2$.}
 \label{fig:mapHS1}
\end{figure}

The antenna equivalent length is computed using NEC4\ \cite{NEC4}, assuming infinite flat ground 
and a radio source placed an infinite distance away from the antenna. 
To evaluate the effect of a more realistic setting, we also perform an alternative calculation: we first compute analytically the effect of the diffraction of radio waves induced by the ground\ \cite{ITU}, and then we compute the voltage at the antenna output via the antenna equivalent length computed by NEC4 in free space. 

\medskip

{\it DAQ simulation.---}  The voltage signals at the antenna output are then filtered numerically through a 50--200~MHz Butterworth filter, mimicking the treatment that will be carried out in reality.  However, in reality, the frequency range used for triggering will be chosen to optimize the signal-to-noise ratio in the local background conditions.

We consider an antenna to be triggered if the peak-to-peak amplitude of the signal exceeds either 75~$\mu$V --- a conservative threshold --- or 30~$\mu$V --- an aggressive threshold.  The conservative threshold corresponds to five times the stationary noise level expected for the \HA, and is achievable with present technology.  The aggressive threshold will be reached by significantly reducing the noise level at antenna output --- 15~$\mu$V rms in the 50--200~MHz band for the present design of the \HA\ (see Section \ref{section:background}) --- or by using advanced triggering algorithms, \eg, based on machine learning\ \cite{FuhrerARENA:2018, Erdmann:2019nie}.  

\begin{figure}[t!]
 \centering
 \includegraphics[width=\columnwidth]{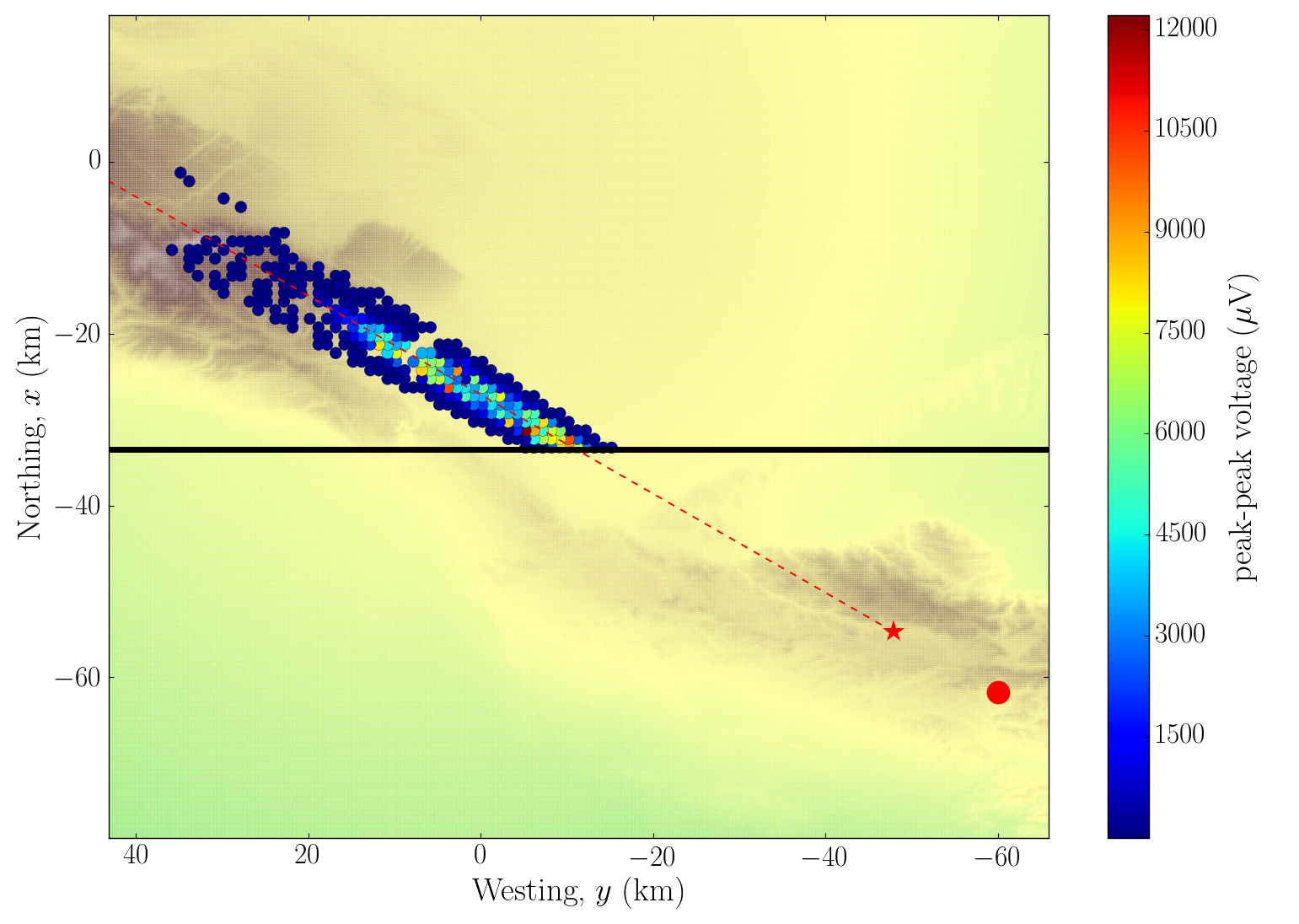}
 \caption{One simulated neutrino event displayed over the ground topography of HotSpot1. The large red circle shows the position of the tau production and the red star, its decay.  The dotted line indicates the shower trajectory.  Circles mark the positions of triggered antennas.  The color code represents the peak-to-peak voltage amplitude of the antennas.  The Southern border of HotSpot1 is indicated with a black line. }
 \label{fig:exEvent}
\end{figure}

The simulation implements also a second-level trigger: the shower is considered to be detected if at least five antennas located within a radius of size $\sqrt{2} d$, with $d = 1$~km the antenna spacing --- are triggered by the same shower; see Section\ \ref{section:array_layout}.  This occurs if the time difference between their triggers is, at most, $\sqrt{2} d /c$.

\begin{figure*}[t!]
 \centering
 \includegraphics[width=1.015\columnwidth]{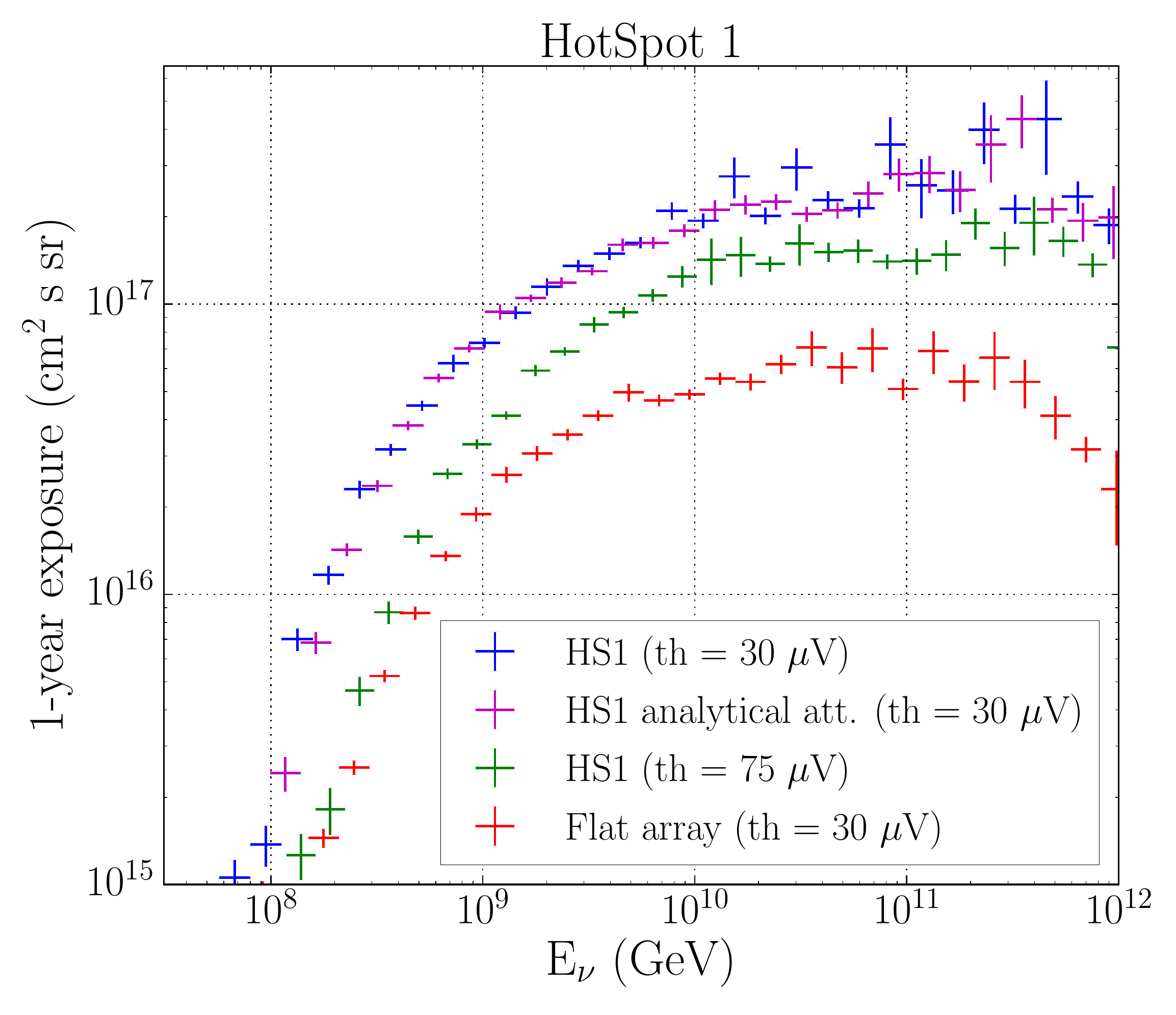}
 \includegraphics[width=\columnwidth]{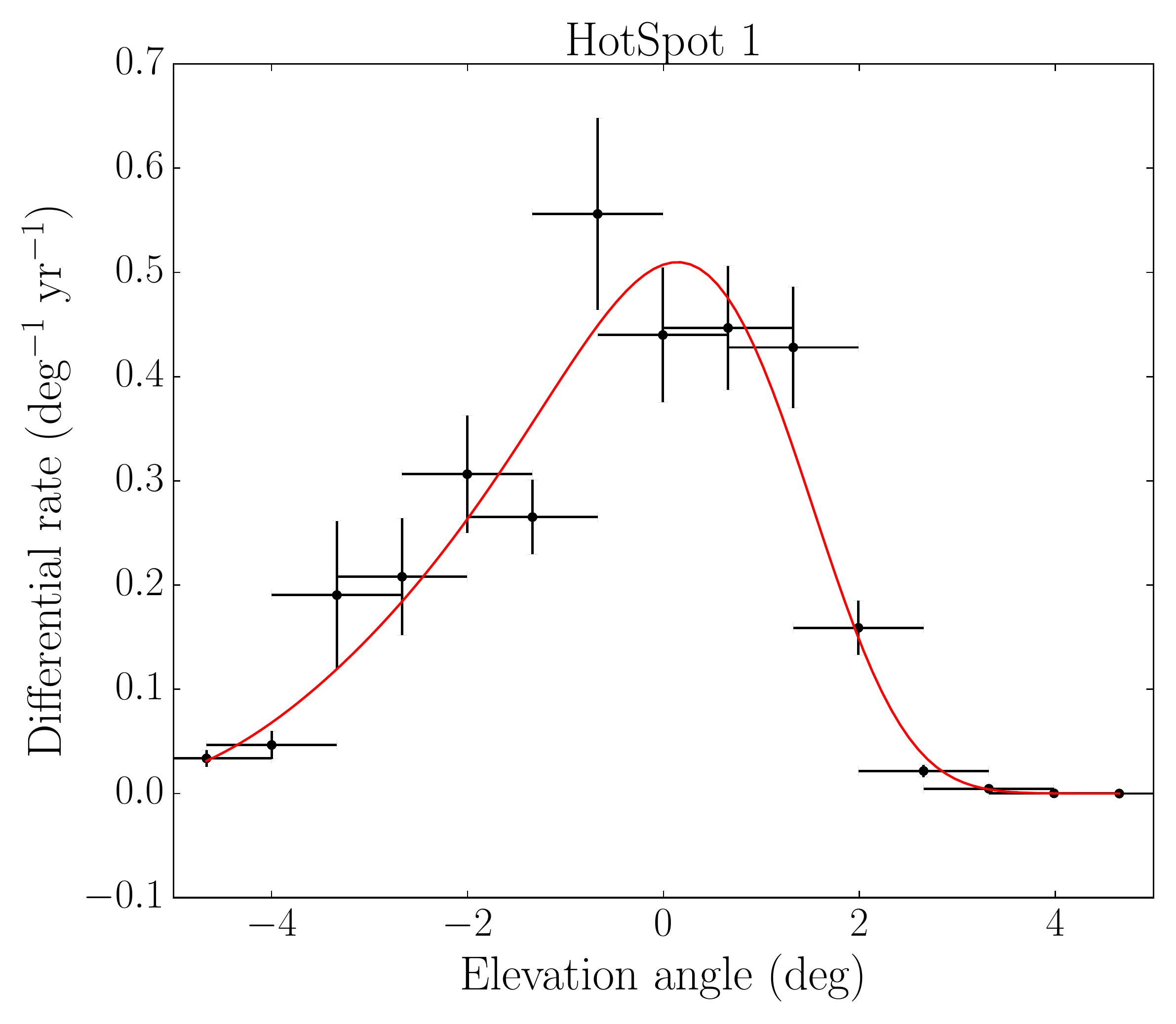}
 \caption{Results of the simulation of a GRAND 10\,000-antenna detector located at HotSpot1.  {\it Left:} Exposure as a function of neutrino energy, using an aggressive (30~$\mu$V) and a conservative (75~$\mu$V) antenna detection threshold, and two choices of wave propagation.  See the main text for details.  {\it Right:} Simulated event rate of neutrino-initiated showers, computed using the aggressive threshold, as a function of elevation angle $\alpha$ for neutrino flux set to the Waxman-Bahcall bound\ \cite{Waxman:1998yy}.  Downgoing trajectories have $\alpha<0^{\circ}$ and up-going trajectories have $\alpha>0^{\circ}$.}
 \label{fig:exposure_nu}
\end{figure*}

\medskip

\paragraph{Predicted sensitivity} 

\smallskip

Figure\ \ref{fig:mapHS1} shows {\it HotSpot 1} (HS1), the 10\,000~km$^2$ area located at the Southern rim of the Tian Shan mountain range for which the simulation chain described above was run.  The simulated detector is composed of 10\,000 \HA{\sc s} deployed on a square-grid layout with 1~km step size.  This region was identified in a preliminary study of GRAND neutrino sensitivity\ \cite{FangGRANDICRC:2017} as a site with favorable topography for neutrino detection, due to having a 100~km-wide basin surrounded by high mountains.  Thus, it could host a GRAND10k sub-array; see Section\ \ref{section:GRANDStages-GRAND10k}.

Figure \ref{fig:exEvent} shows, for illustration, the result of one simulated neutrino-initiated shower.  We simulated 20\,000 air showers initiated by $\nu_\tau$ interactions underground, with the condition that the shower trajectories cross HS1.  For the wave propagation, we used separately the standard treatment --- NEC antenna response simulation with ground --- and the alternative treatment ---analytical computation of the effect of ground and NEC simulation in free space.  We considered the aggressive and conservative detection thresholds separately. 

Figure\ \ref{fig:exposure_nu} shows the results of our simulation in terms of the exposure of the detector and the rate of detected showers, assuming the Waxman-Bahcall bound\ \cite{Waxman:1998yy} as reference neutrino flux.  The effect of the choice of treatment of wave propagation is small, and compatible with statistical fluctuations.  Using the aggressive threshold instead of the conservative threshold increases the effective area roughly by a factor of 2.5. The topography is clearly imprinted on the angular distribution of detected events: the distribution peaks in a narrow window around the horizon, with about 60\% of events having downward-going trajectories.  This shows that mountains play an important role as targets for neutrinos.  Our simulation also shows that using a flat topography significantly degrades performance for neutrino detection: the event rate for a flat array is roughly four times smaller than for HS1.

\begin{figure}[t!]
 \centering
 \includegraphics[width=\columnwidth]{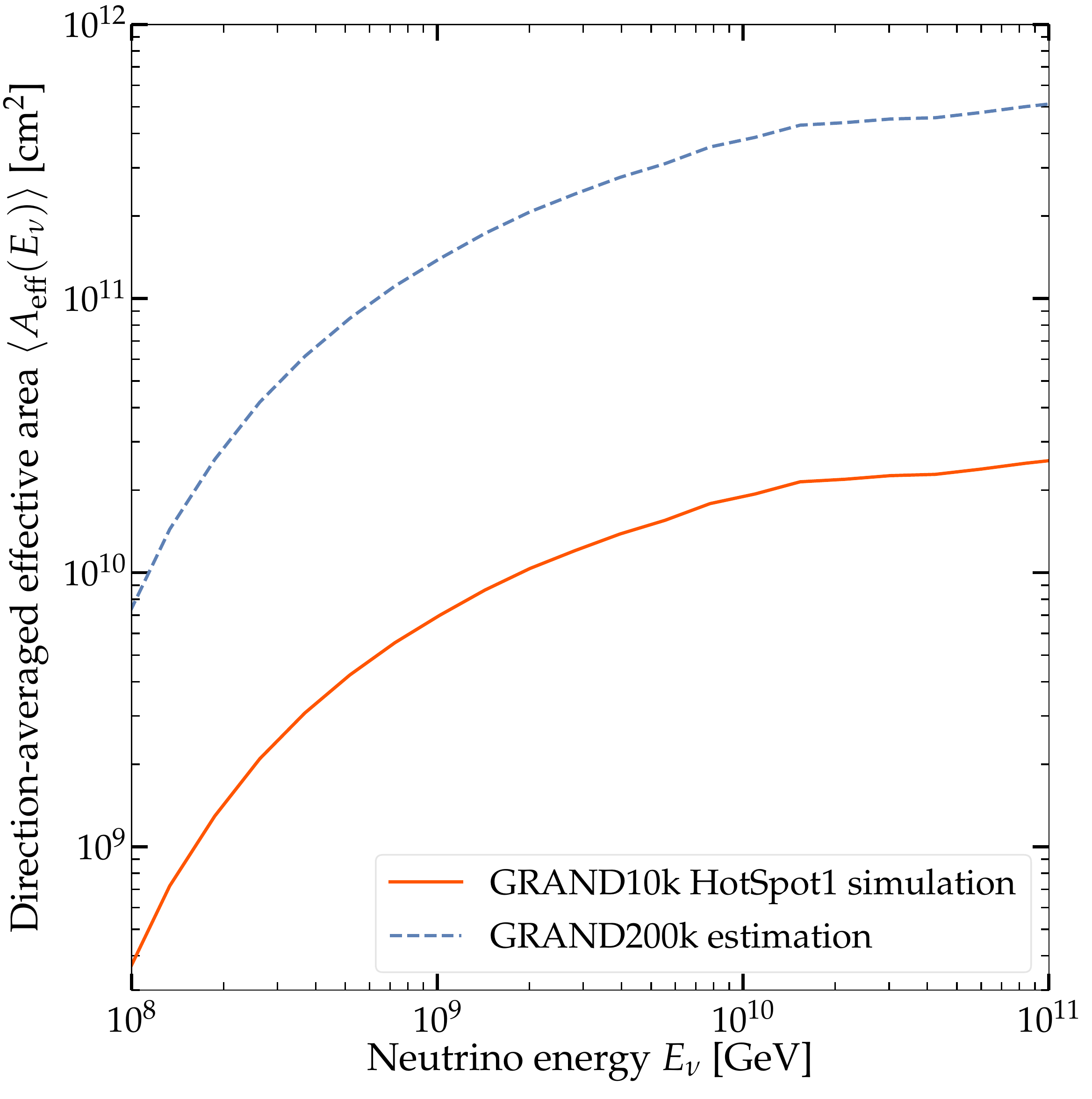}
 \caption{Effective area of GRAND10k HotSpot1, computed from simulations using the aggressive antenna detection threshold (see the left panel of \figu{exposure_nu}), and of GRAND200k, estimated by multiplying the GRAND10k HotSpot1 effective area by a factor of 20.  See the main text for details.}
 \label{fig:aeff}
\end{figure}

Figure\ \ref{fig:aeff} shows the direction-averaged effective area as a function of neutrino energy, for the simulated GRAND10k sub-array at HotSpot1, and for GRAND200k, estimated by scaling up by a factor of 20; see Section\ \ref{section:GRANDStages-GRAND200k}.

Figure \ref{fig:cosmo_neut} shows the resulting 90\% C.L.\ HS1 upper limit on a neutrino spectrum $\Phi_\nu \sim E_\nu^{-2}$.  Assuming background-free detection, we set an upper limit of 2.44 events\ \cite{Feldman:1997qc} per decade of energy during 3 years of exposure.  We make a rough estimate of the sensitivity achieved by combining 20 identical hotspots --- making up GRAND200k (see Section\ \ref{section:GRANDStages-GRAND200k}) --- by assuming a total effective area 20 times larger than that of HS1.  Dedicated simulations accounting for real locations and sizes of the different hotspots will yield a more precise figure.  Using the conservative threshold, the integrated GRAND200k upper limit is $E_\nu^2 \Phi_\nu \approx 1 \cdot 10^{-9}$~GeV~cm$^{-2}$~s$^{-1}$~sr$^{-1}$; using the aggressive threshold, it is $4 \cdot 10^{-10}$~GeV~cm$^{-2}$~s$^{-1}$~sr$^{-1}$.


\subsubsection{Sensitivity to cosmic rays and gamma rays}
\label{section:GRANDDesignPerf-Performance-PhotonsCR}

We ran preliminary simulations to estimate the aperture of GRAND to cosmic rays.  
We used ZHAireS to simulate 1\,200 incoming proton primaries with energies between $10^{9}$ and $10^{10.5}$~GeV and zenith angles between $60^{\circ}$ and $85^{\circ}$, uniformly sampling all azimuth directions.  The simulated detector was a square grid of \HA{\sc s} with 1~km spacing, deployed on a flat, horizontal area of 200~km$^2$, which was then extrapolated to 200\,000\,km$^2$, taking into account border effects.  To compute the response of the \HA{\sc s} to the time-dependent electric field output by ZHAireS, we followed the same procedure as for neutrino-initiated showers.  We consider separately the aggressive and conservative detection thresholds.

\begin{figure}[t!]
 \centering
 \includegraphics[trim = 0.1cm 0 1.5cm 1.0cm, clip=true, width=\columnwidth]{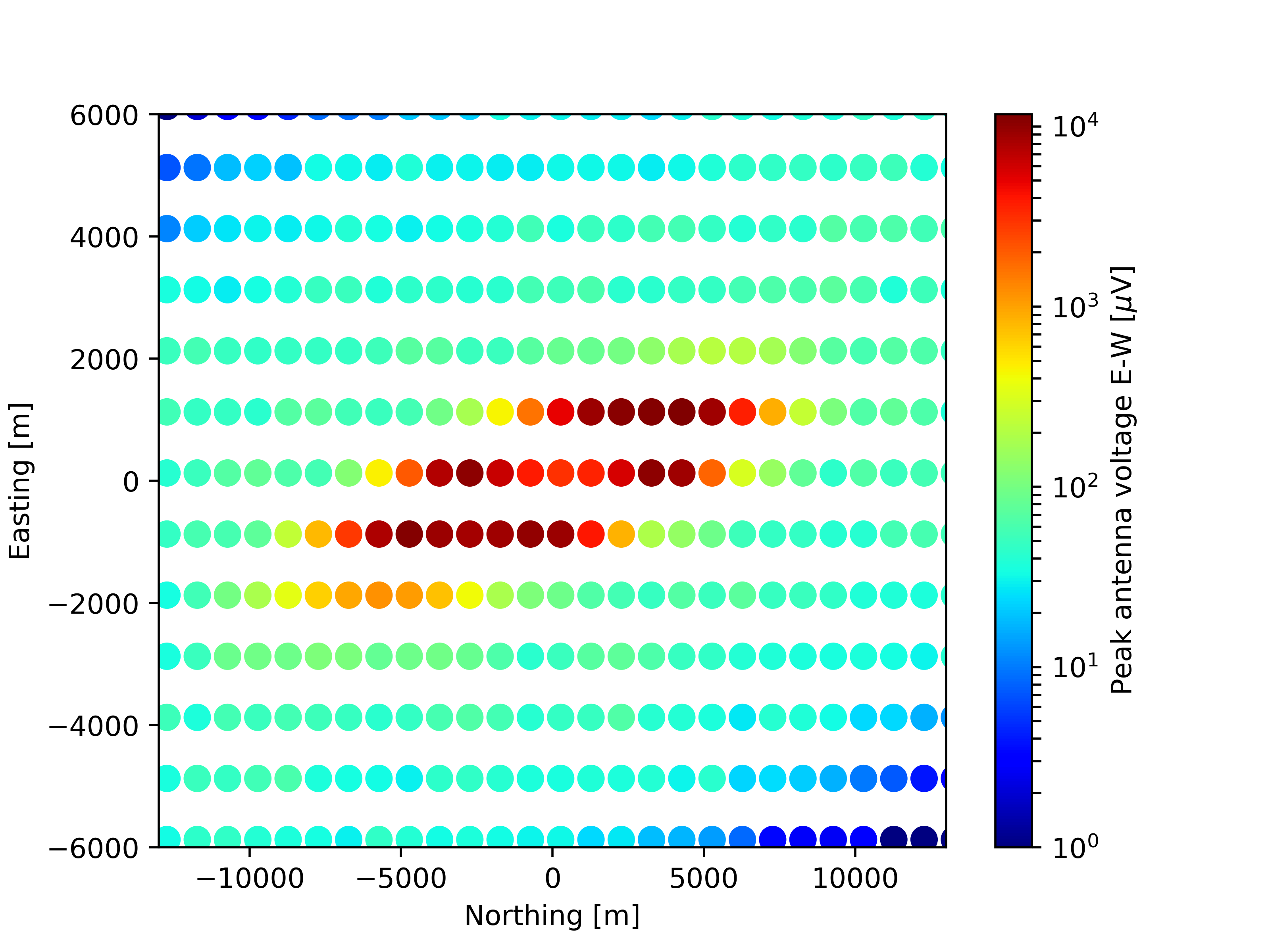}
 \caption{\label{fig:CRevent} Peak-to-peak amplitudes induced on the E-W arms of GRAND \HA{\sc s} by an air shower initiated by a proton with energy $E = 10^{10}$~GeV, zenith angle $\theta_z=80^\circ$, an azimuth direction offset by $10^\circ$  from the South, and a core position (272~m, 127~m).  The amplitude is above threshold for over 100 antennas for the aggressive and conservative detection thresholds.}
\end{figure}

Figure \ref{fig:CRevent} shows, for illustration, the result of one simulation run for a proton of $10^{10}$~GeV and zenith angle of $80^{\circ}$.
The resulting aperture of GRAND200k at $10^9$~GeV is 20\,000~km$^2$~sr in the conservative scenario and 25\,000~km$^2$~sr in the aggressive scenario. Near $10^{10}$~GeV, the effective area is 107\,000~km$^2$~sr in both cases.  The rate of detected UHECR-initiated showers above $10^{10}$~GeV, computed using the UHECR flux from \Ref\ \cite{TheTelescopeArray:2015mgw}, is about 200~events~day$^{-1}$.

Figure\ \ref{fig:phA} shows that the sensitivity of GRAND200k to UHE gamma rays is sufficient to detect them even in the pessimistic case where UHECRs are heavy.  The aperture of GRAND to UHE gamma rays should be similar to UHECRs.  To compute the preliminary sensitivity of GRAND200k to UHE gamma rays, we assumed that the detector is fully efficient to gamma ray-initiated air showers with energies above $10^{10}$~GeV in the zenith range $60^{\circ}$--$85^{\circ}$.  The sensitivity is the Feldman-Cousins upper limit at the $95\%$ C.L., assuming no candidate events, null background, and a UHE gamma-ray spectrum $\propto E^{-2}$.  The assumption of a background-free search is reasonable in the $10^{10}$--$10^{10.5}$~GeV range, even for the conservative hypothesis that GRAND reaches a resolution in $X_{\rm max}$ of only 40~g~cm$^2$ resolution; see Section\ \ref{section:xmax}.


\subsubsection{Performance of shower reconstruction}
\label{section:GRANDDesignPerf-Performance-Reconstruction}

Today, methods to reconstruct the properties of the primary particle that initiates an air shower --- its arrival direction, energy, mass --- based on radio data alone perform comparably to standard methods used in cosmic-ray detection that use ground-level particle and fluorescence data \cite{Aab:2016eeq, Buitink:2016nkf, Carduner:2017mqf, Kostunin:2017rbf}.  However, so far, the performance of the radio-based methods was assessed using dense radio arrays that detect showers with typical zenith angles below $60^{\circ}$.
While the methods are, in principle, applicable to GRAND, the precision with which inclined showers can be reconstructed using a sparse array remains to be studied and the methods optimized.  Below, we present preliminary results from an ongoing dedicated study.

\medskip

\paragraph{Angular resolution}  

\smallskip

We have performed a preliminary estimation of the GRAND angular resolution based on a set of 3\,550 air showers initiated by neutrinos of $10^{10}$~GeV, generated with the simulation chain described in Section\ \ref{sensitivity}, but using ZHAireS instead of radio morphing.  The simulation was run over a simplified, toy-model topography.  For shower production, we assumed a spherical Earth.  For the detection area, we assumed a square grid of \HA{\sc s} with a 1-km spacing, deployed on a flat surface inclined by an elevation angle $\beta$ from the horizontal, facing the shower trajectory and placed at a distance $D$ from the tau decay point.  The size of the detection area is chosen to fully contain the radio footprints; the plane of the detector stops at a height 3000~m a.s.l.  We ran simulations for $\beta= 0^{\circ}, 5^{\circ}, 10^{\circ}, 15^{\circ}, 20^{\circ}$, and $D$ ranging from 20 to 100~km.

We found that 1\,606 simulated showers pass the aggressive trigger condition defined in Section\ \ref{sensitivity}.  Out of these, 1\,370 simulated showers trigger 10 or more antennas; on these, we apply the analysis explained below.  For each antenna, the trigger time is taken as the time when the Hilbert envelope of the signal is maximum.  We smear this time using a Gaussian of width 5~ns to account for the uncertainty on the experimental precision of trigger timing.

The standard method for angular reconstruction infers the arrival direction of a shower from the arrival times and positions of triggered antennas\ \cite{Ardouin:2010gz}, assuming a specific shape for the radio wavefront.  Models and measurements converge on wavefronts being hyperbolic\ \cite{Apel:2014usa, Corstanje:2014waa}, even though some experiments do not observe a significant deviation from a plane wave\ \cite{Carduner:2017mqf}.  At present, our analysis only considers plane waves.  The fit of the simulation results to a plane-wave wavefront is poor, which indicates that the wavefronts in our simulations have a different shape.  Thus, in the future, it is likely that reconstruction will improve significantly by considering a hyperbolic wavefront instead.

\begin{figure}[t!]
 \centering
 \includegraphics[trim = 1.5cm 0.5cm 1.5cm 0.5cm, clip=true, width=\columnwidth]{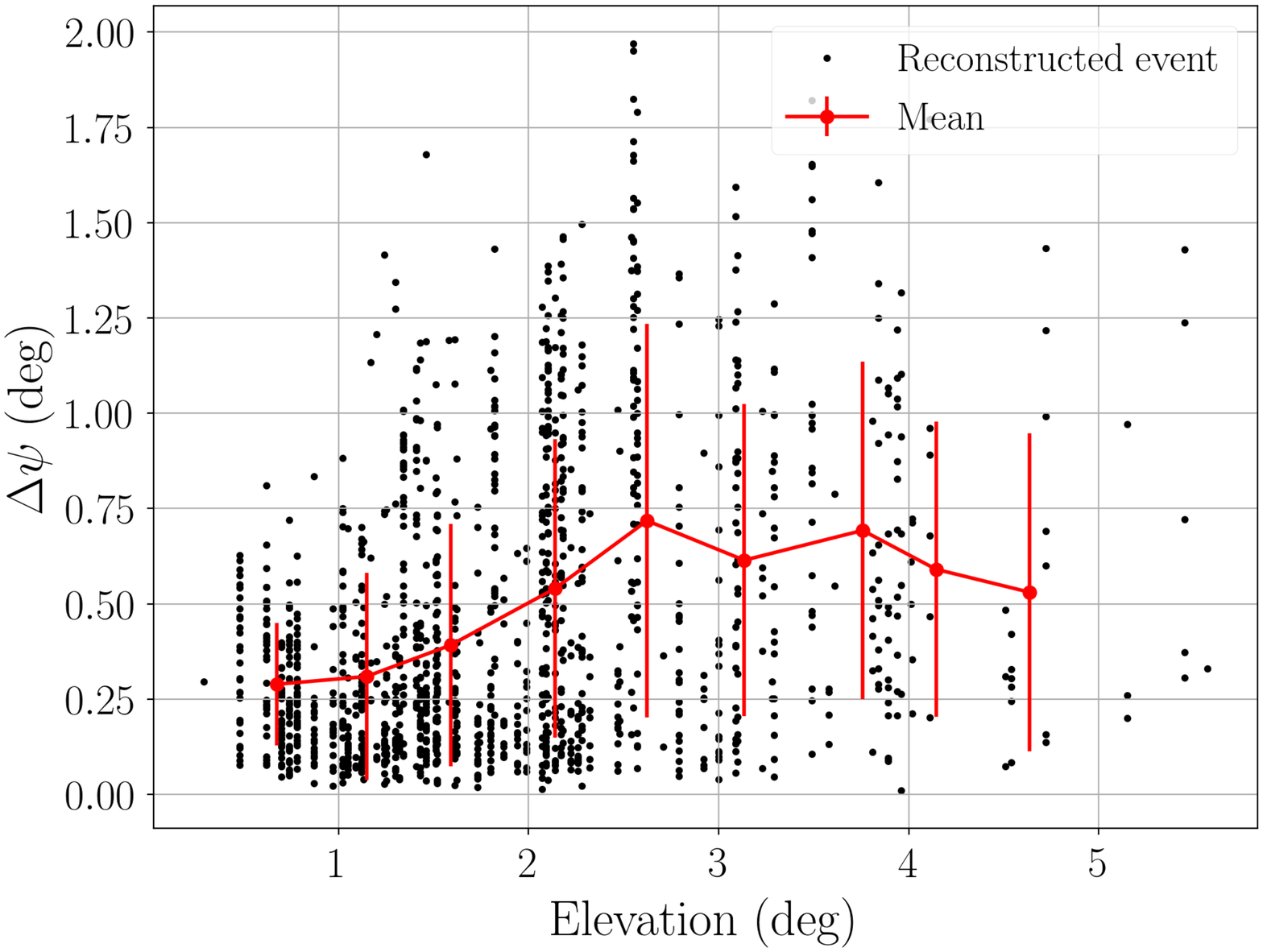}
 \includegraphics[trim = 1.5cm 0.5cm 1.5cm 0.5cm, clip=true, width=\columnwidth]{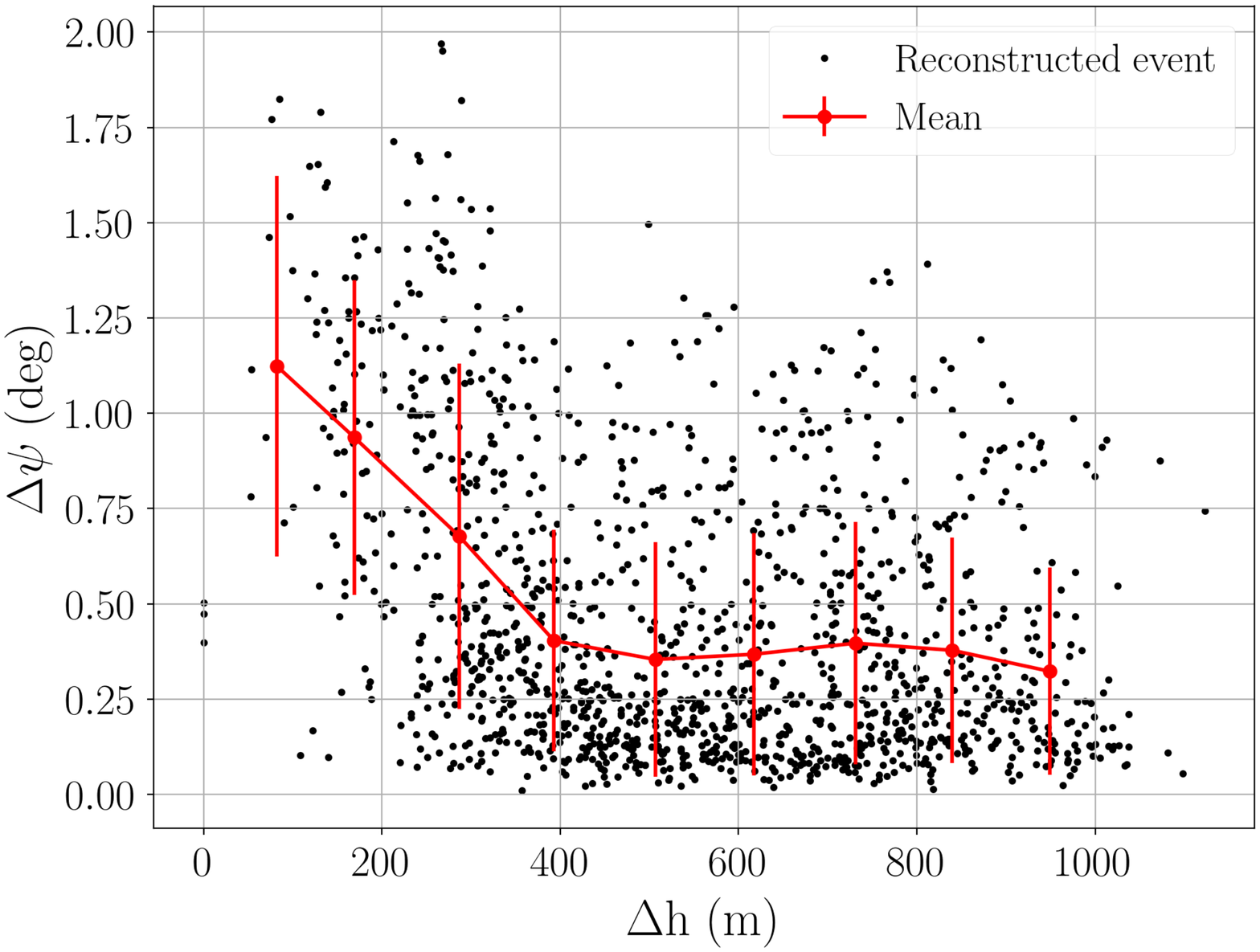}
 \caption{Angular resolution $\Delta \psi$ inferred from 1\,370 simulated air showers detected by a GRAND toy-model array.  See the main text for details.  {\it Top:}  Resolution $\Delta \psi$ as a function of elevation angle with respect to the horizontal.  The error is defined as $\Delta \psi=\arccos(\cos\theta^*\cos\theta + \cos(\phi^*- \phi)\sin\theta\sin\theta^*)$, with ($\theta, \phi$) and ($\theta^*, \phi^*$) the true and reconstructed directions of the shower respectively.  {\it Bottom:} Resolution as a function of height difference between antennas participating in each detected shower.}
 \label{fig:angres}
\end{figure}

Figure\ \ref{fig:angres} shows the angular resolution inferred from the simulations using the plane-wave approximation.  Already these results provide insight on the potential of GRAND.  We achieve an average angular resolution better than 0.5$^{\circ}$, comparable to experimental results obtained in plane-wave reconstruction of non-inclined showers \cite{MartineauHuynh:2012vj}.  This angular resolution applies to showers down to horizontal trajectories, owing to the difference of elevation between triggered antennas, which provides a lever arm for reconstruction.

For UHECRs, because they are deflected by magnetic fields and nominally do not point back at their sources, achieving an angular resolution of about $0.5^{\circ}$ is sufficient.  For neutrinos, a resolution of about $0.1^{\circ}$ would allow to potentially discover UHE neutrino sources and do UHE neutrino astronomy; see Section \ref{section:uhe_neutrinos}.  

Ultimately, the angular resolution is limited by the lever arm ($d$) and the timing resolution of the detector ($\Delta t$).  This sets the scale of the angular uncertainty to $\Delta \psi \sim d/\Delta t$.  However, LOPES, LOFAR, and AERA have shown that there are several other systematics that must be considered, such as the assumed shape of the radio wavefront and the first interaction point of the shower.  

With GRAND we expect to reach a sub-degree angular resolution due to the following:
\begin{itemize}
 \item
  GRAND will use a hyperbolic description of the radio wavefront instead of a planar one.  Doing this has led to an estimated improvement in the precision of the reconstructed direction of origin of the air shower of a factor of 2 in LOPES\ \cite{Apel:2014usa} and a factor of 10 in LOFAR\ \cite{Corstanje:2014waa}. (However, the absence of an absolute determination of this parameter does not allow for a definitive assessment of its precision.)
 \item
  Naively, one expects the angular resolution to degrade at very large zenith angles.  However, our simulations show that this is overcome by the topography of the area of the detector and the differences in elevation between triggered antennas; see \figu{angres}.  Besides, the large footprints of very inclined showers trigger antennas at large distances from the shower axis --- up to several km, see \figu{exEvent} --- and provide a means to determine the shower wavefront and geometry.
 \item
  Precisely knowing the antenna trigger times is crucial, and we expect GRAND to achieve sufficient timing accuracy.  On the one hand, satellite clocks keep improving in accuracy. On the other hand, the required timing accuracy decreases with the lever arm $d$, which is an order of magnitude larger in GRAND compared to previous experiments.  Using calibration measurements similar to those used in AERA\ \cite{PierreAuger:2016zxi}, GRAND will validate the timing accuracy during run-time, possibly by means of calibration beacons.
\end{itemize}
For these reasons, we believe it is reasonable to consider a precision on the reconstructed direction of origin of very inclined showers of $0.2^\circ$ or better.  This will be tested in GRANDProto300; see Section \ref{section:GRANDStages-GRANDProto300}.

\medskip

\paragraph{Energy resolution}  

\smallskip

The radio signal from a shower reflects the energy of its electromagnetic component.  Existing radio arrays achieve $15\%$ precision on the reconstruction of the energy of individual showers.  They do this via two methods; see \Ref\ \cite{Schroder:2016hrv} for a detailed review.  The first method determines the total radiated energy by integrating the measured power over the extension of the footprint and over the duration of the radio pulse\ \cite{Aab:2015vta, Aab:2016eeq}.  The second method uses the radio amplitude at a detector-specific reference distance from the shower axis\ \cite{Kostunin:2015taa, Apel:2014jol, Aab:2015vta}.  These methods should reach a similar performance in GRAND, in particular in showers whose radio footprint is fully contained by the array.  

For UHECRs, this will often be the case, since showers are less inclined and so radio footprints are relatively smaller.  Energy resolution is particularly important to resolve features near the GZK cut-off; see Section \ref{section:uhe_messengers}.  

For neutrinos, the precision on the reconstructed shower energy is less critical, since the uncertainty on the incident neutrino energy is dominated by the processes leading to the shower creation.  Because showers are initiated by the decay of the tau --- which carries an unknown fraction of the energy of the parent neutrino (see Section \ref{section:detection_principle}) --- the energy reconstructed from the radio signal is only a lower bound on the energy of the parent neutrino.  This limitation is akin to that of reconstructing neutrino energy from neutrino-initiated through-going muons born outside IceCube\ \cite{Aartsen:2015rwa, Aartsen:2016xlq}.  For these events, IceCube uses Monte-Carlo based parametric unfolding to reconstruct neutrino energy.  For GRAND, ongoing studies will determine the uncertainty on reconstructed neutrino energy.

\vspace*{-0.6cm}


\subsubsection{Identification of the primary}  
\label{section:xmax}

The measurement of the shower maximum $X_{\rm max}$ is the most robust technique to infer the nature of the primary particle in radio-detection experiments.  In GRAND, high-precision studies of UHECRs will depend on the ability to resolve $X_{\rm max}$.  The ultimate goal is to match the present best accuracy of about 20~g~cm$^{-2}$ in the 10$^{8}$--10$^{11}$~GeV range; see Section \ref{section:uhe_messengers_uhecr}.  This level of performance still has to be demonstrated for sparse radio arrays and for inclined showers, but prospects are encouraging.  With this resolution at the highest energies, it will be possible to distinguish between a light and a heavy primary on an event-to-event basis.
For neutrinos and gamma rays, a lower resolution, of about 40~g~cm$^{-2}$, would be enough to distinguish them from cosmic rays; see Section \ \ref{section:background}.  

To estimate the ability of GRAND to identify the primary, we simulated a sample of UHECR-initiated showers using the simulation chain described in Section\ \ref{sensitivity} --- but using ZHAireS instead of radio morphing ---  and reconstructed their values of $X_{\rm max}$ using the top-down method from \Ref\ \cite{Buitink:2014eqa}.  Below, to illustrate the method, we pick one of the simulated showers as a test shower: a shower from a proton primary of $10^{10}$~GeV, zenith angle $83^\circ$, azimuth angle $40^\circ$, and position of the shower vertex chosen at random.  The simulation yields $X_{\max} = 789$~g~cm$^{-2}$ for the test shower.  The layout of the simulated detector is a square-grid array with an antenna spacing of 1~km deployed over a flat area tilted by $10^\circ$ with respect to the horizontal.  

We start by simulating a companion set of 50 proton-initiated showers and 20 iron-initiated showers, each with identical energy and arrival direction as the test shower, but with a different position of the shower core, chosen at random.  When generating each companion shower, we calculate its value of $X_{\max}$.  Then we compare the radio footprint on the array of the test shower to the footprints of the companion showers, via a least-squares fit, to find the best-fit value of $X_{\max}$ for the test shower.

\begin{figure}[t!]
 \centering
 \includegraphics[trim = 0.65cm 0.5cm 1.65cm 2.0cm, clip=true, width=\columnwidth]{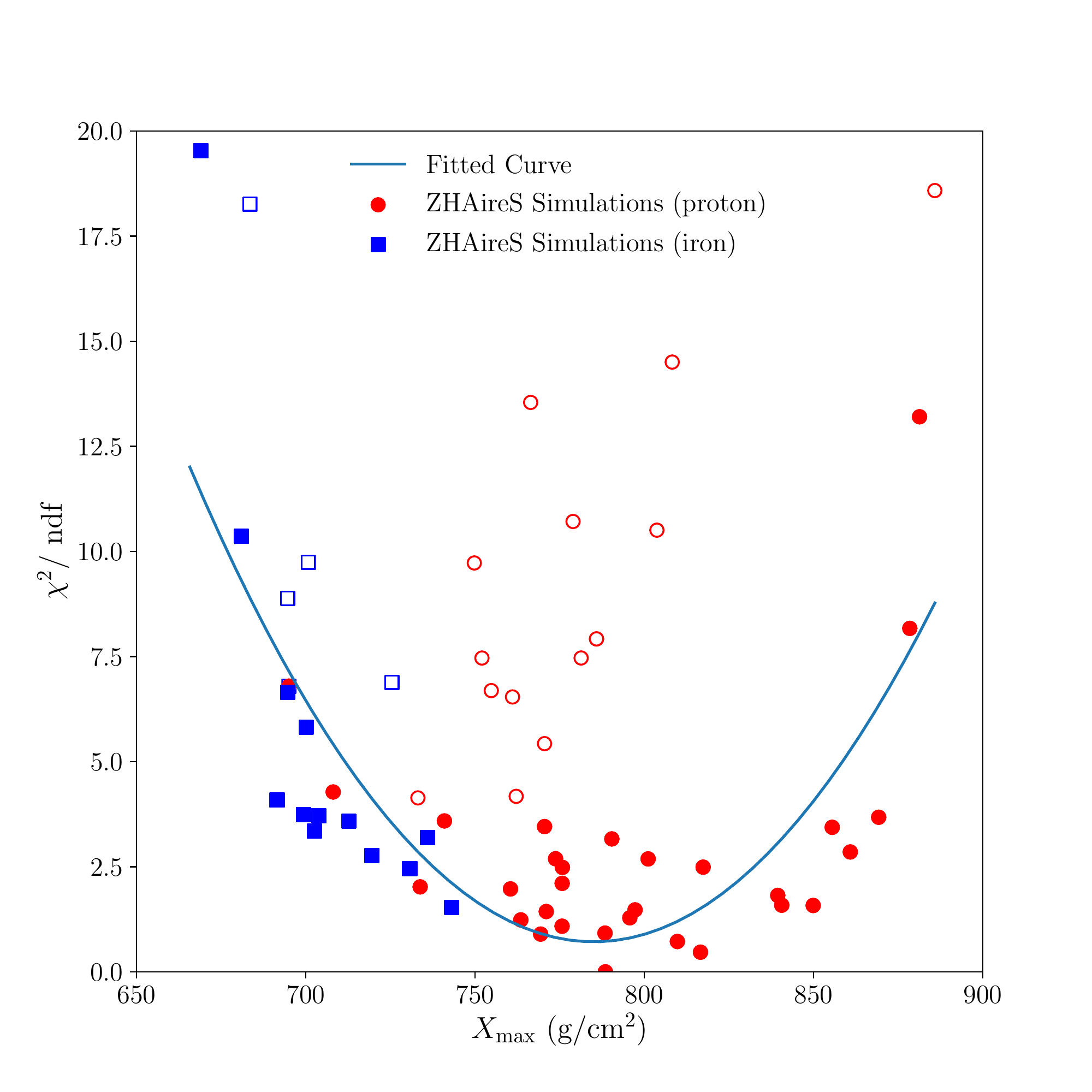}
 \caption{Reconstruction of $X_{\rm max}$ for the test shower used in our example.  Each point represents one of the simulated companion showers against which the test shower is fitted.  See the main text for details.  The $x$-axis is divided into nine bins of $X_{\rm max}$, and the minimum $\chi_{\min}^2$ and standard deviation $\sigma$ is calculated for each.  Only the showers represented by filled symbols --- for which $\chi^2 - \chi_{\min}^2 < \sigma$ --- were used in fitting the parabola from which the best-fit value of $X_{\rm max}$ of the test event is inferred.}
 \label{fig:Xmax_reco}
\end{figure}

Figure\ \ref{fig:Xmax_reco} shows the result of the fit.  The reconstructed value is $X_{\max} = 785$~g~cm$^{-2}$, close to the real value.   
Based on our simulations, we have found that, above $10^{9.5}$~GeV, the shower maximum can already be reconstructed with a precision better than 40~g~cm$^{-2}$, independently of the antenna spacing, as long as the shower triggers 10 or more antennas.  High statistics at these energies and access to further information beyond $X_{\max}$, \eg, the geometry of the Cherenkov ring, will improve the precision.  Recently, it has been shown that also using the pulse shape measured in each antenna can further improve the accuracy in $X_{\max}$\ \cite{Bezyazeekov:2018yjw}.  However, in the present study, companion showers were simulated using the same direction and energy as the test shower.  Future studies will explore the impact of dropping these assumptions, of using a realistic topography, and of including the reconstructed position of the Cherenkov cone in the analysis.

\vspace*{-2cm}

\newpage




\section{Construction stages}
\label{section:construction_stages}

\mybox{{\bf At a glance}}{Tan!70}{grand_brown!20}
{
 \begin{itemize}[leftmargin=*]
  \item
   GRAND will be modular and built in stages, with the array size progressively growing
  \item
   GRANDProto35: 35 antennas plus surface particle detectors; 
   optimize detection efficiency and background rejection
  \item
   GRANDProto300: 300 antennas in 100--300~km$^2$; focus on the detection of horizontal showers, study UHECRs with $10^{7.5}$--$10^9$~GeV
  \item
   GRAND10k: Built at a geographical hotspot where the rate of neutrino-initiated showers is high; integrated neutrino sensitivity of $8 \cdot 10^{-9}$~GeV~cm$^{-2}$~s$^{-1}$~sr$^{-1}$
  \item
   GRAND200k: Made up of 20 separate, independent hotspots; integrated neutrino sensitivity of $4 \cdot 10^{-10}$~GeV~cm$^{-2}$~s$^{-1}$~sr$^{-1}$
 \end{itemize}
}

\begin{figure*}[t!]
 \centering
 \includegraphics[width=\textwidth]{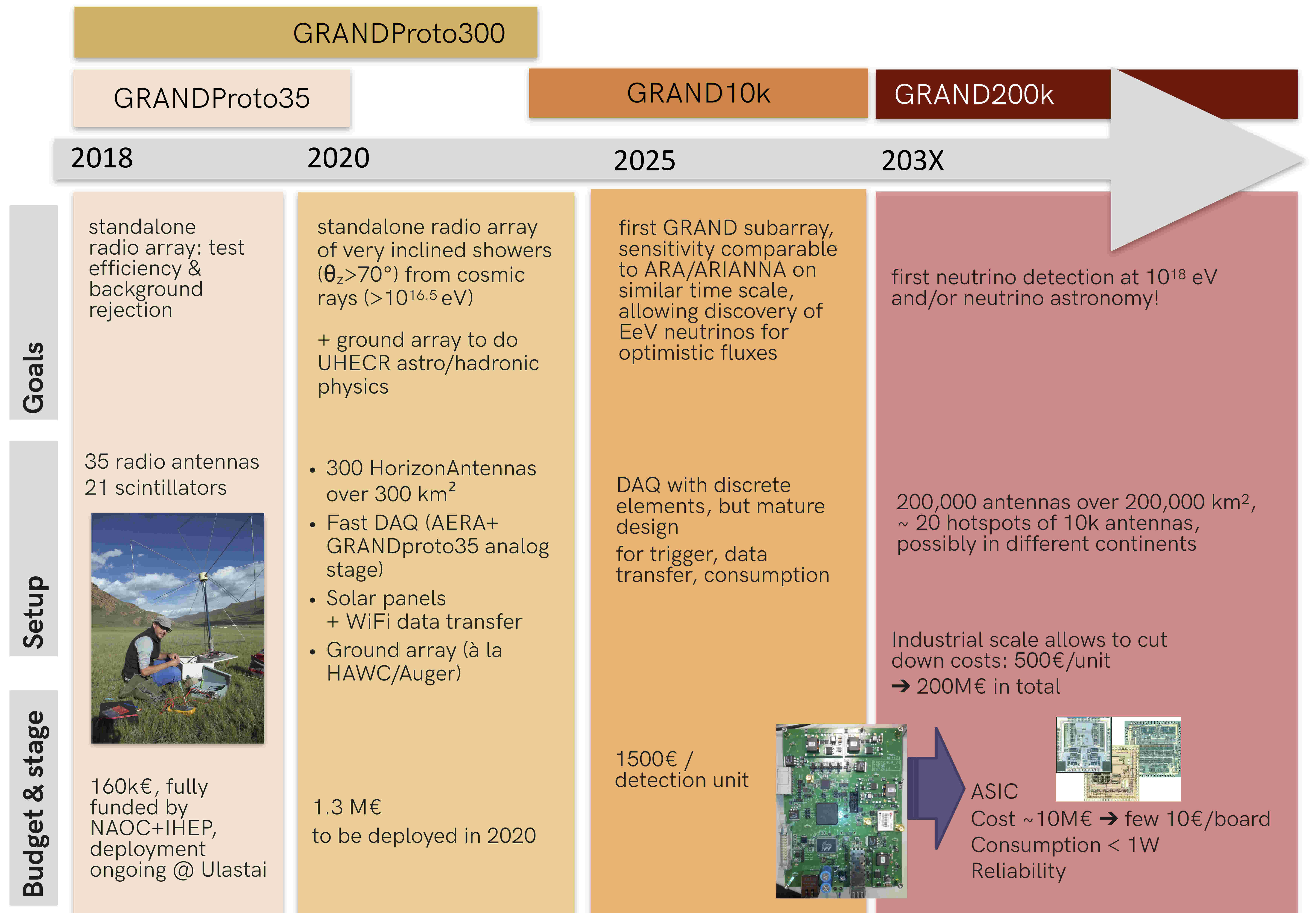}
 \caption{Timeline of the construction stages of GRAND}
 \label{fig:roadmap}
\end{figure*}

GRAND will be a major leap forward in the detection of air showers, in particular, and in astroparticle physics, in general.  Achieving this will require that we deploy and maintain several radio arrays --- each covering 10\,000~km$^2$ --- and autonomously trigger on very inclined air showers, efficiently separate them from the dominant background, and precisely reconstruct the properties of primary UHE particles.  The challenge is significant, but the field is mature and ready to tackle it.  The reward is high, in the form of vast progress in UHE astroparticle physics, radioastronomy, and cosmology.

There is no conceptual obstacle to realizing GRAND; the challenge is purely technical.  The following recent developments lend support to this claim:
\begin{itemize}
 \item
  Experiments like LOFAR \cite{Stappers:2011ni} have shown the feasibility of building and operating arrays of thousands of radio antennas, because antennas are relatively inexpensive, structurally robust and stable, and can be deployed easily, making arrays scalable
 \item
  Features of the radio emission from air showers --- pulse shape\ \cite{Barwick:2016mxm, Charrier:2018fle}, polarization pattern\ \cite{Carduner:2017mqf}, amplitude pattern\ \cite{Nelles:2014dja} --- differ significantly from those of the background, and be used to efficiently discriminate signal from background even under background-dominated conditions\ \cite{Charrier:2018fle}; see Section\ \ref{section:background}
 \item
  Substantial and ongoing progress in data treatment and communication has made it possible to collect large volumes of data reliably across large areas at affordable costs
 \item
  Today, existing radio arrays are able to reconstruct properties of primary UHE particles, from radio data alone, with an accuracy comparable to that achieved with ground-array and fluorescence data; simulations show that this performance should carry over to a sparse array like GRAND; see Section\ \ref{section:GRANDDesignPerf-Performance-Reconstruction}
\end{itemize}

Figure \ref{fig:roadmap} summarizes the construction plans of GRAND.  In order to thoroughly test the instrumental concepts on which GRAND is based, we have chosen a staged construction approach that will progressively validate key steps required to build the final configuration.  Further, to increase the scientific returns of early and intermediate stages, they are designed to tackle science goals by themselves.  

To overcome the technical challenges, GRAND needs to demonstrate the successful integration of the aforementioned developments into the design of a radio array that triggers autonomously on very inclined showers --- initiated by neutrinos --- using affordable, robust, and energy-efficient detection units.  This will be the main technology goal of GRANDProto300, the 300-antenna pathfinder stage of GRAND.  After achieving its goal, GRANDProto300 will be turned into a test bench for the final, optimized design of GRAND.
The following stage, GRAND10k, with 10\,000 antennas, is designed to reach a neutrino sensitivity comparable to that of other potential contemporary detectors.  The final stage, GRAND200k, will replicate GRAND10k arrays in different locations, in order to reach the ultimate target sensitivity of GRAND.  Below, we present each construction stage in detail.


\subsection{GRANDProto35 (2018)}
\label{section:GRANDStages-GRANDProto35}

GRANDProto35\ \cite{Gou:2017ICRC}, the first construction stage, will lay the groundwork for future stages.  Its goal is to reach an efficiency higher than 80\% for the radio-detection of showers and a background rejection that keeps the ratio of false positives to true positives below 10\%.  GRANDProto35 builds on the experience from TREND\ \cite{Ardouin:2010gz, LeCoz:2017ICRC, Charrier:2018fle}, and is deployed at the same site, in the Tian Shan mountains in the XinJiang province of China.  GRANDProto35 is presently in commissioning phase. 

\begin{figure}[t!]
 \centering
 \includegraphics[width=\columnwidth]{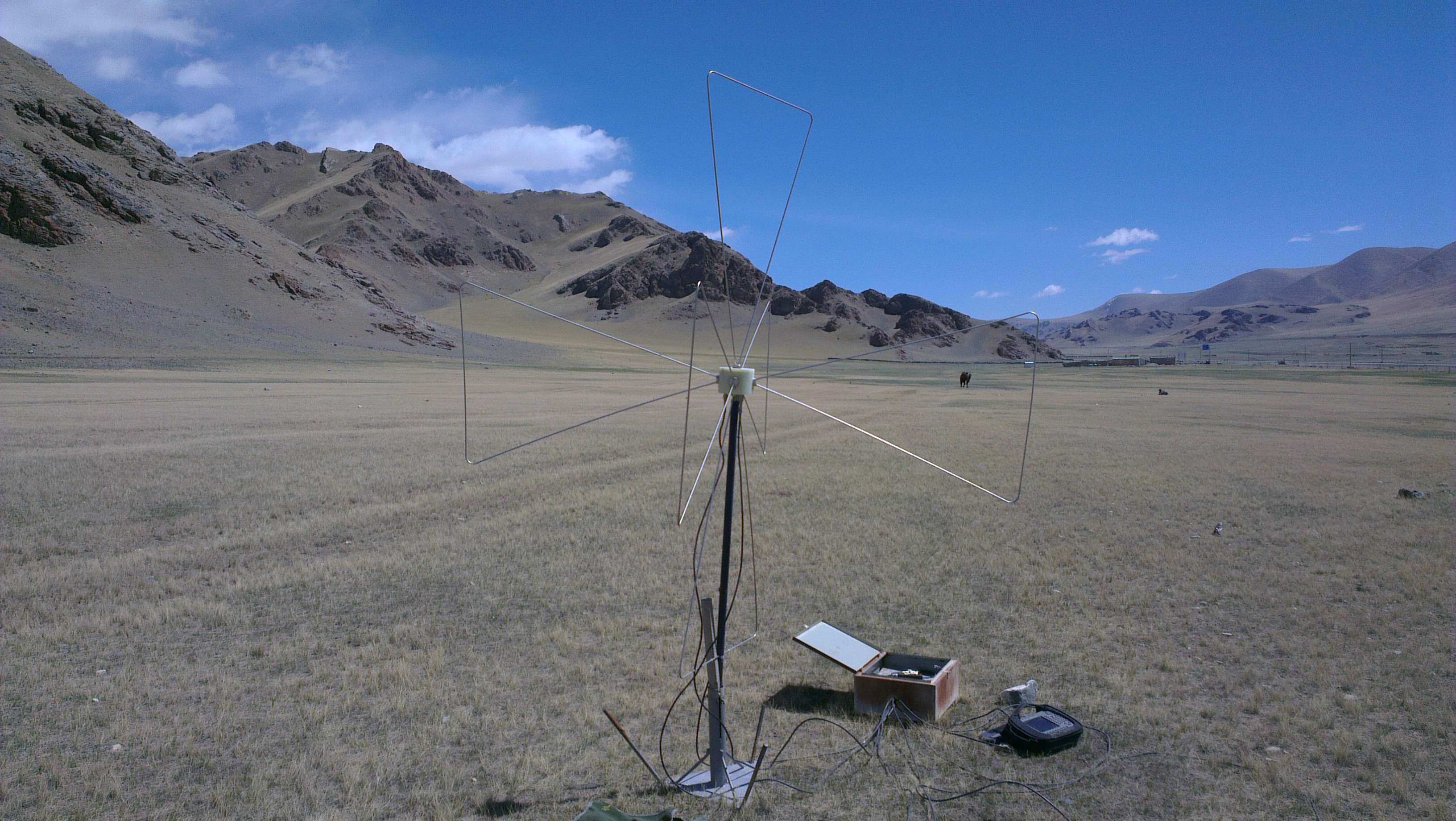}
 \caption{One of the antennas used in GRANDProto35, deployed at the construction sites of GRANDProto35\ \cite{Gou:2017ICRC} and TREND\ \cite{Ardouin:2010gz, LeCoz:2017ICRC}, in the Tian Shan mountains of China.  Photo by Olivier Martineau-Huynh.}
 \label{fig:site_photos}
\end{figure}

Figure\ \ref{fig:site_photos} shows one of the GRANDProto35 detection units deployed on-site.  The array consists of 35 radio-detection units, each with a bow-tie antenna inspired by the ones first designed for CODALEMA\ \cite{Ardouin:2009zp, Charrier:2012zz} and later used in AERA\ \cite{Abreu:2012pi}.  GRANDProto35 antennas contain an additional vertical arm to sample all three polarization directions.  The radio-detection units are deployed on the infrastructure of the 21 CentiMeter Array (21CMA)\ \cite{Zheng:2016}, in a rectangular grid 800~m long in the East-West axis and 2\,400~m long in the North-South axis. This setup is optimized for the detection of air showers coming from the North, the direction from which their radio emission is most intense, due to the geomagnetic effect being maximal.

The combined information from the three antenna arms completely determines the signal polarization.  Polarization will be used to separate air showers from the background, as their radio emission has a characteristic polarization pattern on the ground, owing to the interplay of the radio emission mechanisms; see Section \ref{section:antenna_design}.  Polarization plus additional shower-identification methods developed by TREND\ \cite{Ardouin:2010gz} should effectively reject the background.

\begin{figure}[t!]
 \centering
 \includegraphics[trim = 2.5cm 0.5cm 2.0cm 2.0cm, clip=true, width=\columnwidth]{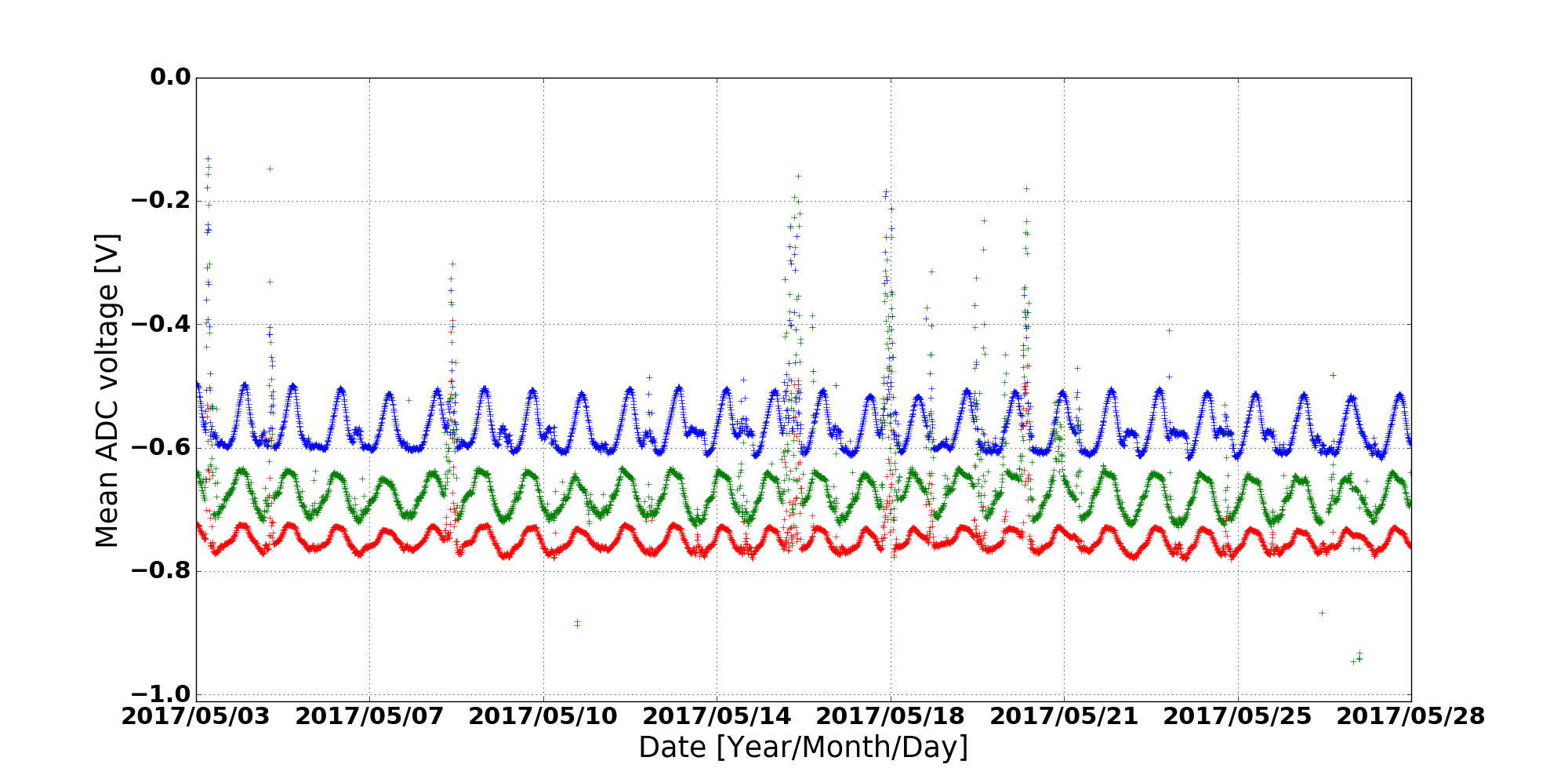}
 \caption{Monitoring measurement of the mean voltage of the signal in a GRANDProto35 radio-detection unit during a period of 25 days at the array site, for West-East (blue), South-North (green), and vertical (red) antenna-arm channels.  The periodic fluctuations correspond to the daily transit of the Galactic plane in the antenna field of view. Figure taken from \Ref\ \cite{Gou:2017ICRC}.}
 \label{fig:GP35} 
\end{figure}

Figure\ \ref{fig:GP35} shows the signal recorded during 25 days by one GRANDProto35 units deployed at the array site.  In the data acquisition (DAQ) chain, first, the raw signals collected by the antennas are filtered through a sharp 30--100 MHz~passive analog filter.  Then, the trigger condition is evaluated by comparing the filtered signals to a threshold value set remotely by a human operator.  The envelopes of the signals that satisfy the trigger condition are digitized at 50~million samples per second with a 12-bit ADC.  A 3 $\times$ 3.6~$\mu$s subset of the digitized signals is sent to the central DAQ via optical fiber, together with a GPS tag of the trigger time.  We have tested that this DAQ system achieves 100\% detection efficiency for trigger rates up to 20~kHz\ \cite{Gou:2017ICRC}.  Therefore, it can record all transient signals under standard background conditions at the array site, which will significantly improved the air-shower detection efficiency compared to TREND.  Figure\ \ref{fig:GP35} shows that the DAQ system is stable in the long run under real conditions.

To cross-check the efficiency of the antennas, GRANDProto35 also includes an autonomous surface array of particle detectors, co-located with the antenna array.  It consists of scintillator tiles of dimensions $70.7\ \text{cm} \times 70.7\ \text{cm} \times 2\ \text{cm}$ connected to Hamamatsu R7725 PMTs.  To match the expected radio-detection efficiency, the scintillators are mounted on a support structure that is tilted North.
This setup optimizes the effective area to showers inclined by about $40^\circ$ coming from the North.  The DAQ chain and trigger logic of the scintillator array are fully independent from the radio array: analog signals collected by the PMTs are sent via optical fiber to the central DAQ, where they are digitized in real time.  If a pulse is observed simultaneously in the signals from three or more scintillators, the signals are written to disk.  Offline comparison between scintillator and radio data will quantify the radio-detection efficiency and the contamination by background events.

\subsection{GRANDProto300 (2020)}
\label{section:GRANDStages-GRANDProto300}

GRANDProto300 will be an array of 300 antennas deployed over 300~km$^2$.  It will act as pathfinder for the later, larger stages of GRAND.  Its main goal is to demonstrate the viability of detection principle of GRAND.  This means demonstrating that it is possible, from radio data alone, to trigger on nearly horizontal air showers, separate them from the background, and reconstruct the properties of the primary particles with a precision similar to standard techniques used for cosmic-ray detection.  Because GRANDProto300 will not be large enough to detect cosmogenic neutrinos, the viability will be tested using instead air showers initiated by very inclined UHECRs, thus providing an opportunity to do cosmic-ray science.

\medskip

{\it Site.---} Eight candidate sites have been surveyed to host GRANDProto300 in the Chinese provinces of XinJiang, Inner Mongolia, Yunnan, and Gansu. Six of them comply with the requirements for radio-quietness; see Section \ref{section:array_layout}.  We are presently in the last phase of the site selection process, evaluating additional parameters such as ease of access, availability of infrastructure, support by local authorities, and possible extension to the GRAND10k stage.  A decision will be made before the end of 2018.

\begin{figure}[t!]
 \begin{center}
  \includegraphics[trim = 0cm 0.4cm 1.9cm 0.8cm, clip=true, width=\columnwidth]{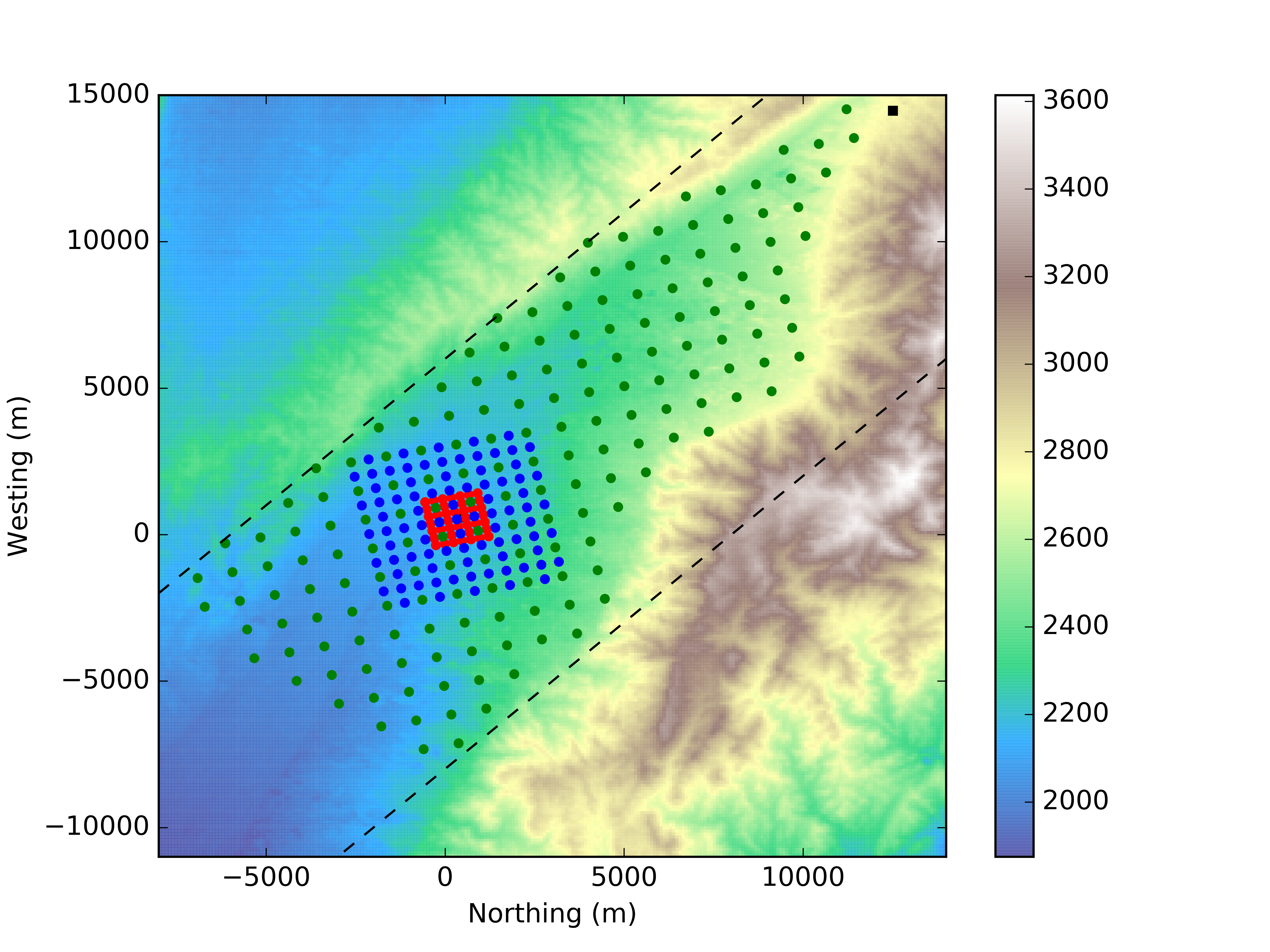}
 \end{center}
 \caption{\label{fig:layoutYiwu}Possible layout for the GRANDProto300 radio array at one of the candidate sites.  The layout includes 135 antennas deployed on a 1-km square grid (green), with two denser in-fills, one containing 116 antennas on a  of 500-m spacing (blue), and one containing 49 antennas on a 250-m spacing (red).}
\end{figure}

\medskip

{\it Layout.---} GRANDProto300 will consist of 300 antennas. It will be the largest radio array for autonomous air-shower detection, almost 10 times larger than GRANDProto35 and twice as large as the present phase of AERA\ \cite{Abreu:2012pi}.  The baseline layout is a square grid with a 1~km inter-antenna spacing, just as for later stages; see Section \ref{section:array_layout}.

Figure\ \ref{fig:layoutYiwu} shows one of the possible layouts considered for GRANDProto300.  A denser in-fill will improve statistics down to shower energies of $\sim$$10^{7.5}$~GeV and test the dependence of the array performance as a function of the density of detection units.  The exact layout of the array will be defined through dedicated simulations, taking into account the science goals of the experiment, presented below, and the physical properties of the selected site.

\medskip

{\it Antennas.---} The antenna used in GRANDProto300 will be a first version of the \HA, which is designed to improve the sensitivity close to the horizon; see Section \ref{section:antenna_design}.  Based on the experience in GRANDProto300, the antenna design will be optimized for the next stages.  

\begin{figure}[t!]
 \centering
 \includegraphics[width=\columnwidth]{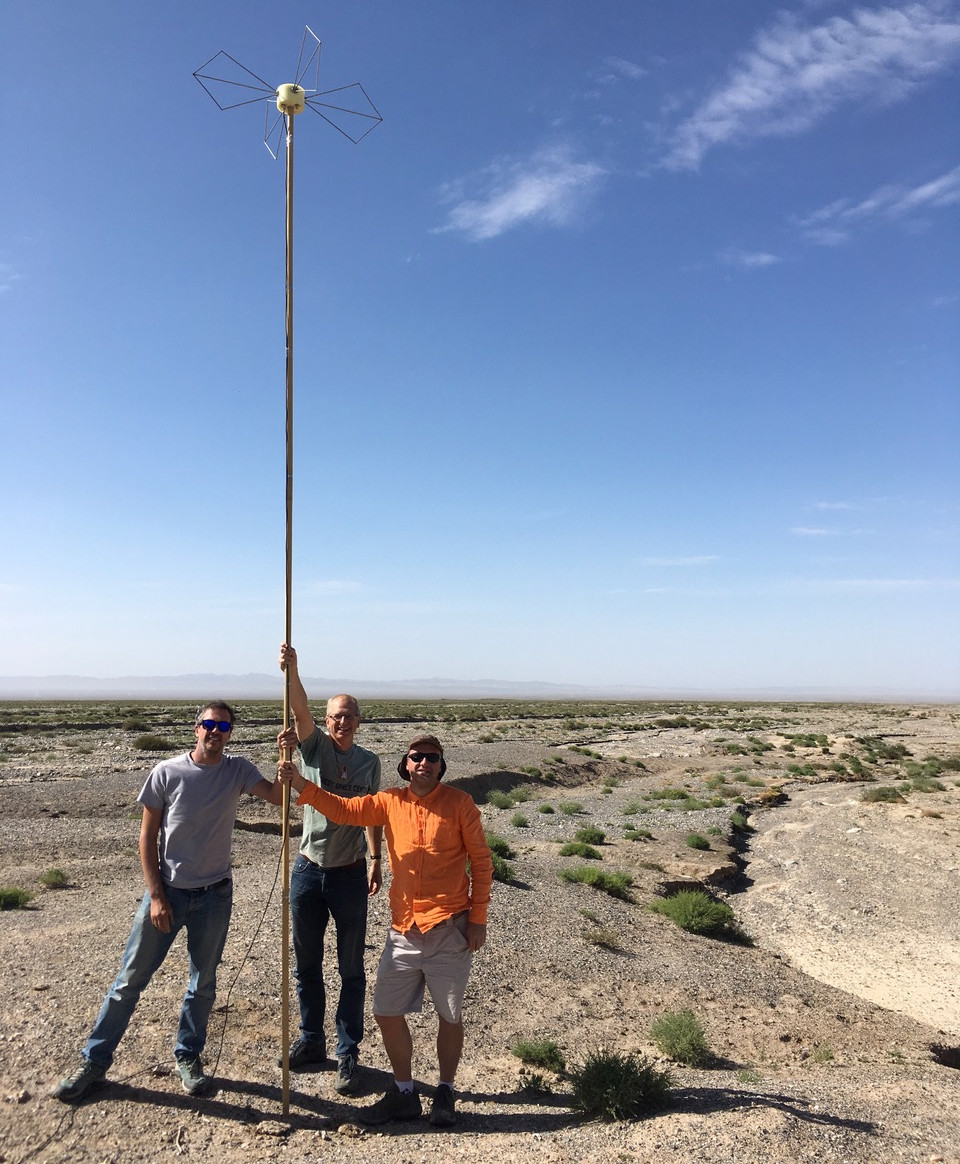}
 \hspace*{0.2cm}
 \caption{A prototype of the \HA\, in test during a GRANDProto300 site survey. Photo by Feng Yang.}
 \label{fig:HApic}
\end{figure}

Figure\ \ref{fig:HApic} shows a prototype version of the \HA\ which was successfully tested during site surveying in summer of 2018.

\medskip

{\it DAQ.---} In a first phase, the GRANDProto300 DAQ system will be based on full sampling, for each trigger, of a $\sim$3~$\mu$s subset of the signals from the X, Y, and Z antenna channels using 14 bits at a rate of 500~million samples per second, following passive analog filtering in the 50--200~MHz band.  Adjustable digital notch filters will allow to reject continuous-wave emitters that may appear in this band.  We will transfer data via WiFi, which allows for a throughput of 38~MB~s$^{-1}$ per antenna, sufficient for our needs.

Three consecutive levels are designed to progressively reduce the background:
\begin{itemize}
 \item 
  The zeroth-level trigger (T0) acts locally at the detection units.  It directly compares each channel amplitude to an adjustable threshold value.  T0 events are time-stamped with ns precision using a local GPS unit\footnote{GPS-based Timing Considerations with u-blox 6 GPS receivers --- Application Note: \href{https://www.u-blox.com/}{https://www.u-blox.com/}} 
 \item
  The first-level trigger (T1) acts locally at the detection units.  It evaluates the duration and structure of the time traces, and the signal polarization.  T1 time-stamps are sent to the central DAQ for evaluation.  In our estimates, we assume that the bandwidth needed for this will be 1~MB~s$^{-1}$.  Measurements performed during the GRANDProto35 R\&D phase show that a large fraction of background events could exhibit a relatively constant polarization at fixed antenna locations, a feature that could help to identify them; see \Fig\ 5 in \Ref\ \cite{Gou:2017ICRC}.
 \item
  The second-level trigger (T2) acts on the time-stamps sent by the T1 triggers and searches for simultaneous detection in neighboring antennas.  If such a detection is found, the DAQ pulls the time traces of the participating detection units.  Extrapolation from TREND results yields a T2 rate of only a few mHz for GRANDProto300\ \cite{Martineau-Huynh:2017bpw}.  For each T2 issued, a 2~$\mu$s sample is collected for the 3 channels of each of the detection units involved in the event.
\end{itemize}

The design of the DAQ will also allow to perform searches of fast radio bursts and giant radio pulses (GP); see Sections\ \ref{section:FRB} and \ref{section:Giant_radio_pulses}.  To achieve this, the power spectral density (PSD) of the signals must be computed in the 100--200~MHz range with a 25~kHz resolution every 10~ms.  After subtracting the Galactic radio background from the PSD --- \ie, spectrum whitening --- the signal can be digitized using a single byte for each frequency value.  Consequently, the corresponding data rate for one antenna is 200~KB~s$^{-1}$.

FPGAs in the detection units will allow us to treat the signal to make EoR measurements (see Section\ \ref{section:EoR}), \ie, removal of RFI, foreground subtraction, and spectrum whitening. The resulting signal will be averaged at the detection unit level on time scales of minutes or longer, and sent to the DAQ, thus representing a minor contribution to the total data volume.

In a second phase, the GRANDProto300 DAQ system will evolve with the next construction stage --- GRAND10k --- in mind.  In this phase, sophisticated data treatment techniques --- \eg, adaptive filtering and machine learning\ \cite{Erdmann:2019nie, FuhrerARENA:2018} --- will be tested at the level of the detection unit.

\medskip

{\it Monitoring and calibration.---}  Amplitude calibration will use the well-known background sky emission, modulated daily, as in TREND\ \cite{LeCoz:2017ICRC}. Timing calibration will use airplane radio tracks, as in AERA\ \cite{PierreAuger:2016zxi}.  Further, some antennas could be used in transmitting mode to calibrate simultaneously amplitude and timing.  Trigger rates, power consumption, battery level, air pressure, and temperature at each unit will be collected periodically to monitor the status of the array, jointly with the PSD information, for a limited cost in terms of data transmission.

\medskip

{\it Power supply.---}  The consumption of one detection unit is estimated at 10~W.  Thus, a 100-W solar panel, coupled to a battery, should allow for its continuous operation.

\medskip

{\it Mechanics.---}  The mechanical integration of the detection unit is under study.  The antenna weighs under 2~kg, the full unit weighs under 30~kg, and the typical size of a 100-W solar panel is 0.7~m$^2$.  Thus, we can adopt standard solutions used in electric power distribution and deploy the units on 5~m-tall wooden utility poles buried 1~m underground.  In addition, by placing the electronics atop a pole, we reduce the ecological footprint of the detector and the risk of damage by wild fauna and cattle.

\medskip

{\it Ground array.---} We plan to complement the GRANDProto300 radio array with a stand-alone, autonomous ground array of particle detectors deployed at the same location.  We present the science motivation for this array below.  One possibility is to use water-Cherenkov tanks similar to those in Auger and HAWC, but optimized for the detection of inclined showers.  Based on the performance of Auger\ \cite{Aab:2017njo} and HAWC\ \cite{Abeysekara:2017yqc}, we estimate deploying 500 units over the 200~km$^2$ area of the array.  The exact design and layout of the array requires a dedicated study.  The ground array will be synchronized in time with the radio array, so that they can simultaneously detect the electromagnetic and muon components of showers.

\medskip

{\it Science program.---}  The objectives of GRANDProto300 are not limited to validating the radio-detection technique for very inclined showers.  Integral to this stage is also an appealing science program in cosmic rays, gamma rays, radioastronomy, and cosmology.

For cosmic rays, the hybrid detection strategy of GRANDProto300 will measure separately the shower components\ \cite{Holt:2017dyo}.  The electromagnetic component will be measured by the radio array and the muon component will be measured by the ground array of particle detectors.  This is possible because, for very inclined showers, muons are the only particles that reach the ground\ \cite{Zas:2005zz}.
The number of muons --- and also the depth of shower maximum, inferred from radio --- is correlated with the mass of the primary cosmic ray\ \cite{Aab:2016vlz}, while the energy in radio is correlated with the energy of the primary cosmic ray\ \cite{Aab:2016eeq}.  

\begin{figure}[t!]
 \centering
 \includegraphics[width=\columnwidth]{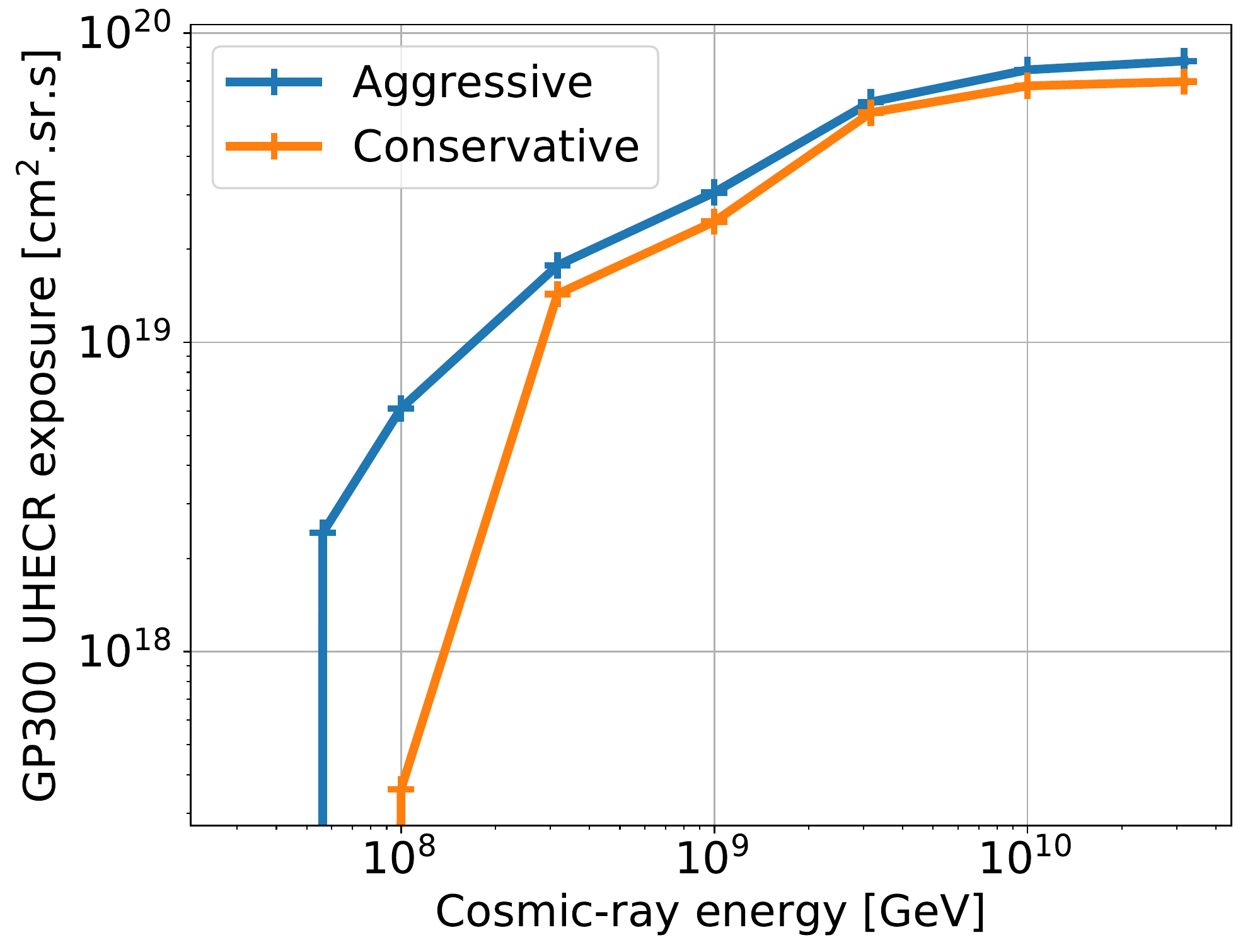}
 \hspace*{0.2cm}
 \caption{Simulated 1-year UHECR exposure for the GRANDProto300 array, assuming the layout in \figu{layoutYiwu}. The aggressive and conservative thresholds correspond to a minimum peak-to-peak amplitude of 30 and 75~$\mu$V, respectively, simultaneously measured in at least five units; see Section \ref{section:GRANDDesignPerf-Performance-NuDetection}.  Event rates are, respectively, $1.2 \cdot 10^6$ and $2.5 \cdot 10^6$ events per year. }
 \label{fig:GP300_Spectrum}
\end{figure}

Figure\ \ref{fig:GP300_Spectrum} shows that, after only one year of operation, GRANDProto300 will have recorded more than $10^5$ UHECRs with $10^{7.5}$--$10^{9}$~GeV.  At these energies, cosmic rays diffuse on cosmic magnetic fields, so no directional excess is expected.  Indeed, in the Southern Hemisphere, Auger has not found large-scale anisotropies at these energies.  However, TA has not monitored the Northern Hemisphere with the same exposure.  Therefore, GRANDProto300 is in a privileged position to discover or constrain the existence of a low-energy, large-scale Northern-Hemisphere anisotropy of size $10^{-5}$, should it exist, thus discovering or constraining the existence of nearby UHECR sources.

The precise measurement of energy and mass composition, and the large event statistics of GRANDProto300 place it in a privileged position to study the transition between Galactic and extragalactic cosmic rays.  The transition is expected to occur between $10^{8}$ and $10^{9}$~GeV\ \cite{Nagano:2000ve, LetessierSelvon:2011dy, Grenier:2015egx, Dawson:2017rsp}.
GRANDProto300 will be able to infer the distributions of arrival directions of light and heavy primaries separately, and their variation as a function of energy.

GRANDProto300 is a suitable setup to test alternative radio-based methods for finding the identity of the primary\ \cite{Billoir:2015cua} and the validity of different hadronic interaction models.  So far, measurements made by cosmic-ray experiments in the energy range of GRANDProto300 have found intriguing discrepancies\ \cite{AbuZayyad:1999xa, Fomin:2016kul, Dembinski:2017zkb, Bogdanov:2018sfw, Bellido:2018toz}. 

Above $10^{9.5}$~GeV, there is currently a significant excess in the number of muons measured, compared to the number expected from air-shower simulations\ \cite{Aab:2014pza}.  Below $10^8$~GeV, there is no clear evidence of this discrepancy\ \cite{Fomin:2016kul, Dembinski:2017zkb, Bogdanov:2018sfw}.  In-between, because the mass composition is changing from heavy to light\ \cite{AbuZayyad:1999xa, Plum:2018fig}, the number of muons should instead decrease.  The discrepancy is difficult to solve because most of these experiments measure only the muon component.  Therefore, they deduce the shower energy indirectly\ \cite{Bellido:2018toz} or using simulations that depend on the choice of hadronic interaction model\ \cite{Fomin:2016kul, Dembinski:2017zkb, Bogdanov:2018sfw}.  
Because the hybrid GRANDProto300 detector will independently measure the electromagnetic component --- which depends less strongly on hadronic interaction models --- and the muon component, it could disentangle the differences in shower development due to different choices of hadronic interaction models and mass compositions, thus alleviating one of the main sources of uncertainty when inferring the mass composition of UHECRs in ground arrays\ \cite{Aab:2016enk, Aab:2017cgk}.

The GRANDProto300 ground array of particle detectors will be used as a veto to search for air showers initiated by UHE gamma rays.  In showers initiated by gamma rays with $\theta_z \geq 65^{\circ}$, the electromagnetic component is dominant and is fully absorbed by the atmosphere before reaching the ground.  In comparison, in showers initiated by cosmic rays, surviving muons will reach the particle detectors.  The performance of this setup in separating cosmic rays from gamma rays is promising.  Preliminary simulations indicate that the separation will be close to 100\% for $65^{\circ} \leq \theta_z \leq 85^{\circ}$ and energies above $10^{9}$~GeV.  In a sample of 10\,000 showers at these energies, collected in 2 years, if no gamma-ray events are identified, then we would place a limit on the fraction of gamma ray-initiated showers of 0.03\% at the 95\%~C.L., while the current best limit is 0.1\%\ \cite{Niechciol:2017vqf}.

Finally, GRANDProto300 will search for FRBs, GPs, and study the EoR; see Section\ \ref{section:cosmology_radioastronomy}.  A simple down-scaling from the final GRAND sensitivity computed in the simulations in Section\ \ref{section:FRB} results in a 750~Jy sensitivity threshold for GRANDProto300 in the 100--200~MHz band.  This would allow for detection of GPs from the Crab.  The nearly full-sky field of view and 100\% live-time of the observatory could allow for full-sky surveys of sources of similar intensity across the sky.  For measurements of the 21-cm signature from the EoR, already 30 antennas would be enough to reach the required sensitivity; see Section\ \ref{section:EoR}.


\subsection{GRAND10k (2025)}
\label{section:GRANDStages-GRAND10k}

GRAND10k will be the first large sub-array of GRAND, and the first construction stage that is sensitive to UHE neutrinos.  It will consist of 10\,000 antennas deployed over a 10\,000~km$^2$ area carefully selected for its suitability for the detection of neutrino-initiated air showers --- a {\it hotspot}.

Simulations show that an area of 10\,000~km$^2$ centered on the southern rim of the Tian Shan mountains --- labeled {\it HotSpot 1} --- would yield an integrated sensitivity to UHE neutrinos of $8 \cdot 10^{-9}$~GeV~cm$^{-2}$~s$^{-1}$~sr$^{-1}$ after 3 years, assuming an aggressive detection threshold; see Section \ref{sensitivity}.  GRAND10k will be able to probe flux models of cosmogenic neutrinos made by light UHECRs (see Section \ref{section:uhe_neutrinos}), with a sensitivity comparable to the planned final configurations of ARA and ARIANNA.

GRAND10k will be the largest UHECR detector built, with an area 3 times larger than Auger or the planned TA$\times$4\ \cite{Kido:2016kep}, and an aperture twice that of Auger --- around 12\,000~km$^2$~sr for energies above $10^{10}$~GeV.  The field of view of GRAND10k overlaps with Auger and TA$\times$4, which allows for cross-checking with both.  Further, the radioastronomy and cosmology measurements available to GRANDProto300 will be improved in GRAND10k.
  
The design of GRAND10k will be informed by GRANDProto300, with further optimization of power consumption --- aiming a 5~W per detection unit --- triggers, and data transfer --- \eg, using WiMax or a smart mesh network.
Data handling in GRAND10k will build on the precise measurement of the rate and features of background events in GRANDProto300.  The strategy will likely include on-board treatment of the triggered signals, in order to optimally select, already at the detection unit, what data must be transmitted for offline analysis.  These developments will be tested on GRANDProto300, which will gradually turn from an instrument dedicated to the goals detailed in the previous section to a test bench for GRAND10k.


\subsection{GRAND200k (2030s)}
\label{section:GRANDStages-GRAND200k}

GRAND200k will be the full planned configuration of GRAND.  Following a modular design, it will consist of 20 independent arrays of 10\,000 antennas each --- replicas of GRAND10k --- built at separate geographical locations that are hotspots for neutrino detection.  Combined, they will total 200\,000 antennas covering 200\,000~km$^2$.  We do not expect important design changes compared to GRAND10k; the antennas, electronics, triggers, and data collection will have been validated at that stage or earlier.  GRAND200k will address all of the physics goals from Sections\ \ref{section:uhe_messengers} and \ref{section:cosmology_radioastronomy}, including reaching sensitivities to cosmogenic neutrino fluxes of the order of $10^{-10}$~GeV~cm$^{-2}$~s$^{-1}$~sr$^{-1}$.  Below, we give preliminary remarks on the technical aspects of this setup, though by necessity they cannot be too specific at this early stage.

The main challenge in building GRAND200k is its scale, in terms of cost, deployment, and maintenance.  The appropriate response to this challenge lies in the size of the project itself: the scale of the project forces us to adopt an industrial approach to building GRAND200k.

For the electronics, developing a fully integrated application-specific integrated circuit (ASIC) board --- an expensive solution when building only a few thousand boards --- is likely the cheapest solution to build 200\,000 units, while providing reduced power consumption --- a factor of 10 is typical --- and increased reliability.

A precise and standardized procedure will have to be defined for detector transportation and installation, and factors linked to detector aging have to be carefully identified.  For this purpose, the expertise acquired during previous construction stages will be crucial.


\subsection{Data policy}
\label{section:data_policy}

We aim to provide public access to detected events recorded by GRAND after a reasonable amount of time to allow the astroparticle community at large to interact with and benefit from our results.  We plan to work with pointing electromagnetic telescopes to provide alerts to perform multi-messenger physics.  We will implement a long-term storage policy to ensure the longevity of the data.


\subsection{Outreach}
\label{section:outreach}

To ensure the long-term viability of the detector, it is key to have the support of the local community living in the site where the detector will be deployed.  Ultimately, the community should consider hosting GRAND to be a source of pride and an asset.  To achieve this, public outreach in the hosting community has proven to be an effective strategy in large-scale experiments such as Auger\ \cite{Timmermans:2017ICRC}.


\section{Summary}
\label{section:Summary}

The Giant Radio Array for Neutrino Detection (GRAND) aims to solve the long-standing mystery of the origin of ultra-high-energy cosmic rays (UHECRs), \ie, cosmic rays with energies above $10^8$~GeV.  To achieve this, GRAND will look for impulsive radio signals in the 50--200~MHz range, emitted in the atmosphere by extensive air showers (EAS) initiated by UHECRs and by UHE gamma rays and neutrinos that are born from UHECR interactions.  

GRAND will be the largest UHE observatory.  Its design is scalable and modular.  
Because radio antennas are relatively inexpensive, robust, and easy to maintain, they are suitable to instrument the large areas needed to contain the extended radio footprints of EAS on the ground, reach sensitivity to potentially tiny fluxes of UHE neutrinos and gamma rays and, and collect large UHECR statistics.  

The ambitious design and goals of GRAND represent the culmination of substantial progress experienced by the field of radio-detection of air showers over the past years.  Experiments have demonstrated that arrays of radio antennas can autonomously detect showers and that radio data alone is sufficient to reconstruct the properties of the primary particles that initiate them.  GRAND will gear these advances, for the first time, toward detecting UHE neutrinos. 

GRAND has a staged construction plan, designed not only to progressively validate experimental techniques, but also to achieve important science goals in themselves. 
The first prototype stage, GRANDProto35, is an array of 35 antennas and scintillators, and is currently being deployed in the Tian Shan mountains, in China.  Following that, GRANDProto300 (2020), the pathfinder of the project, will consist in 300 antennas and a ground array of particle detectors.  Its technology goal will be to demonstrate the autonomous radio-detection of very inclined air showers with high efficiency and background rejection.  Its science goals will be to study the transition from Galactic to extragalactic cosmic rays with large statistics and to tackle the muon deficit problem by independently measuring the electromagnetic and muon components of showers.  GRAND10k (2025), will be the first large sub-array of GRAND.  It will consist of 10\,000 antennas deployed over 10\,000~km$^2$, making it already the largest ground-based UHE observatory built.  Its neutrino sensitivity will be comparable to that of envisioned upgrades of existing in-ice radio neutrino detectors, and sufficient to detect cosmogenic neutrinos if their flux is close to their current upper limit.   Finally, GRAND200k (2030s) will realize the full potential of GRAND.  It will be made up of several separate replicas of GRAND10k, deployed at different locations, instrumenting a total area of 200\,000~km$^2$.

In the detection of UHECRs, GRAND200k will have an effective area 10 times that of Auger.  It is expected to reach a resolution in $X_{\max}$ of 20~g~cm$^{-2}$, comparable to that of particle and fluorescence shower detectors, and sufficient to make precise studies of cosmic-ray mass composition.  Already with GRANDProto300 it will tackle important open issues in UHECRs.  In later stages, the high  statistics collected will resolve small-scale anisotropies and features near the high-energy end of their spectrum.

In the detection of UHE gamma rays, GRAND200k will be sensitive to a fraction of gamma ray-initiated showers down to 0.03\% in 2 years of operation, a factor-of-3 improvement over current limits.

In the detection of UHE neutrinos, GRAND will push back the energy frontier a thousand-fold, to the EeV scale.  GRAND200k will be sensitive to a diffuse neutrino flux of $4 \cdot 10^{-10}$~GeV~cm$^{-2}$~s$^{-1}$~sr$^{-1}$ for the first time, in 3 years of operation, enough to detect the long-sought cosmogenic neutrinos even if their flux is at the level of pessimistic predictions.  If their flux is higher, GRAND10k could already detect them, and GRAND200k could collect in excess of 100 events in 3 years.  Further, GRAND could discover the first point sources of UHE neutrinos, owing to its angular resolution of a fraction of a degree and large sky coverage.  In doing so, it would kickstart UHE neutrino astronomy, an essential component of multi-messenger astronomy.

GRAND will also achieve important goals in radioastronomy and cosmology.  By monitoring about 80\% of the sky every day, it will be sensitive to astrophysical radio transients --- fast radio bursts and giant radio pulses --- and potentially collect an unprecedented number of them.   It will probe whether fast radio bursts extend down to 200~MHz and below.  It will record giant pulses from the Crab pulsar above 5~Jy at 200~MHz with a rate of about 200 per day.  By incoherently adding antenna signals, GRAND will be able to measure the global 21-cm signature of the beginning of the epoch of reionization.

Currently, numerical and experimental work is ongoing on technological development and background rejection strategies in GRAND.  We have designed and tested an antenna with higher sensitivity towards the horizon.  We have developed an end-to-end, sophisticated numerical simulation chain tailored to GRAND --- from incoming neutrino, through shower development, to radio-detection and reconstruction under real topographical conditions.  Future simulations will improve the results presented here.

At the beginning of the era of multi-messenger astroparticle physics, the ambitious science goals and design of GRAND place it in a privileged position to become one of the leading instrument in the ultra-high-energy range.



\section*{Acknowledgements}
\label{sec:Acknowledgements}
\addcontentsline{toc}{section}{Acknowledgements}

The GRAND project is supported by the APACHE grant No.\ ANR-16-CE31-0001 of the French Agence Nationale de la Recherche, 
the France-China Particle Physics Laboratory, the China Exchange Program from the Royal Netherlands Academy of Arts and Sciences and the Chinese Academy of Sciences, 
the Key Projects of Frontier Science of the Chinese Academy of Sciences under grant No.\ QYZDY-SSW-SLH022, 
the Strategic Priority Research Program of Chinese Academy of Sciences under grant No.\ XDB23000000, and the National Key R\&D Program of China under grant No.\ 2018YFA0404601.  
RAB is supported by grant No.\ 2017/12828-4, S\~{a}o Paulo Research Foundation (FAPESP).  
MB is partially supported from NSF grants PHY-1404311 and PHY-1714479.
MB and PD are supported by Danish
National Research Foundation (DNRF91), Danmarks Grundforskningsfond grant No.\ 1041811001, and {\sc Villum Fonden} grant No.\ 13164.  
WCJ is supported by grant No.\ 2015/15735-1, S\~{a}o Paulo Research Foundation (FAPESP).  
QBG is supported by the Natural Science Foundation of China grant No.11375209.
KDV is supported by the Flemish Foundation for Scientific Research (FWO-12L3715N -- K.\ D.\ de Vries). 
CT is supported by the Netherlands Organisation for Scientific Research (NWO).
XPW is supported by the Key Projects of Frontier Science of Chinese Academy of Sciences, grant No.\ QYZDY-SSW-SLH022, and the Strategic Priority Research Program of Chinese Academy of Sciences, grant No.\ XDB23000000. 
JLZ is supported by the  Natural Science Foundation of China, grant No.\ 11505213, ``Data analysis for radio detection array at 21CMA base.''  
We thank Markus Ahlers, Daniel Ardouin, Johannes Bl\"umer, Jordan Hanson, Andreas Haungs, Naoko Kurahashi, Pascal Lautridou, Fran\c{c}ois Montanet, Angela Olinto, Andres Romero-Wolf, Subir Sarkar, Abigail Vieregg, and Stephanie Wissel 
for useful discussion and comments on the manuscript; Feng Yang and the staff in Ulastai; 
the engineers who developed the GRANDProto35 electronics, Julien Coridian, Jacques David, Olivier Le Dortz, David Martin, and Patrick Nayman;
and the groups of Zhang Fushun and Guo Lixin, who built the GRANDProto35 antennas and the \HA\ prototypes.
The GRAND neutrino simulations were run through the France-Asia Virtual Organisation on the IN2P3 computing center, the GRIF-LPNHE computing grid, the IHEP computing center.  
Part of the simulations was performed on the computational resource ForHLR I funded by the Ministry of Science, Research and the Arts Baden-W\"urttemberg and DFG ("Deutsche Forschungsgemeinschaft").  
The GRAND cosmic-ray simulations were run on the Horizon Cluster, hosted by the Institut d'Astrophysique de Paris.  
The SRTMGL1 (v3) topographical data used in this study were retrieved from the online USGS EarthExplorer and NASA Earthdata Search tools, 
courtesy of the NASA EOSDIS Land Processes Distributed Active Archive Center (LP DAAC), USGS/Earth Resources Observation and Science (EROS) Center, Sioux Falls, South Dakota.  
Figures \ref{fig:grand_propagation} and \ref{fig:grand_det_principle} were designed for the GRAND Collaboration by Ingrid Delgado.



\pagestyle{bib}
%


\end{document}